\documentclass[11pt]{article}
\usepackage{setspace}

\usepackage{titling}
\usepackage{booktabs}
\usepackage{array}
\usepackage{tikz}
\usetikzlibrary{arrows.meta}
\usepackage{natbib}
\usepackage{calc}
\usepackage{subcaption}
\usepackage[a4paper, margin=1in]{geometry}
\usepackage{rotating}
\usepackage{booktabs}
\usepackage{mathrsfs}
\usepackage{fancyhdr}
\usepackage{amsfonts}
\usepackage{amstext}
\usepackage[T1]{fontenc}     
\usepackage{lmodern}         
\usepackage{amsmath,amsthm} 
\usepackage{amssymb}
\usepackage{algorithm}
\usepackage{algpseudocode}
\usepackage{adjustbox}
\usepackage{hyperref}
\hypersetup{ hidelinks }
\hypersetup{breaklinks=true}
\usepackage{bookmark}
\usepackage{setspace}
\usepackage{longtable}
\usepackage{float}
\usepackage{threeparttable}
\usepackage{fullpage}
\usepackage{times}
\usepackage[utf8]{inputenc}
\usepackage{units}

\usepackage{nicefrac}
\usepackage{color}
\usepackage{tabularx}
\usepackage{pdflscape}

\newtheorem{theorem}{Theorem}

\newtheorem{corollary}{Corollary}
\newtheorem{lemma}{Lemma}
\newtheorem{remark}{Remark}
\newtheorem{proposition}{Proposition}
\newtheorem{definition}{Definition}

\pretitle{\begin{center}\large\bfseries}
\posttitle{\end{center}}
\preauthor{\begin{center}
\large \lineskip 0.5em%
\begin{tabular}[t]{c}}
\postauthor{\end{tabular}\par\end{center}}
\predate{\begin{center}\large}
\postdate{\par\end{center}\vskip 0.5em}

\newcommand{\HRule}{\rule{\linewidth}{0.5mm}}
\newcommand{\BRule}{\rule{\linewidth}{0.25mm}}
\title{\HRule \\[0.1mm]
Kernel Two-Sample Testing via Directional Components Analysis\thanks{Li acknowledges support from the Research Development Fund (RDF-23-02-022) of the Xi'an Jiaotong--Liverpool University. Song acknowledges support from the National Natural Science Foundation of China (Grant Numbers 72373007 and 72333001).} \\[0.1mm]
\BRule}

\author{Rui Cui\thanks{cuirui.econ@outlook.com} \\ \small Guangzhou Nanfang College
\and 
Yuhao Li\thanks{yuhao.li@xjtlu.edu.cn} \\ \small Xi'an Jiaotong--Liverpool University
\and 
Xiaojun Song\thanks{sxj@gsm.pku.edu.cn} \\ \small Peking University
}

\date{} 

\begin{document}
\maketitle
\begin{abstract}
    Standard kernel two-sample tests, such as those based on the Maximum Mean Discrepancy (MMD), aggregate squared differences across all directions in a Reproducing Kernel Hilbert Space (RKHS). However, in finite samples, trailing directional components are noisy, which degrades test power. We propose a novel kernel-based test that resolves this by truncating the spectral decomposition of the MMD, retaining only the well-estimated leading eigen-directions. By aggregating these robust components, our method achieves superior power and robustness, particularly in high-dimensional and unbalanced settings. Furthermore, we introduce a computationally efficient parametric bootstrap procedure for approximating critical values, which is theoretically justified and significantly faster than permutation-based alternatives. Extensive simulations and empirical studies demonstrate that our method maintains strict Type I error control while delivering higher power than existing MMD-based tests.

\end{abstract}

\vspace{1em} 


\pagenumbering{arabic} \setcounter{page}{1} \setcounter{footnote}{0}

\section{Introduction}
\label{sec:intro}

We propose a novel kernel-based two-sample test that leverages the spectral decomposition of the maximum mean discrepancy (MMD) to identify and utilize well-estimated directional components in the reproducing kernel Hilbert space (RKHS). This approach is motivated by the observation that finite-sample estimation quality varies significantly across directional components: those associated with leading eigen-directions are estimated more reliably than those corresponding to smaller eigenvalues. By focusing on these high-quality components, the proposed test achieves higher power and improved robustness. Furthermore, we develop a computationally efficient, theoretically justified parametric bootstrap procedure to approximate critical values, offering significant speedups over permutation-based alternatives.

Given two probability distributions $P$ and $Q$ defined on a Hausdorff topological space $\mathcal{Z}$, the two-sample testing problem seeks to determine whether $H_0: P = Q$ versus $H_1: P \neq Q$, based on independent random samples $\{x_i\}_{i=1}^n$ and $\{y_j\}_{j=1}^m$ drawn from $P$ and $Q$, respectively. Throughout this paper, we assume the sample proportion $\hat{p} = n/N$ converges to $p \in (0,1)$ as the total sample size $N = n+m \to \infty$.

Classical nonparametric tests, such as the Kolmogorov--Smirnov and Cramér--von Mises tests, are widely used for univariate data, with multivariate extensions extensively studied \citep{weiss1960two, bickel1969distribution}. Recently, kernel-based approaches have gained popularity for two-sample testing due to their flexibility and strong power in detecting diverse distributional differences \citep{gretton2012kernel, gretton2012optimal, sejdinovic2013equivalence, chatterjee2025boosting}.

Kernel-based methods embed probability distributions into an RKHS $\mathcal{H}$ using a characteristic kernel $k$, enabling distribution comparison via their mean embeddings. Formally, let $\mu_P = \mathbb{E}_{X}[k(X,\cdot)]$ and $\mu_Q = \mathbb{E}_{Y}[k(Y,\cdot)]$ be the mean embeddings of $P$ and $Q$, respectively. If $k$ is characteristic, the mean map is injective, ensuring $\mu_P = \mu_Q$ if and only if $P = Q$. Consequently, the two-sample problem reduces to testing $H_0: \mu_P - \mu_Q = 0$ in the RKHS. The MMD \citep{gretton2012kernel} is defined as the squared RKHS norm of this difference: $\text{MMD}^2(P, Q) = \|\mu_P - \mu_Q\|_{\mathcal{H}}^2$.

Although MMD enjoys desirable theoretical properties, its finite-sample performance can be poor. We argue that this inefficiency partly arises because the standard MMD estimator aggregates squared norms across all RKHS directions, including those with poor estimation quality. Spectrally, the mean embedding difference decomposes along the eigenfunctions of the kernel integral operator with respect to the mixture distribution $\rho = p P + (1-p) Q$. The estimation quality of these directional components is governed by the eigengap: leading components (large eigenvalues) are estimated reliably, whereas trailing components are noisy.

To address this, we propose a test statistic focusing on a subset of well-estimated directional components. Instead of summing over all dimensions, our statistic truncates the spectral expansion at dimension $d$, capturing only the reliably estimated leading components. We provide rigorous theoretical justification for this truncation by deriving finite-sample error bounds for eigenfunction and directional-component estimation using the kernel's spectral properties. Furthermore, we establish the asymptotic null distribution of the proposed statistic, which converges to a weighted sum of independent chi-squared variables. This closed-form characterization facilitates a computationally efficient parametric bootstrap for critical values, offering significant speedups over permutation-based methods while maintaining valid Type I error control.

Beyond improved finite-sample performance, we investigate the method's robustness to the choice of kernel. While standard kernel tests can be sensitive to hyperparameters, population-level directional components are stable under perturbations to the kernel parameters. By restricting the test to leading spectral components with large eigengaps, our statistic inherits this stability, rendering it less sensitive to specific kernel choices than full-rank MMD tests.

It is worth noting that recent literature has recognized that the standard MMD is suboptimal in finite samples. To mitigate this inefficiency, \cite{hagrass2024spectral} proposed a spectral regularized kernel two-sample test (referred to herein as SpectraRegu). While both SpectraRegu and our method leverage the spectral decomposition of the kernel operator to improve test power, they diverge fundamentally in their mechanistic approaches, theoretical foundations, and computational implementations. Specifically, SpectraRegu is rooted in inverse problem theory and minimax statistics. It views the MMD's failure as an ``ill-posed'' inversion problem and fixes it by applying a continuous spectral regularizer that incorporates covariance information. Our method, however, is rooted in integral operator learning theory and dimensionality reduction. It views the MMD's failure as a ``variance inflation'' problem caused by noisy tail components and fixes it by applying a hard truncation that retains only well-estimated leading components. Similar ideas of spectral regularization are also applied in goodness-of-fit testing problems \citep{balasubramanian2021optimality,sang2025integral}.

The remainder of this paper is organized as follows. Section~\ref{sec:prelim} establishes the mathematical and operator-theoretic foundations. Section~\ref{sec:directional} introduces directional components, their estimation, and asymptotic properties. Section~\ref{sec:test} details the test statistic construction, its asymptotic properties, and a data-driven truncation dimension $d$ selection method. Section~\ref{sec:kernel_choice} presents a perturbation analysis demonstrating the test's insensitivity to kernel choice. Section~\ref{sec:simulation} reports extensive simulations and empirical applications validating our method's finite-sample performance. Finally, Section~\ref{sec:conclusion} concludes with a discussion of potential extensions.

The code for this paper is available at \url{https://github.com/yuhaoli-academic/TwoSample_KDCA}.
\section{Preliminaries}
\label{sec:prelim}

This section establishes the mathematical and operator-theoretic foundations underpinning our analyses. Standard references include \cite{rosasco2010learning}, \cite{steinwart2012mercer}, and \cite{hsing2015theoretical}. Throughout, we assume the following conditions:

\textbf{A1. Independence and Identical Distribution.} The samples $\{x_i\}_{i=1}^n$ and $\{y_j\}_{j=1}^m$ are i.i.d. draws from $P$ and $Q$, respectively, and the two samples are mutually independent. 

\textbf{A2. Characteristic Kernel.} The kernel $k$ is characteristic (i.e., the mean embedding map $\mu: P \mapsto \mu_P$ is injective), establishing the equivalence between $H_0: P=Q$ and $H_0: \mu_P = \mu_Q$.

\textbf{A3. Boundedness and Continuity.} The kernel $k$ is bounded, $\sup_{z \in \mathcal{Z}} k(z,z) \le \bar{k} < \infty$, and continuous. These two conditions imply that the corresponding RKHS $\mathcal{H}$ is compactly embedded in $C(\mathcal{Z})$, the space of continuous functions on $\mathcal{Z}$ with compact-open topology, and the associated integral operator admits a discrete spectrum \citep{aronszajn1950theory}.

\subsection{Integral and Empirical Operators}

Let $\mathcal{Z}$ be a locally compact separable metric space with Borel probability measure $\rho = p P + (1-p) Q$. Let $k: \mathcal{Z} \times \mathcal{Z} \to \mathbb{R}$ be a bounded, continuous and measurable kernel with associated reproducing kernel Hilbert space (RKHS) $\mathcal{H}$. We define the integral operator $L_k: L^2(\rho) \to L^2(\rho)$ by $(L_k f)(x) = \int_{\mathcal{Z}} k(x, y) f(y) \, d\rho(y)$, where $L^2(\rho)$ is the space of square-integrable functions with norm $\|f\|_{\rho} = \left( \int_{\mathcal{Z}} |f(z)|^2 \, d\rho(z) \right)^{1/2}$. The uniform kernel bound is $\bar{k} = \sup_{z \in \mathcal{Z}} k(z,z) < \infty$.

Given an i.i.d. sample $\{Z_i\}_{i=1}^N \sim \rho$, we construct the scaled Gram matrix $K \in \mathbb{R}^{N \times N}$ with entries $K_{\alpha \beta} = k(Z_\alpha, Z_\beta)/N$. While $K$ naturally estimates $L_k$, they act on fundamentally different spaces. To bridge this gap, we introduce the integral operator $T_{\mathcal{H}}: \mathcal{H} \to \mathcal{H}$ and its empirical counterpart $T_N: \mathcal{H} \to \mathcal{H}$:
\begin{equation*}
T_{\mathcal{H}} = \mathbb{E}\left[k(Z,\cdot) \otimes_{\mathcal{H}} k(Z,\cdot)\right] = \int_{\mathcal{Z}} k(z,\cdot) \otimes_{\mathcal{H}} k(z,\cdot) \, d\rho(z), \quad T_N = \frac{1}{N} \sum_{\alpha=1}^{N} k(z_\alpha,\cdot) \otimes_{\mathcal{H}} k(z_\alpha,\cdot),
\end{equation*}
where for a Hilbert space $\mathcal{A}$, $(a \otimes_{\mathcal{A}} b) c = \langle b, c \rangle_{\mathcal{A}} a$ for all $a,b,c \in \mathcal{A}$. Unlike $L_k$, $T_{\mathcal{H}}$ has both its domain and range in $\mathcal{H}$.

Next, define the restriction operator $R_N: \mathcal{H} \to \mathbb{R}^N$ and its adjoint $R_N^*: \mathbb{R}^N \to \mathcal{H}$ by $R_N f = \left(f(z_1), \dots, f(z_N)\right)^\top$ and $R_N^* u = \frac{1}{N} \sum_{\alpha=1}^{N} u_\alpha k(z_\alpha, \cdot)$ for $u = (u_1, \dots, u_N)^\top$. Let $\langle \cdot, \cdot \rangle_N$ denote the scaled canonical inner product on $\mathbb{R}^N$, i.e., $\langle u, v \rangle_N = \frac{1}{N} \sum_{\alpha=1}^N u_\alpha v_\alpha$. For any $f \in \mathcal{H}$ and $u \in \mathbb{R}^N$, the adjoint property is verified by:
\begin{equation*}
\langle R^*_N u, f \rangle_{\mathcal{H}} = \left\langle \frac{1}{N} \sum_{\alpha=1}^{N} u_\alpha k(z_\alpha, \cdot), f \right\rangle_{\mathcal{H}} = \frac{1}{N} \sum_{\alpha=1}^{N} u_\alpha f(z_\alpha) = \langle u, R_N f \rangle_N.
\end{equation*}
These definitions yield the factorizations $T_N = R_N^* R_N$ and $K = R_N R_N^*$. Analogously, define the inclusion operator $R_{\mathcal{H}}: \mathcal{H} \to L^2(\rho)$ mapping a function to its equivalence class, and its adjoint $R_{\mathcal{H}}^*: L^2(\rho) \to \mathcal{H}$ by $R_{\mathcal{H}} f = f$ and $R_{\mathcal{H}}^* f = \int_{\mathcal{Z}} k(z,\cdot) f(z) \, d\rho(z)$. These satisfy $T_{\mathcal{H}} = R_{\mathcal{H}}^* R_{\mathcal{H}}$ and $L_k = R_{\mathcal{H}} R_{\mathcal{H}}^*$.

\subsection{Singular Value Decomposition of Operators}

The following results detail the spectral relationships between population and empirical operators.

\begin{lemma}
\label{svd_population}
The following facts hold for the population operators.
\begin{enumerate}
    \item The operators $L_k$ and $T_{\mathcal{H}}$ are positive, self-adjoint, compact, and trace-class. Consequently, $\sigma(L_k), \sigma(T_{\mathcal{H}}) \subset [0,\bar{k}]$, where $\sigma(\cdot)$ denotes the spectrum.
    \item The non-zero spectra of $L_k$ and $T_{\mathcal{H}}$ coincide. If $\lambda$ is a non-zero eigenvalue with associated eigenfunctions $\phi$ (in $L^2(\rho)$) and $e$ (in $\mathcal{H}$), normalized to unit norm, then 
    \begin{equation*}
        \phi(z) = \lambda^{-1/2} e(z) \text{ for } \rho\text{-a.e. } z \in \mathcal{Z}, \quad e(z) = \lambda^{-1/2} \int_{\mathcal{Z}} k(z,z') \phi(z') \, d\rho(z') \text{ for all } z \in \mathcal{Z}.
    \end{equation*}      
    \item The spectral decompositions are given by:
    \begin{equation*}
        L_k = \sum_{i \in I} \lambda_i \phi_i \otimes_{\rho} \phi_i, \quad T_{\mathcal{H}} = \sum_{i \in I} \lambda_i e_i \otimes_{\mathcal{H}} e_i,
    \end{equation*}
    where $\{\phi_i\}_{i \in I}$ forms an orthonormal basis (ONB) of $\operatorname{Ker}(L_k)^\perp$, and $\{e_i\}_{i \in I}$ forms an ONB of $\operatorname{Ker}(T_{\mathcal{H}})^\perp$.    
\end{enumerate}
\end{lemma}
\noindent Here, $\operatorname{Ker}(A) = \{f \in \operatorname{Dom}(A) : Af = 0\}$ denotes the kernel (null space) of an operator $A$.

\begin{lemma}
\label{svd_empirical}
The following facts hold for the empirical operators.
\begin{enumerate}
    \item The operator $T_N$ is finite rank, self-adjoint, and positive semi-definite. The matrix $K$ is conjugate symmetric and positive semi-definite. The spectrum $\sigma(T_N)$ contains finitely many non-zero elements, all in $[0,\bar{k}]$. 
    \item The non-zero spectra of $K$ and $T_N$ coincide. If $\hat{\lambda}$ is a non-zero eigenvalue with associated eigenvector $\hat{\phi}$ (in $\mathbb{R}^N$) and eigenfunction $\hat{e}$ (in $\mathcal{H}$), normalized to unit norm, 
    \begin{equation*}
        \hat{\phi} = \hat{\lambda}^{-1/2} \bigl(\hat{e}(z_1), \dots, \hat{e}(z_N)\bigr)^\top, \quad \hat{e} = \hat{\lambda}^{-1/2} \frac{1}{N} \sum_{\alpha=1}^{N} \hat{\phi}_\alpha k(z_\alpha,\cdot),
    \end{equation*}      
    where $\hat{\phi}_\alpha$ is the $\alpha$-th component of $\hat{\phi}$.
    \item The spectral decompositions are given by $K = \sum_{i=1}^{r} \hat{\lambda}_i \hat{\phi}_i \otimes_N \hat{\phi}_i$ and $T_N = \sum_{i=1}^{r} \hat{\lambda}_i \hat{e}_i \otimes_{\mathcal{H}} \hat{e}_i$, 
    where $r = \operatorname{rank}(K)$, $\{\hat{\phi}_i\}_{i=1}^{r}$ forms an ONB of $\operatorname{Ker}(K)^\perp \subset \mathbb{R}^N$, and $\{\hat{e}_i\}_{i=1}^{r}$ forms an ONB of $\operatorname{Ker}(T_N)^\perp \subset \mathcal{H}$.    
\end{enumerate}
\end{lemma}

\begin{remark}
From Lemma \ref{svd_empirical}, the restriction operator and its adjoint admit the 
finite-dimensional SVD representations: $R_N = \sum_{i=1}^{r} \sqrt{\hat{\lambda}_i} \, \hat{\phi}_i \otimes_{\mathcal{H}} \hat{e}_i$ and $R_N^* = \sum_{i=1}^{r} \sqrt{\hat{\lambda}_i} \, \hat{e}_i \otimes_N \hat{\phi}_i$.
\end{remark}

\subsection{Bounds on Spectral Projections}

We now present finite-sample probabilistic bounds for spectral projection estimation.

\begin{lemma}
\label{bound_projections}
Let $v$ be an integer and $d$ be the sum of the multiplicities of the first $v$ distinct eigenvalues of $L_k$, such that $\lambda_1 \geq \lambda_2 \geq \dots \geq \lambda_d > \lambda_{d+1}$. Let $r = \operatorname{rank}(K)$ and $\{\hat{\phi}_i\}_{i=1}^{r}$ be the eigenvectors corresponding to the non-zero eigenvalues of $K$ in decreasing order. Denote by $\{\hat{e}_i\}_{i=1}^{r} \in \mathcal{H}$ the corresponding Nyström extensions (Lemma \ref{svd_empirical}). 

Given $\tau > 0$, if the sample size $N$ satisfies $N > 128 \bar{k}^2 \tau/(\lambda_d - \lambda_{d+1})^2$,
then with probability at least $1 - 2e^{-\tau}$, we have
\begin{equation*}
\sum_{j=1}^d \sum_{\substack{i \ge d+1 \\ T_{\mathcal{H}}e_i \neq 0}} |\langle e_i, \hat{e}_j \rangle_{\mathcal{H}}|^2 + \sum_{j=d+1}^r \sum_{i=1}^d |\langle e_i, \hat{e}_j \rangle_{\mathcal{H}}|^2 \le \left\|\hat{P}_d - P_d\right\|_{\mathrm{HS}}^2 \le \frac{32 \bar{k}^2 \tau}{N (\lambda_d - \lambda_{d+1})^2},  
\end{equation*}  
where $P_d = \sum_{i=1}^d e_i \otimes_{\mathcal{H}} e_i$, $\hat{P}_d = \sum_{i=1}^d \hat{e}_i \otimes_{\mathcal{H}} \hat{e}_i$, and $\| \cdot \|_{\mathrm{HS}}$ denotes the Hilbert--Schmidt norm.
\end{lemma}

\begin{remark}
The two terms on the left-hand side of Lemma \ref{bound_projections} admit clear geometric interpretations. The first term, $\sum_{j=1}^d \sum_{i \ge d+1} |\langle e_i, \hat{e}_j \rangle_{\mathcal{H}}|^2$, quantifies the ``leakage'' from the estimated leading subspace into the orthogonal complement, representing the Nyström extension estimation error. The second term, $\sum_{j = d+1}^r \sum_{i=1}^d |\langle e_i, \hat{e}_j \rangle_{\mathcal{H}}|^2$, quantifies the ``pollution'' of the leading subspace by trailing empirical eigenfunctions. The lemma demonstrates that both error sources are jointly controlled by $N$ and the eigengap $\lambda_d - \lambda_{d+1}$.
\end{remark}

\begin{remark}
When the projection operators $P_i$ and $\hat{P}_i$ are associated with a simple eigenvalue $\lambda_i$, the Hilbert--Schmidt bound for each individual component reduces to:
\begin{equation*}
\left\|P_i - \hat{P}_i\right\|_{\mathrm{HS}}^2 = 2\left(1 - \langle e_i, \hat{e}_i \rangle_{\mathcal{H}}^2\right) \ge 2\left(1 - \langle e_i, \hat{e}_i \rangle_{\mathcal{H}}\right) = \left\|e_i - \hat{e}_i \right\|_{\mathcal{H}}^2 = O_p(N^{-1}).
\end{equation*}
\end{remark}
\section{Directional Components}
\label{sec:directional}

The MMD between $P$ and $Q$ is defined as the distance between their mean embeddings in $\mathcal{H}$: $\text{MMD}^2(P, Q) = \|\mu_P - \mu_Q\|_{\mathcal{H}}^2 = \sum_{i \ge 1} \langle \mu_P - \mu_Q, e_i \rangle_{\mathcal{H}}^2 = \sum_{i \ge 1} d_i^2$, where $\{e_i\}_{i \ge 1}$ is an ONB of $\mathcal{H}$ consisting of eigenfunctions of $T_{\mathcal{H}}$. Specifically, $\{e_i\}_{i \in I_+}$ and $\{e_i\}_{i \in I_0}$ form ONBs of $\text{Ker}(T_{\mathcal{H}})^\perp$ and $\text{Ker}(T_{\mathcal{H}})$, respectively, with $I_+ \cup I_0 = \mathbb{N}$. We refer to $d_i = \mathbb{E}[e_i(X)] - \mathbb{E}[e_i(Y)]$, $X \sim P, Y \sim Q$ 
as the $i$-th directional component of the MMD. Consequently, the MMD equals the $\ell^2$-norm of the directional component sequence. As Lemma \ref{bound_projections} shows, controlling the approximation error of Nyström eigenfunction extensions requires a sufficiently large eigengap and a sufficiently large sample size. For fixed sample sizes, this control is reliable only for leading distinct eigenvalues; thus, we focus on estimating directional components along these directions.

\subsection{Analysis of Directional Components when Eigenvalues are Simple}

We first present estimation results for $d_i$ assuming the corresponding eigenvalue $\lambda_i$ is simple (i.e., multiplicity one). Here, $e_i$ is unique up to a sign, allowing direct estimation of $d_i$. The case of eigenvalues with multiplicity greater than one is deferred to Section 3.2. We construct the empirical counterpart of $d_i$ by replacing population expectations with empirical means and population eigenfunctions with their Nyström extensions:
\begin{equation*}
\hat{d}_i = \frac{1}{n} \sum_{\alpha=1}^n \hat{e}_i(x_\alpha) -  \frac{1}{m} \sum_{\beta=1}^m \hat{e}_i(y_\beta) = \langle \hat{\mu}_P - \hat{\mu}_Q, \hat{e}_i \rangle_{\mathcal{H}},
\end{equation*}
where $\hat{\mu}_P = n^{-1}\sum_{\alpha=1}^n k(x_\alpha, \cdot)$ and $\hat{\mu}_Q =m^{-1}\sum_{\beta=1}^m k(y_\beta, \cdot)$ are the empirical mean embeddings. By Lemma \ref{svd_empirical}(2), we obtain
\begin{equation*}
\hat{e}_i(x_\alpha) = \hat{\lambda}_i^{-1} \frac{1}{N} \sum_{\gamma=1}^N k(x_\alpha, z_\gamma) \hat{e}_i(z_\gamma), \quad \hat{e}_i(y_\beta) = \hat{\lambda}_i^{-1} \frac{1}{N} \sum_{\gamma=1}^N k(y_\beta, z_\gamma) \hat{e}_i(z_\gamma),
\end{equation*}
with $N = n + m$, $z_\gamma = x_\gamma$ for $\gamma \le n$, and $z_\gamma = y_{\gamma-n}$ for $\gamma > n$.

We now investigate the impact of replacing population eigenfunctions with their empirical Nyström extensions. Decomposing $\hat{d}_i$ yields:
\begin{equation*}
\hat{d}_i = \langle \hat{\mu}_P - \hat{\mu}_Q, e_i \rangle_{\mathcal{H}} + \langle \hat{\mu}_P - \hat{\mu}_Q, \hat{e}_i - e_i \rangle_{\mathcal{H}} \stackrel{H_0}{=} \langle (\hat{\mu}_P-\mu) - (\hat{\mu}_Q-\mu), e_i \rangle_{\mathcal{H}} + \langle (\hat{\mu}_P-\mu) - (\hat{\mu}_Q-\mu), \hat{e}_i - e_i \rangle_{\mathcal{H}}.
\end{equation*}
Under $H_0: \mu_P = \mu_Q = \mu$, the first term captures the estimation error from replacing population mean embeddings with empirical counterparts, while the second stems from both the Nyström extension approximation error and the mean embedding estimation error.

To characterize the asymptotic distribution of $\hat{d}_i$, we require the following functional central limit theorem \citep{hsing2015theoretical}.

\begin{theorem}
\label{clt}
If $\mathbb{E}\|k(Z,\cdot)\|_{\mathcal{H}} < \infty$, then under $H_0$, $\hat{\mu}_P \xrightarrow{a.s.} \mu$ and $\hat{\mu}_Q \xrightarrow{a.s.} \mu$. Furthermore, if $\mathbb{E}\|k(Z,\cdot)\|_{\mathcal{H}}^2 < \infty$, then $\sqrt{n} \left(\hat{\mu}_P - \mu\right) \xrightarrow{d} \xi_1$ and $\sqrt{m} \left(\hat{\mu}_Q - \mu\right) \xrightarrow{d} \xi_2$, 
where $\xi_1$ and $\xi_2$ are i.i.d.\ Gaussian random elements in $\mathcal{H}$ with mean zero and covariance operator $\Sigma = \mathbb{E}\left[(k(Z,\cdot) - \mu) \otimes_{\mathcal{H}} (k(Z,\cdot) - \mu)\right]$.
\end{theorem}

\begin{remark}
The boundedness of $k$ implies both moment conditions in Theorem \ref{clt}. Specifically, $\|k(Z,\cdot)\|_{\mathcal{H}}^2 = k(Z,Z) \le \bar{k}$, ensuring $\mathbb{E}\|k(Z,\cdot)\|_{\mathcal{H}} < \infty$ and $\mathbb{E}\|k(Z,\cdot)\|_{\mathcal{H}}^2 < \infty$. 
In addition, Theorem \ref{clt} implies $\|\hat{\mu}_P - \mu\|_{\mathcal{H}} = O_p(n^{-1/2}) = O_p(N^{-1/2})$ and $\|\hat{\mu}_Q - \mu\|_{\mathcal{H}} = O_p(m^{-1/2}) = O_p(N^{-1/2})$.
\end{remark}

\begin{lemma}
\label{residual}
Under $H_0$, $\langle \hat{\mu}_P - \hat{\mu}_Q, \hat{e}_i - e_i \rangle_{\mathcal{H}} = O_p(N^{-1})$, whereas under a fixed alternative, $\langle \hat{\mu}_P - \hat{\mu}_Q, \hat{e}_i - e_i \rangle_{\mathcal{H}} = O_p(N^{-1/2})$.
\end{lemma}

Lemma \ref{residual} demonstrates that under $H_0$, the second term in the decomposition of $\hat{d}_i$ is asymptotically negligible, substantially simplifying the derivation of its null distribution. Under a fixed alternative, this term is of the same order as the first and cannot be ignored, leading to a more complex asymptotic distribution discussed later.

\begin{theorem}
\label{null_distribution_single}
Suppose Assumptions A1-A3 hold and $\lambda_i$ is a simple eigenvalue. Under $H_0$,
\begin{equation*}
\sqrt{N} \hat{d}_i \xrightarrow{d} \mathcal{N}\left(0, \frac{1}{p(1-p)} \langle \Sigma e_i, e_i \rangle_{\mathcal{H}}\right),
\end{equation*}
where $p = \lim_{n,m \to \infty} n/(m+n)$ and $\Sigma$ is the covariance operator defined in Theorem \ref{clt}.
\end{theorem}

\begin{remark}
The variance term $\langle \Sigma e_i, e_i \rangle_{\mathcal{H}} = \mathbb{E}\left[(e_i(Z) - \mathbb{E}[e_i(Z)])^2\right] = \text{Var}(e_i(Z))$ yields a natural consistent estimator for the variance of $\sqrt{N}\hat{d}_i$ under $H_0$:
\begin{equation*}
\widehat{\text{Var}}(\sqrt{N} \hat{d}_i) = \frac{1}{\hat{p}(1-\hat{p})} \frac{1}{N-1} \sum_{\alpha=1}^N \left(\hat{e}_i(z_\alpha) - \frac{1}{N} \sum_{\gamma=1}^N \hat{e}_i(z_\gamma)\right)^2,
\end{equation*}
where $\hat{p} = n/N$.
\end{remark}

\subsection{Analysis of Directional Components when Eigenvalues are Repeated}

We now consider the general case involving the first $v$ distinct eigenvalues of $L_k$, with $d$ denoting the sum of their multiplicities. Let $\{e_i\}_{i=1}^d$ be the corresponding eigenfunctions of $T_{\mathcal{H}}$. Here, the subspace $E$ spanned by $\{e_i\}_{i=1}^d$ is identifiable only up to an orthogonal transformation. Let $\hat{E}$ be the subspace spanned by empirical eigenfunctions $\{\hat{e}_i\}_{i=1}^d$. We estimate the projection of mean embeddings onto $E$ by projecting empirical mean embeddings onto $\hat{E}$. A suitably rotated version of $\{\hat{e}_i\}_{i=1}^d$ converges to $\{e_i\}_{i=1}^d$ (see Appendix \ref{orthogonal transformation}). Throughout, we assume assumption below holds:

\textbf{A4. Spectral Gap.} A positive spectral gap $\lambda_d - \lambda_{d+1} > 0$ exists at the truncation point $d$. 

The following results characterize the asymptotic distribution of the estimated directional components under $H_0$, and the consistency of the parametric bootstrap procedure for approximating critical values.

\begin{theorem}
\label{null_distribution_multiple}
Let Assumptions A1-A4 hold. Under $H_0$, there exists a sequence of random orthogonal matrices $O_N \in \mathbb{R}^{d \times d}$ such that, defining $\tilde{d}_{N,i} = \sum_{j=1}^d (O_N)_{ji} \left\langle \hat{\mu}_P - \hat{\mu}_Q, \hat{e}_j \right\rangle_{\mathcal{H}}$ for $i=1,\dots,d$, 
we have $\sqrt{N}(\tilde{d}_{N,1},\dots,\tilde{d}_{N,d}) \xrightarrow{d} \sqrt{\frac{1}{p(1-p)}}(\langle \xi,e_1\rangle_{\mathcal{H}},\dots,\langle \xi,e_d\rangle_{\mathcal{H}})$. Consequently,
\begin{equation*}
D_N^d = N \sum_{i=1}^d \left\langle \hat{\mu}_P - \hat{\mu}_Q, \hat{e}_i \right\rangle_{\mathcal{H}}^2 \xrightarrow{d} \sum_{i=1}^d \frac{\sigma_i}{p(1-p)} \chi_{1,i}^2,
\end{equation*}
where $\{\chi_{1,i}^2\}_{i=1}^d$ are independent chi-squared variables with one degree of freedom, and $\sigma_1 \ge \cdots \ge \sigma_d$ are the eigenvalues of the matrix representation $\Sigma_d$ of the restricted covariance operator $P_d \Sigma P_d$ with respect to the basis $\{e_i\}_{i=1}^d$, i.e., $(\Sigma_d)_{ij} = \langle \Sigma e_i, e_j \rangle_{\mathcal{H}}$. Here, $P_d = \sum_{i=1}^d e_i \otimes_{\mathcal{H}} e_i$ is the orthogonal projection onto the span of $\{e_i\}_{i=1}^d$.
\end{theorem}

\begin{lemma}
    \label{lm:consistency_cov_matrix}
    Let Assumptions A1-A4 hold, and let $ \{\hat{e}_i\}_{i=1}^d $ be the Nystr\"om extensions of leading empirical eigenfunctions. Define empirical covariance matrix $\hat{\Sigma}_d \in \mathbb{R}^{d\times d}$ by $ (\hat{\Sigma}_d)_{ij} = \sum_{\alpha=1}^N (\hat{e}_i(z_\alpha - \bar{e}_i)) (\hat{e}_j(z_\alpha - \bar{e}_j)) / (N-1) $, with $ \bar{e}_i = \sum_{\alpha=1}^N \hat{e}_i(z_\alpha) / N $. Then, there exists a sequence of random orthogonal matrices $ O_N \in \mathbb{R}^{d\times d} $ such that $ \tilde{\Sigma}_d=O_N \hat{\Sigma}_d O_N^\top \xrightarrow{p} \Sigma_d $ as $ N\to\infty $. Consequently, the eigenvalues of $ \hat{\Sigma}_d $ are consistent estimators of those of $ \Sigma_d $.    
\end{lemma}

\begin{lemma}
    \label{lm:parametric_bootstrap_consistency}
    Let Assumptions A1-A4 hold, and let $ \hat{\Sigma}_d $ be the empirical covariance matrix defined in Lemma \ref{lm:consistency_cov_matrix}, and let $ \hat{\Gamma}_d = \hat{\Sigma}_d / (\hat{p}(1-\hat{p})) $. For a given sample, generate bootstrap replicates as $ W^{(b)} \sim \mathcal{N}(\boldsymbol{0}, \hat{\Gamma}_d), b=1,\dots,B $. Denote the conditional distribution of $ \|W^{(b)}\|^2 $ as $ \hat{F}_N = \mathbb{P}(\|W^{(b)}\|^2 \le x \mid z_1,\ldots,z_N) $, then $ \sup_{x\in \mathbb{R}} |\hat{F}_N(x) - F(x)| \to 0 $ in probability as $ N \to \infty $, where $ F $ is the null distribution of $ D_N^d $.     
\end{lemma}

Computing the full eigen-decomposition of the Gram matrix $K$ via SVD requires $O(N^3)$ operations, which is prohibitive for large $N$. However, since only the top $d$ eigenpairs are needed, efficient algorithms like the Lanczos method reduce the cost to $O(N^2 d)$. In practice, when $d$ is small, the computational overhead is minimal.
\section{Test Statistic, Power Analysis, and Selection of $d$}
\label{sec:test}

Our proposed test statistic is an unbiased version of $D_N^d$. We detail its construction, asymptotic null distribution, power properties under fixed and local alternatives, and the practical selection of the truncation parameter $d$.

\subsection{Test Statistic and Its Asymptotic Distribution under the Null}

Recall the Gram matrix $K$ with entries $K_{\alpha\beta} = k(z_\alpha, z_\beta)/N$. Consider a scaled, spectrally truncated version of $K$:
\begin{equation*}
K^{(d)} = N \sum_{i=1}^d \hat{\lambda}_i \hat{\phi}_i \otimes_N \hat{\phi}_i = N \sum_{i=1}^d \hat{\lambda}_i u_i u_i^\top = \begin{pmatrix}
    K_{XX}^{(d)} & K_{XY}^{(d)} \\
    K_{YX}^{(d)} & K_{YY}^{(d)}
\end{pmatrix},
\end{equation*}
where $\hat{\lambda}_1 \ge \dots \ge \hat{\lambda}_d > 0$ are the leading $d$ eigenvalues of $K$ (corresponding to the first $v$ distinct population eigenvalues) and $\{u_i\}_{i=1}^d$ are the corresponding eigenvectors, using $\hat{\phi}_i = \sqrt{N} u_i$ due to the $\otimes_N$ normalization. The unbiased test statistic is then constructed as:
\begin{equation*}
\tilde{D}_N^d = \frac{N}{n(n-1)} \left(\mathbf{1}_n^\top K_{XX}^{(d)} \mathbf{1}_n - \operatorname{Tr}(K_{XX}^{(d)}) \right) + \frac{N}{m(m-1)} \left(\mathbf{1}_m^\top K_{YY}^{(d)} \mathbf{1}_m - \operatorname{Tr}(K_{YY}^{(d)}) \right) - \frac{2N}{nm} \mathbf{1}_n^\top K_{XY}^{(d)} \mathbf{1}_m.
\end{equation*}

The following lemma provides an alternative expression for $\tilde{D}_N^d$, facilitating subsequent asymptotic analysis.

\begin{lemma}
\label{alternative_expression}
Let $H_N = I_N - \frac{1}{N}\mathbf{1}_N\mathbf{1}_N^{\top}$ be the centering matrix and $\bar{K}^{(d)} = H_N K^{(d)} H_N$. Then,
\begin{equation*}
\tilde{D}_N^d = \frac{N}{n(n-1)}\bigl(\mathbf{1}_n^{\top}\bar{K}_{XX}^{(d)}\mathbf{1}_n - \operatorname{Tr}(\bar{K}_{XX}^{(d)})\bigr) + \frac{N}{m(m-1)}\bigl(\mathbf{1}_m^{\top}\bar{K}_{YY}^{(d)}\mathbf{1}_m - \operatorname{Tr}(\bar{K}_{YY}^{(d)})\bigr) - \frac{2N}{nm}\mathbf{1}_n^{\top}\bar{K}_{XY}^{(d)}\mathbf{1}_m.
\end{equation*}
\end{lemma}

\begin{remark}
\label{rk:biased_and_unbiased}
Let $\hat{\mu} = \frac{1}{N} \sum_{\gamma=1}^{N} k(z_{\gamma}, \cdot)$ and $\hat{\mu}_i = \langle \hat{\mu}, \hat{e}_i \rangle_{\mathcal{H}}$ for $i = 1, \dots, d$. The biased test statistic, $D_N^d$, expands as:
\begin{align*}
D_N^{d} &= N \sum_{i=1}^{d} \left\langle (\hat{\mu}_{P} - \hat{\mu}) - (\hat{\mu}_{Q} - \hat{\mu}), \hat{e}_i \right\rangle_{\mathcal{H}}^2= N \sum_{i=1}^{d} \left( \frac{1}{n} \sum_{\alpha=1}^{n} (\hat{e}_i(x_{\alpha}) - \hat{\mu}_i) - \frac{1}{m} \sum_{\beta=1}^{m} (\hat{e}_i(y_{\beta}) - \hat{\mu}_i) \right)^2 \\
&= N \sum_{i=1}^{d} \left( \frac{1}{n^2} \mathbf{1}_n^{\top} \bar{K}_{XX}^{(d)} \mathbf{1}_n + \frac{1}{m^2} \mathbf{1}_m^{\top} \bar{K}_{YY}^{(d)} \mathbf{1}_m - \frac{2}{nm} \mathbf{1}_n^{\top} \bar{K}_{XY}^{(d)} \mathbf{1}_m \right) \\
&\approx \tilde{D}_N^d + \frac{N}{n^2} \operatorname{Tr}(\bar{K}_{XX}^{(d)}) + \frac{N}{m^2} \operatorname{Tr}(\bar{K}_{YY}^{(d)}).
\end{align*}
The difference $D_N^{d} - \bigl(\tilde{D}_N^d + N / n^2 \operatorname{Tr}(\bar{K}_{XX}^{(d)}) + N / m^2 \operatorname{Tr}(\bar{K}_{YY}^{(d)})\bigr)$ is $O(1/N)$.
\end{remark}

Combining Theorem \ref{null_distribution_multiple} with Remark \ref{rk:biased_and_unbiased} yields the asymptotic null distribution of $\tilde{D}_N^d$.

\begin{theorem}
\label{null_distribution_test_statistic}
Let Assumptions A1-A4 hold. Under the null hypothesis,
\begin{equation*}
\tilde{D}_N^d \xrightarrow{d} \sum_{i=1}^d \frac{\sigma_i}{p(1-p)} (\chi_{1,i}^2-1),
\end{equation*}
where $p = \lim_{n,m \to \infty} n / (m+n)$, and $\sigma_1 \ge \cdots \ge \sigma_d$ are the eigenvalues of the restricted covariance matrix $\Sigma_d$ defined in Theorem~\ref{null_distribution_multiple}.
\end{theorem}

\subsection{Power Analysis under Fixed and Local Alternatives}

We evaluate the power of $\tilde{D}_N^d$ under both fixed and Pitman-type local alternatives.

\begin{theorem}
\label{th:fixed-alternative}
Let Assumptions A1-A4 hold. Let $\lambda_1 > \lambda_2 > \cdots > \lambda_v$ be the first $v$ distinct eigenvalues of $T_{\mathcal{H}}$ with multiplicities $m_1, \dots, m_v$, and let $d = \sum_{k=1}^v m_k$. Denote by $P_k$ the orthogonal projection onto the eigenspace corresponding to $\lambda_k$, and let $P_d = \sum_{k=1}^v P_k$ project onto the span of the first $d$ eigenfunctions. Under the fixed alternative $H_1 \colon \mu_P = \mu + g$ and $\mu_Q = \mu$, where $g \in \mathcal{H}$ is a fixed non-zero function satisfying $\sum_{k=1}^v \|P_k g\|_{\mathcal{H}}^2 \neq 0$, we have
\begin{equation*}
\sqrt{N} \Bigl( \frac{1}{N} \tilde{D}_N^d - \sum_{k=1}^v \|P_k g\|_{\mathcal{H}}^2 \Bigr) \xrightarrow{d} \mathcal{N}\bigl(0, \sigma_{H_1}^2\bigr),
\end{equation*}
where the asymptotic variance is $\sigma_{H_1}^2 = \frac{1}{p} \operatorname{Var}_P(\xi_P(X)) + \frac{1}{1-p} \operatorname{Var}_Q(\xi_Q(Y))$. The influence functions are defined as
\begin{equation*}
\xi_P(x) = 2 P_d g(x) + 2p \sum_{k=1}^v P_k g(x) w_k(x), \quad \xi_Q(y) = -2 P_d g(y) + 2(1-p) \sum_{k=1}^v P_k g(y) w_k(y),
\end{equation*}
with $w_k(z) = \sum_{l > v} P_l g(z)/(\lambda_l - \lambda_k)$, where $P_l$ denotes the spectral projection onto the eigenspace of the $l$-th distinct eigenvalue $\lambda_l$. 
\end{theorem}

\begin{corollary}
Let Assumptions A1-A4 hold. Under the fixed alternative with $\sum_{k=1}^v \|P_k g\|_{\mathcal{H}}^2 \neq 0$, it follows that for any $t \in \mathbb{R}$, $\mathbb{P}(\tilde{D}_N^d \ge t) \to 1$ as $N \to \infty$.
\end{corollary}

We next investigate the power under a Pitman-type local alternative.

\begin{theorem}
\label{th:local_alternative}
Let Assumptions A1-A4 hold. Assume the local alternative $H_{1N}:\mu_{P}= \mu + g/\sqrt{N}$ and $\mu_Q = \mu$, where $g\in\mathcal H$ satisfies $\sum_{i=1}^d\langle g,e_i\rangle_{\mathcal H}^2 \neq 0$.  
Suppose there exists a bounded measurable function $a:\mathcal Z\to\mathbb R$ with $\int a\,dP=0$ such that for all sufficiently large $N$ the distribution $P_N$ has density $1+N^{-1/2}a$ with respect to $P$; this yields $g = \int k(\cdot,z)a(z)\,dP(z)$.  
Let Assumptions~1--5 hold.  Then
\[
\tilde D_N^d \xrightarrow{d}
\sum_{i=1}^d \left( \frac{1}{\sqrt{p(1-p)}}\langle \xi,e_i\rangle_{\mathcal H} + \langle g,e_i\rangle_{\mathcal H} \right)^2
- \frac{1}{p(1-p)}\sum_{i=1}^d \langle \Sigma e_i,e_i\rangle_{\mathcal H},
\]
where $\xi$ is the Gaussian random element defined in Theorem \ref{null_distribution_single}.
\end{theorem}

\subsection{Discussion on Truncation Dimension and Power Improvement}
\textbf{Fixed vs. Diverging Truncation Dimension.} 
It is worthwhile to discuss whether $d$ should be fixed or allowed to diverge with the sample size (i.e., $d = d(N) \to \infty$ as $N \to \infty$). In the diverging regime, provided the kernel satisfies Assumptions 1--3 in \cite{sun2005mercer}\footnote{For example, when one has the Gaussian kernel and the $ \mathcal{Z} = \mathbb{R}^q $, where $ q $ is a fixed and finite integer, Assumptions 1--3 in \cite{sun2005mercer} are satisfied.}, the sum in Theorem \ref{null_distribution_test_statistic} extends over increasing dimensions, naturally converging to the limiting null distribution of the classical, full-rank MMD statistic. Correspondingly, the preceding corollary becomes a standard omnibus statement. Despite this theoretical appeal, a diverging $d(N)$ is problematic in practice. By Lemma \ref{bound_projections}, the approximation error of the Nyström eigenfunction extensions is fundamentally governed by the eigengap $(\lambda_d - \lambda_{d+1})^2$. Since kernel eigenvalues typically decay rapidly, trailing eigengaps vanish, requiring prohibitively large $N$ for accurate eigenfunction estimation. Including these poorly estimated higher-order components introduces excessive variance, severely degrading finite-sample power. Thus, treating $d$ as a fixed, moderate truncation level is substantially more beneficial, enforcing a robust bias-variance trade-off via well-estimated leading eigenfunctions. The data-driven strategy for selecting this fixed $d$ is detailed next.

\textbf{Power Improvement under Concentrated Signals.} 
The proposed test statistic $\tilde{D}_N^d$ can significantly outperform the full-rank MMD when the signal is concentrated in the leading $d$ directional components, i.e., $\|g\|_{\mathcal{H}}^2 = \sum_{i=1}^v \|P_i g\|_{\mathcal{H}}^2$. In such cases, both the full-rank MMD and $\tilde{D}_N^d$ capture the same signal strength, but the asymptotic null variance of $\tilde{D}_N^d$ is proportional to $\sum_{i=1}^d \sigma_i^2$ while that of the full-rank MMD is proportional to $\sum_{i \ge 1} \sigma_i^2$. The former is strictly smaller, yielding a higher signal-to-noise ratio and thus greater power.  

On the other hand, if the signal is dispersed across trailing components, the proposed test statistic would only have trivial power, while the full-rank MMD may theoretically capture the signal. However, in this scenario, the full-rank MMD's power is also severely compromised by the large null variance contributed by noisy tail components. In short, it is our belief that under this case, both tests would perform poorly in finite samples. In practice, signals are often concentrated in leading components, as shown in our simulation studies, making $\tilde{D}_N^d$ a more powerful choice in many real-world applications.

\subsection{A Data-Driven Method for the Selection of $d$}
\label{sec:selection_d}

Given fixed $N$, a natural approach selects $d$ by substituting the unknown population eigengap $\lambda_d - \lambda_{d+1}$ with its empirical estimate $\hat{\lambda}_d - \hat{\lambda}_{d+1}$, increasing $d$ until the sample size condition $N > 128 \bar{k}^2 \tau/(\hat\lambda_d - \hat\lambda_{d+1})^2$ in Lemma~\ref{bound_projections} fails. However, this strategy is fundamentally unstable: if $\lambda_d = \lambda_{d+1}$, the lower bound $128 \bar{k} \tau / (\hat{\lambda}_d - \hat{\lambda}_{d+1})^2$ diverges. A small empirical eigengap may reflect genuine eigenvalue multiplicity rather than a narrow true gap, causing erratic selection.

To circumvent this and directly target test power, we propose a data-driven selection criterion based on the asymptotic signal-to-noise ratio (SNR). Under $H_0$, Theorem~\ref{null_distribution_test_statistic} establishes $\tilde{D}_N^d \xrightarrow{d} \sum_{i=1}^d \gamma_i (\chi_{1,i}^2 - 1)$ where $\gamma_i = \sigma_i / (p(1-p))$. Since $\operatorname{Var}(\chi^2_{1,i}) = 2$, the asymptotic null variance is $\operatorname{Var}_{H_0}(\tilde{D}_N^d) = 2 \sum_{i=1}^d \gamma_i^2 = 2 \operatorname{tr}(\Gamma_d^2)$, with $\Gamma_d = \Sigma_d / (p(1-p))$. Under a fixed alternative, let $\boldsymbol{d}_{1:d} = (d_1, \dots, d_d)^\top$ denote the vector of the first $d$ population directional components; the mean shift scales as $\mathbb{E}_{H_1}[\tilde{D}_N^d] - \mathbb{E}_{H_0}[\tilde{D}_N^d] \propto N \|\boldsymbol{d}_{1:d}\|_2^2$. The asymptotic SNR governing the power is therefore
\[ \Delta(d) = \frac{(\mathbb{E}_{H_1}[\tilde{D}_N^d] - \mathbb{E}_{H_0}[\tilde{D}_N^d])^2}{\operatorname{Var}_{H_0}(\tilde{D}_N^d)} \propto N\frac{\|\boldsymbol{d}_{1:d}\|^4}{\operatorname{tr}(\Gamma_d^2)}. \]

Let $\tilde{D}_{N'}^d$ denote the unbiased statistic computed on an independent training set of size $N'$, and $\hat{\Gamma}_d = \hat{\Sigma}_d / (\hat{p}(1-\hat{p}))$ the estimated asymptotic covariance. We define the selection criterion for candidate $d$ as $\widehat{\text{SNR}}(d) =\tilde{D}_{N'}^d / \sqrt{\operatorname{tr}(\hat{\Gamma}_d^2)}$. The selected truncation level is $\hat{d} = \arg\max_{1 \leq d \leq \bar{d}} \widehat{\text{SNR}}(d)$, with $\bar{d}$ a prespecified upper bound (see Algorithm~\ref{alg:select_d} in Appendix~\ref{app:truncation_dim_selection}). Appendix \ref{app:truncation_dim_selection} also provides theoretical guarantees for this selection procedure, showing that it consistently identifies the optimal $d$ maximizing the true SNR under fixed alternatives, while maintaining valid Type I error control under the null.

While population eigenvalues $\sigma_i$ decay rapidly, empirical estimates $\hat{\gamma}_i$ exhibit $O_p(N^{-1/2})$ fluctuations. In the spectral tail where $\gamma_i \approx 0$, these fluctuations dominate, causing $\hat{\gamma}_i$ to plateau. The nonlinear trace $\sum_{i=1}^d \hat{\gamma}_i^2$ amplifies inflated estimates, mirroring the actual inflation of null variance when entering the noise floor. Consequently, $\tilde{D}_{N'}^d / \sqrt{\operatorname{tr}(\hat{\Gamma}_d^2)}$ peaks where true signal ends and estimation noise begins, enforcing a statistically principled bias-variance cutoff without eigengap thresholds.

The computational complexity of Algorithm~\ref{alg:select_d} remains highly efficient. Computing the eigendecomposition and full covariance once, then evaluating the criterion in $O(d^2)$ per candidate (dominated by the trace of a $d \times d$ product), renders the selection cost negligible compared to the $O(N^2 d)$ eigensolver.

Once $\hat{d}$ is selected via Algorithm~\ref{alg:select_d}, we apply the identical procedure to the test dataset to compute $\tilde{D}_N^{\hat{d}}$ and its $p$-value. This framework bypasses unreliable eigengap estimation, eliminates bootstrap overhead during selection, and provides a principled, efficient alternative to heuristics.

Related approaches leveraging the kernel spectrum for null distribution approximation exist, notably \cite{gretton2009fast}. Our method differs in two key aspects. First, we use the spectrum to jointly determine $d$ and estimate the null distribution, whereas \cite{gretton2009fast} relies solely on spectral information for null estimation without adaptive truncation. Second, we employ parametric bootstrap resampling based on the raw Gram matrix $K$, in contrast to \cite{gretton2009fast}, which uses the spectrum of the centered Gram matrix $H_N K H_N$ to approximate the full-rank MMD null distribution.
\section{Robustness to Kernel Parameter Choice}
\label{sec:kernel_choice}

In this section, we formalize the robustness of the proposed test to the choice of kernel parameter via perturbation analysis. We systematically analyze the stability of population quantities, the finite-sample test statistic, and the limiting null distributions under perturbations of the kernel parameter $\theta$.

A fundamental challenge in analyzing kernel sensitivity is that the RKHS $\mathcal{H}_\theta$ inherently depends on $\theta$. Comparing geometric quantities across different Hilbert spaces lacks a canonical interpretation and significantly complicates standard perturbation arguments. To circumvent this, we leverage the fact that for a bounded kernel $k_\theta$, $\mathcal{H}_\theta$ is compactly embedded into $L^2(\rho)$ \citep{steinwart2012mercer}. This allows us to map the analysis from the family of varying spaces $\{\mathcal{H}_\theta\}$ into a single, fixed function space.

Specifically, let $h = \frac{dP}{d\rho} - \frac{dQ}{d\rho} \in L^2(\rho)$ be the Radon--Nikodym derivative representing the distributional difference. By Lemma \ref{svd_population}, the population test statistic can be equivalently expressed entirely within $L^2(\rho)$. Rather than evaluating the norm in $\mathcal{H}_\theta$, the summation of squared directional components translates to the squared $L^2(\rho)$-norm of $h$ projected onto the leading spectral blocks of the integral operator $L_\theta$ associated with $k_\theta$. Since $L^2(\rho)$ is independent of $\theta$, this formulation provides a unified geometric foundation, allowing us to evaluate stability by tracking perturbations in the spectral projections of the fixed function $h$ in $L^2(\rho)$. We detail this $L^2(\rho)$-based analysis below.

Consider a family of kernels $\{k_{\theta}\}$ indexed by a parameter vector $\theta$, and define the corresponding integral operator $L_{\theta}: L^2(\rho) \to L^2(\rho)$ as $L_{\theta}f(z) = \int k_{\theta}(z,z') f(z') \,d\rho(z')$. Denote its eigenvalues by $\lambda_{1}(\theta) \ge \lambda_{2}(\theta) \ge \dots$, and let $\{\phi_{i,j}(\theta)\}_{j=1}^{m_i}$ be an ONB of the eigenspace corresponding to the $i$-th largest eigenvalue $\lambda_i(\theta)$. The contribution of the $i$-th spectral block to the squared MMD is given by
\begin{equation*}
C_i(\theta) = \sum_{j=1}^{m_i} \left(\sqrt{\lambda_i(\theta)}\langle \phi_{i,j}(\theta), h \rangle_{\rho}  \right)^2 = \lambda_i(\theta) \| P_i(\theta) h \|_{\rho}^2,
\end{equation*}
where $ P_i(\theta) = \sum_{j=1}^{m_i}\phi_{i,j}(\theta) \otimes_{\rho} \phi_{i,j}(\theta)$ is the orthogonal projection onto the eigenspace associated with $ \lambda_i(\theta) $. The population test statistic is $ D^d(\theta) = \sum_{i=1}^{v} C_i(\theta) $. We analyze how $C_i(\theta)$ changes under small perturbations of $\theta$. 

Throughout this section, we assume the following regularity condition on the kernel family:

\textbf{A5. Kernel Smoothness.} The kernel mapping $\theta \mapsto k_\theta(z, \cdot)$ is Lipschitz continuous with respect to $\theta$ in the $L^2(\rho)$ norm, uniformly over $z \in \mathcal{Z}$. That is, there exists a constant $0<L < \infty$ such that for any $z \in \mathcal{Z}$, $\|k_{\theta'}(z, \cdot) - k_\theta(z, \cdot)\|_{\rho} \le L \|\theta' - \theta\|$.

\subsection{Perturbation of Population Quantities}
\label{subsec:perturbation_population}

We first analyze the sensitivity of the population-level directional components. The following lemma establishes the Lipschitz continuity of the integral operator with respect to the kernel parameter so that small changes in $\theta$ lead to small changes in the operator.

\begin{lemma}
\label{lm:integral_perturb}
Assume Assumption A5 holds. Then, the integral operator $L_\theta : L^2(\rho) \to L^2(\rho)$ is Lipschitz continuous in operator norm: $\|L_{\theta'} - L_\theta\|_{\mathrm{op}} = O(\|\theta' - \theta\|)$.
\end{lemma}

\begin{theorem}
\label{th:space_bound}
Let Assumptions A3 and A5 hold for $k_\theta$. Let $\lambda_i(\theta)$ be a distinct eigenvalue of $L_\theta$ with multiplicity $m_i$, and define the spectral gap $\eta_i = \frac{1}{2}\min_{l \neq i} |\lambda_l(\theta) - \lambda_i(\theta)|$. If the perturbation satisfies $\|\delta L\|_{\mathrm{op}} < \eta_i$, the following bounds hold:
\begin{enumerate}
    \item \textbf{(Projection Stability)} The perturbation of the spectral projection operator is bounded by: $\| P_i(\theta') - P_i(\theta)\|_{\mathrm{op}} = O\left(\|\theta' - \theta\| / \eta_i\right)$.
    \item \textbf{(Eigenvalue Stability)} For any perturbed eigenvalue $\lambda_{i,j}(\theta')$ corresponding to the $j$-th eigenfunction in the $i$-th block: $|\lambda_{i,j}(\theta') - \lambda_i(\theta)| = O(\|\theta' - \theta\|)$.
    \item \textbf{(Contribution Stability)} The change in the $i$-th spectral block's contribution to the squared MMD, defined as $C_i(\theta) = \lambda_i(\theta) \|P_i(\theta) h\|_{\rho}^2$, is bounded by: $|C_i(\theta') - C_i(\theta)| = O\!\left(\lambda_i \|\theta' - \theta\| / \eta_i\right) + O(\|\theta' - \theta\|)$.
\end{enumerate}
\end{theorem}
Theorem~\ref{th:space_bound} indicates that for leading eigenvalues characterized by large spectral gaps $\eta_i$, the contribution $C_i(\theta)$ is highly stable under small perturbations of $\theta$. Consequently, since $ D^d(\theta) = \sum_{i=1}^{v} C_i(\theta) $, Theorem \ref{th:space_bound} implies $ |D^d(\theta') - D^d(\theta)| = O\!\left(\|\theta' - \theta\| / \eta_i\right) + O(\|\theta' - \theta\|) $.

\subsection{Perturbation of the Test Statistic}
\label{subsec:perturbation_statistic}

We now extend the perturbation analysis to the finite-sample test statistic, beginning with the smoothness of the empirical Gram matrix. 

\begin{lemma}
\label{lm:gram_perturb}
Let Assumptions A3 and A5 hold for $k_\theta$, and let $ K(\theta) \in \mathbb{R}^{N \times N} $ be the scaled Gram matrix with entries $ K(\theta)_{\alpha\beta} = k_{\theta}(z_\alpha, z_\beta) / N $. Then, $\| K(\theta') - K(\theta)\|_{\mathrm{op}} = O(\|\theta' - \theta\|)$.
\end{lemma}

Let $\Pi_d(\theta)$ denote the orthogonal projection matrix onto the subspace spanned by the leading $d$ empirical eigenvectors of $K(\theta)$, and define the empirical spectral gap as $\hat{\eta}_d(\theta) = (\hat{\lambda}_d(\theta) - \hat{\lambda}_{d+1}(\theta))/2$. Conditional on the sample, applying Theorem \ref{th:perturbation_space} in Appendix~\ref{app:perturbation_results}, if $\| K(\theta') - K(\theta)\|_{\mathrm{op}} < \hat{\eta}_d(\theta)$, we obtain $\| \Pi_d(\theta) - \Pi_d(\theta')\|_{\mathrm{op}} = O\left(\| \theta - \theta'\| / \hat{\eta}_d(\theta)\right)$.

Let $\tilde{D}_N^d(\theta)$ be the unbiased test statistic parameterized by $\theta$. As established in Remark \ref{rk:biased_and_unbiased}, it decomposes into a quadratic form and trace correction terms: $\tilde{D}_N^d(\theta) =  D_N^d(\theta) -\frac{N}{n(n-1)} \mathrm{Tr}(\bar{K}_{XX}^{(d)}(\theta))-\frac{N}{m(m-1)} \mathrm{Tr}(\bar{K}_{YY}^{(d)}(\theta))$. Let $\delta \in \mathbb{R}^N$ be the vector with entries $\delta_\alpha = 1/n$ if $z_\alpha \sim P$, and $\delta_\alpha = -1/m$ if $z_\alpha \sim Q$. Note $\delta$ is independent of $\theta$. The biased statistic is then $D_N^d(\theta) = N \delta^\top\Pi_{d}(\theta) \bigl(N K(\theta)\bigr) \Pi_{d}(\theta) \delta$.

\begin{theorem}
\label{th:statistic_bound}
Under the conditions of Lemma \ref{lm:gram_perturb}, the following holds:
\begin{enumerate}
    \item \textbf{(Conditional Bound)} For any fixed dataset such that $L\|\theta' - \theta\| < \hat{\eta}_d(\theta)$, we have:
    \begin{equation*}
    |\tilde{D}_N^d(\theta) - \tilde{D}_N^d(\theta')| = O\left( \frac{\| \theta - \theta'\|}{\hat{\eta}_d(\theta)}\right) + O(\|\theta - \theta'\|).
    \end{equation*}
    \item \textbf{(Unconditional Bound)} Assuming the population spectral gap $\eta_d(\theta) = \frac{1}{2}(\lambda_d(\theta) - \lambda_{d+1}(\theta)) > 0$, we have:
    \begin{equation*}
    |\tilde{D}_N^d(\theta) - \tilde{D}_N^d(\theta')| = O_p\left(\frac{\| \theta - \theta'\|}{\eta_d(\theta)} \right)+ O(\|\theta - \theta'\|).
    \end{equation*}
\end{enumerate}
\end{theorem}

\subsection{Perturbation of the Limiting Distributions}
\label{subsec:perturbation_dist}

To quantify the discrepancy between the limiting null distributions of $\tilde{D}^d_N(\theta)$ and $\tilde{D}^d_N(\theta')$, we employ the Wasserstein-1 distance, $W_1$. By Theorem~\ref{null_distribution_test_statistic}, these limiting distributions are centered weighted sums of independent chi-squared random variables:
\begin{equation*}
F_\theta \sim \sum_{i=1}^d \frac{\sigma_i(\theta)}{p(1-p)}(\chi^2_{1,i}-1), \quad F_{\theta'} \sim \sum_{i=1}^d \frac{\sigma_i(\theta')}{p(1-p)}(\chi^2_{1,i}-1).
\end{equation*}
The weights $\sigma_i(\theta)$ are the eigenvalues of the restricted RKHS covariance matrix $\Sigma_d(\theta)$ with entries $(\Sigma_d(\theta))_{ij} = \text{Cov}(e_i(Z), e_j(Z))$, where $e_i \in \mathcal{H}_\theta$. The Wasserstein-1 distance between $F_\theta$ and $F_{\theta'}$ is $W_1(F_\theta, F_{\theta'}) = \inf_{\gamma \in \Gamma(F_\theta, F_{\theta'})} \mathbb{E}_{(U,V) \sim \gamma}[\|U - V\|] \le \sqrt{ \mathbb{E} \left[ \sum_{i=1}^d \left(\sigma_i(\theta) - \sigma_i(\theta')\right) (\chi^2_{1,i} - 1) \right]^2 } /(p(1-p))$, where $\Gamma(F_\theta, F_{\theta'})$ is the set of all joint distributions with marginals $F_\theta$ and $F_{\theta'}$.

Bounding $W_1(F_\theta, F_{\theta'})$ requires bounding the difference between the weight vectors $\sigma(\theta)$ and $\sigma(\theta')$, which depends on the operator norm difference between $\Sigma_\theta$ and $\Sigma_{\theta'}$. A fundamental technical challenge arises here: these two covariance operators act on distinct Hilbert spaces $\mathcal{H}_\theta$ and $\mathcal{H}_{\theta'}$, rendering their operator norm difference ill-defined. To circumvent this, we map the perturbation analysis into the fixed, parameter-independent space $L^2(\rho)$.

Define the $L^2(\rho)$ feature map $\Phi_\theta : \mathcal{Z} \to L^2(\rho)$ by $\Phi_\theta(z) = k_\theta(z, \cdot)$, and let the $L^2(\rho)$ mean embedding be $\nu_\theta = \mathbb{E}[\Phi_\theta(Z)] = L_\theta \mathbf{1}$, where $\mathbf{1}$ is the constant function. We define the $L^2(\rho)$ covariance operator $\boldsymbol{C}_\theta : L^2(\rho) \to L^2(\rho)$ as $ \boldsymbol{C}_\theta = \mathbb{E} \left[ (\Phi_\theta(Z) - \nu_\theta) \otimes_{\rho} (\Phi_\theta(Z) - \nu_\theta) \right]. $ 
Since $\boldsymbol{C}_\theta$ and $\boldsymbol{C}_{\theta'}$ share the same domain and codomain, $\|\boldsymbol{C}_{\theta'} - \boldsymbol{C}_\theta\|_{\mathrm{op}}$ is well-defined. We now verify that $\boldsymbol{C}_\theta$ is a well-defined covariance operator and, crucially, relate it to the RKHS covariance matrix $\Sigma_d(\theta)$.

\begin{proposition}
\label{prop: l_2_covariance}
Let Assumptions A3 and A5 hold for $k_\theta$. Then the $L^2(\rho)$ covariance operator $\boldsymbol{C}_\theta$ is a self-adjoint, positive semi-definite, trace-class operator on $L^2(\rho)$ that admits the explicit decomposition: $\boldsymbol{C}_\theta = L_\theta^2 - \nu_\theta \otimes_{\rho} \nu_\theta$. Furthermore, let $P_d(\theta)$ be the orthogonal projection onto the span of the first $d$ eigenfunctions $\{\phi_i(\theta)\}_{i=1}^d$ of $L_\theta$, and let $\Lambda_d^{-1/2}(\theta)$ be the diagonal operator acting on this span defined by $\Lambda_d^{-1/2}(\theta) \phi_i(\theta) = \lambda_i^{-1/2}(\theta) \phi_i(\theta)$. Define the operator $ T_\theta = \Lambda_d^{-1/2}(\theta) P_d(\theta) \boldsymbol{C}_\theta P_d(\theta) \Lambda_d^{-1/2}(\theta). $  Let $\Sigma_d(\theta)$ be the $d \times d$ RKHS covariance matrix defined in Theorem~\ref{null_distribution_multiple}. Then $\Sigma_d(\theta)$ is precisely the matrix representation of $T_\theta$ in the orthonormal basis $\{\phi_i(\theta)\}_{i=1}^d$. Consequently, by Theorem~\ref{th:spectral_equivalence} in Appendix~\ref{app:matrix_representation}, the asymptotic variance weights $\sigma_i(\theta)$ are exactly the eigenvalues of the operator $T_\theta$.
\end{proposition}

Because the weights $\sigma_i(\theta)$ are the eigenvalues of $T_\theta$, we can bound their perturbation on the operators $T_\theta$ and $T_{\theta'}$. This requires bounding $\|T_{\theta'} - T_\theta\|_{\mathrm{op}}$, which in turn relies on the Lipschitz continuity of $\boldsymbol{C}_\theta$.

\begin{lemma}
\label{lm:l2_cov_bound}
Let Assumptions A3 and A5 hold for $k_\theta$. Then, the $L^2(\rho)$ covariance operator is Lipschitz continuous in operator norm: $\|\boldsymbol{C}_{\theta'} - \boldsymbol{C}_\theta\|_{\mathrm{op}} = O(\|\theta' - \theta\|)$.
\end{lemma}

We now possess the necessary machinery to quantify the discrepancy between the limiting null distributions. By applying Theorem~\ref{th:perturbation_eigenvalue} to the operators $T_\theta$ and $T_{\theta'}$, we can bound the $\ell_2$-norm of the weight differences, which directly bounds the Wasserstein-1 distance.

\begin{theorem}
\label{th:perturbation_distribution}
Let $F_\theta$ and $F_{\theta'}$ be the limiting null distributions of $\tilde{D}^d_N(\theta)$ and $\tilde{D}^d_N(\theta')$, respectively. Under the conditions of Lemma~\ref{lm:l2_cov_bound}, and assuming the population spectral gap satisfies $\eta_d = \frac{1}{2}(\lambda_d(\theta) - \lambda_{d+1}(\theta)) > 0$, the Wasserstein-1 distance between $F_\theta$ and $F_{\theta'}$ is bounded by:
\begin{equation*}
W_1(F_\theta, F_{\theta'}) = O\left( \frac{\|\theta - \theta'\|}{\eta_d} \right) + O(\|\theta - \theta'\|).
\end{equation*}
\end{theorem}

It is imperative to highlight that the entire perturbation framework established above fundamentally rests on a single, critical assumption: the discrepancy between the operators (or empirical Gram matrices) induced by varying kernel parameters must be strictly smaller than the corresponding spectral gaps. Mathematically, this requires $\|\delta L\|_{\mathrm{op}} < \eta_i$ at the population level and $\|K(\theta') - K(\theta)\|_{\mathrm{op}} < \hat{\eta}_d$ at the sample level. Under this regime, the standard perturbation theorems apply, yielding the unified conclusion that the stability of the population contributions, the test statistic, and the limiting null distributions all scale inversely with the spectral gap--specifically, at a rate of $O(\|\theta - \theta'\| / \eta)+ O(\|\theta - \theta'\|)$.

\section{Simulation and Empirical Study}
\label{sec:simulation}

\subsection{Simulation Studies}

We evaluate the finite-sample performance of $\tilde{D}_N^d$ under both fixed and data-driven truncation levels, comparing it against state-of-the-art kernel 
tests. Unless specified, all methods use the Gaussian kernel $k(z,z') = \exp(-\|z-z'\|_2^2/\sigma)$ for $\sigma > 0$. The simulations comprise three parts: (i) balanced sample comparisons; (ii) unbalanced sample comparisons; and (iii) robustness to kernel bandwidth. Additional results are deferred to the Online Appendix.

\paragraph{Data Generating Processes.}
We consider four null and five alternative scenarios. Let $X \in \mathbb{R}^{n \times q}$ and $Y \in \mathbb{R}^{m \times q}$, where $q \in \{50, 100, 500, 1000\}$ is the data dimension. The null distributions are:
\begin{align*}
    \mathrm{DGP}_{01}:&\; X_{ij}, Y_{ij} \overset{\mathrm{i.i.d.}}{\sim} \mathcal{N}(0,1), &
    \mathrm{DGP}_{02}:&\; X_{ij}, Y_{ij} \overset{\mathrm{i.i.d.}}{\sim} t_5, \\
    \mathrm{DGP}_{03}:&\; X_{ij}, Y_{ij} \overset{\mathrm{i.i.d.}}{\sim} \chi^2_3, &
    \mathrm{DGP}_{04}:&\; X_{ij}, Y_{ij} \overset{\mathrm{i.i.d.}}{\sim} \mathrm{Poisson}(1).
\end{align*}
We focus on the balanced setting $m = n = 100$ in the main text; unbalanced results ($n = 100, m = 10$) are in the Online Appendix. For power evaluation, we set $q_{\mathrm{frac}} = \lfloor 0.1q \rfloor$ so that 10\% of dimensions deviate between samples:

\begin{enumerate}
    \item \textbf{$\mathrm{DGP}_1$ (Location-Scale).} $X_{ij} \sim \mathcal{N}(0,1)$. $Y$ is identical except its first $q_{\mathrm{frac}}$ columns are $\mathcal{N}(\mu, \sigma^2)$. We test $(\mu, \sigma^2) \in \{(0.05, 0.5), (0.1, 1.3), (-0.05, 0.6)\}$.
    \item \textbf{$\mathrm{DGP}_2$ ($t$-Distribution).} $X$ concatenates $X_1 \in \mathbb{R}^{n \times (q - q_{\mathrm{frac}})}$ ($\mathcal{N}(0,1)$) and $X_2 \in \mathbb{R}^{n \times q_{\mathrm{frac}}}$ ($t_{df}$). $Y_{ij} \sim \mathcal{N}(0,1)$. We test $df \in \{3, 5, 10\}$.
    \item \textbf{$\mathrm{DGP}_3$ (Mixture).} Rows of $X$ are $0.8 X_{1i} + 0.2 X_{2i}$ ($X_{1i} \sim \mathcal{N}(0, \Sigma_0)$, $\Sigma_0 = (0.5^{|i-j|})$, $X_{2i} \sim t_5$). Rows of $Y$ are $0.8 Y_{1i} + 0.2 Y_{2i}$ ($Y_{1i} \sim \mathcal{N}(a, b\Sigma_0)$, $Y_{2i} \sim t_3$). We test $(a, b) \in \{(-0.05, 0.8), (0, 1.1), (0.05, 1.05)\}$.
    \item \textbf{$\mathrm{DGP}_4$ (Scale-Only).} A special case of $\mathrm{DGP}_1$ with $\mu = 0$. We test $\sigma^2 \in \{0.6, 0.8, 1.3\}$.
    \item \textbf{$\mathrm{DGP}_5$ (Location-Only).} A special case of $\mathrm{DGP}_1$ with $\sigma^2 = 1$. We test $\mu \in \{-1.0, 0.6, 1.3\}$.
\end{enumerate}

\paragraph{Comparisons under Balanced Sample Size.}
We set $d=1$ for the proposed test, denoted \texttt{TMMD(d=1)}, motivated by the Gaussian kernel's eigenvalue decay rate and the spectral gap condition in Theorem~\ref{th:perturbation_space} (see Appendix~\ref{app:spectral_gap}). Competing MMD-based tests include: the standard quadratic-time \texttt{MMD}; \texttt{MMDAgg} \citep{schrab2023mmd}, aggregating over multiple kernels; \texttt{MMDAggInc} \citep{schrab2022efficient}, an efficient variant using incomplete U-statistics; \texttt{MMD-FUSE} \citep{biggs2023mmd}, combining normalized MMD values via a weighted soft maximum; \texttt{Mahalanobis-MMD} \citep{chatterjee2025boosting}, using a Mahalanobis distance; \texttt{Martingale-MMD} \citep{chatterjee2025martingale}, based on a martingale interpretation; \texttt{BMMD} \citep{zaremba2013b}, averaging subset MMD statistics; \texttt{xMMD} \citep{shekhar2022permutation}, a studentized sample-splitting test; \texttt{SpectrumMMD} \citep{gretton2009fast}, using kernel spectrum eigenvalues; and \texttt{SpectraRegu} \citep{hagrass2024spectral}, incorporating covariance via spectral regularization.

To compare data-driven truncation, we include the sample-splitting test of \cite{sutherland2016generative}, in which the bandwidth maximizes an approximate power function. Both oracle tests double the sample size, using half for learning and half for testing, ensuring equal effective test sizes. These represent the ``best'' truncated and full-rank MMD versions. We denote our data-driven test as \texttt{TMMD-Oracle} (candidate bound $\bar{d}=5$\footnote{Performance is robust to $\bar{d}$; see Appendix~\ref{app:d_bar_selection}.}) and Sutherland's as \texttt{MMD-Oracle}.

Size performance under the null (Figure~\ref{fig:size_balanced}, Appendix~\ref{app:size}) shows that most of the tests evaluated at the 5\% nominal level are well calibrated. \texttt{TMMD(d=1)} and \texttt{TMMD-Oracle} maintain rejection rates between 3.6\% and 6.3\% across dimensions and null DGPs, whereas \texttt{SpectrumMMD} and \texttt{Mahalanobis-MMD} are severely conservative.

\begin{figure}[H]
    \centering
    \begin{subfigure}{\textwidth}
        \centering
        \includegraphics[width=1.0\textwidth]{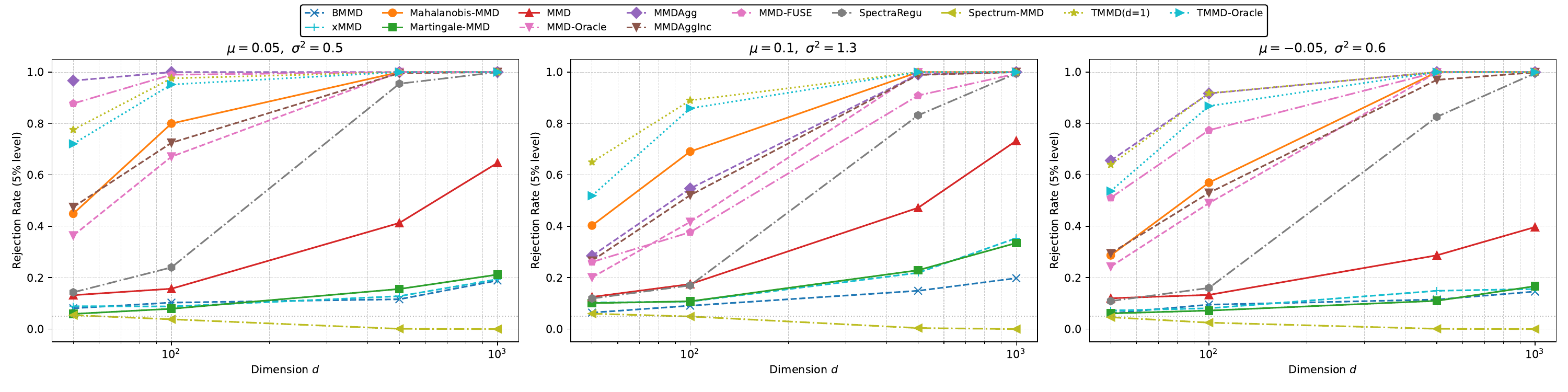}
        \caption{$\mathrm{DGP}_1$}
    \end{subfigure}
    \vspace{0.25cm}
    \begin{subfigure}{\textwidth}
        \centering
        \includegraphics[width=1.0\textwidth]{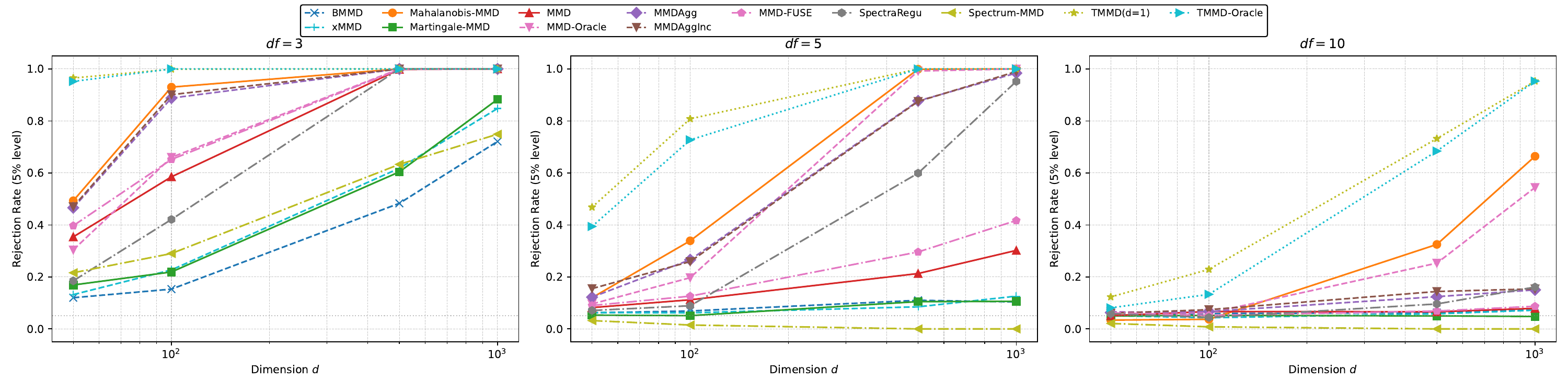}
        \caption{$\mathrm{DGP}_2$}
    \end{subfigure}
    \vspace{0.25cm}
    \begin{subfigure}{\textwidth}
        \centering
        \includegraphics[width=1.0\textwidth]{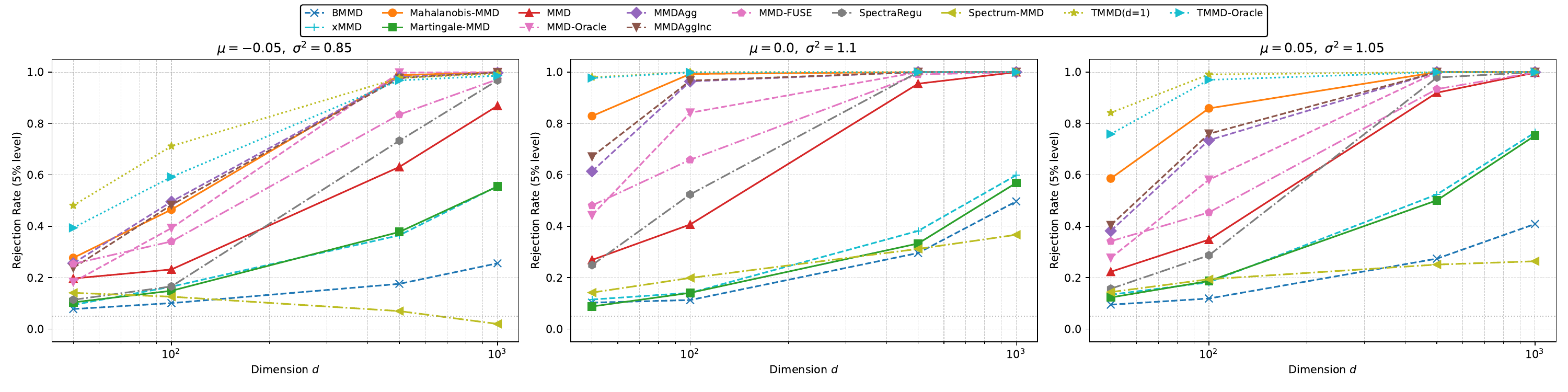}
        \caption{$\mathrm{DGP}_3$}
    \end{subfigure}  
    \caption{Empirical power under Set1---Set3 with balanced sample size ($n = m = 100$).}
    \label{fig:power_balanced_1_3}
\end{figure}

\begin{figure}[H]
    \centering
    \begin{subfigure}{\textwidth}
        \centering
        \includegraphics[width=1.0\textwidth]{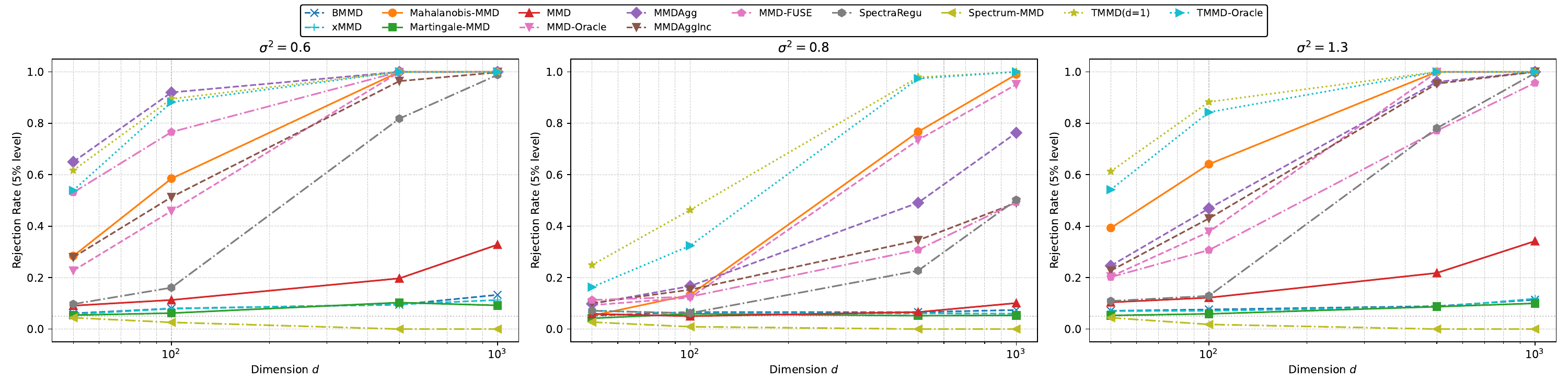}
        \caption{$\mathrm{DGP}_4$}
    \end{subfigure}
    \vspace{0.25cm}
    \begin{subfigure}{\textwidth}
        \centering
        \includegraphics[width=1.0\textwidth]{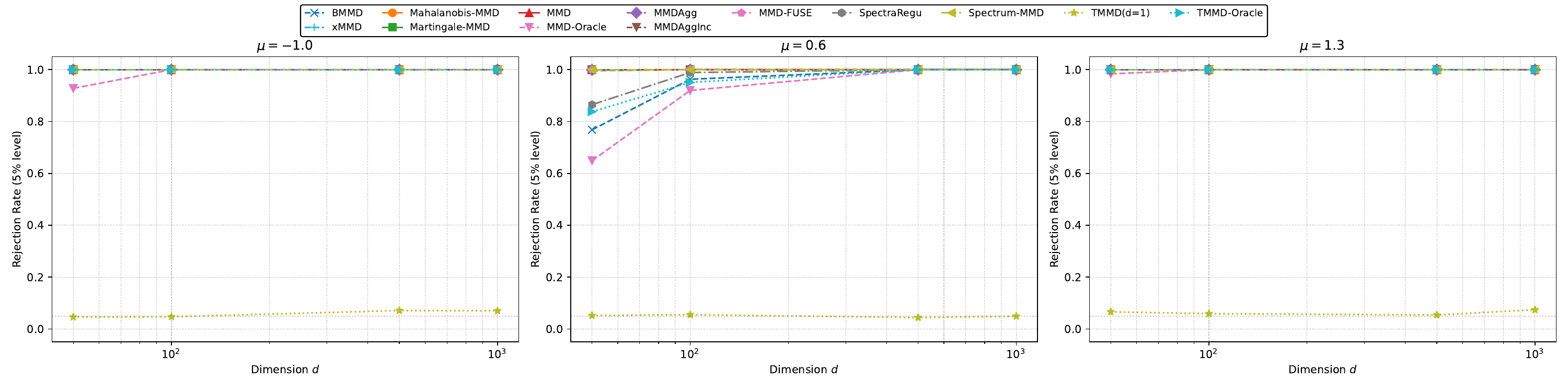}
        \caption{$\mathrm{DGP}_5$}
    \end{subfigure}
    \caption{Empirical power under Set4---Set5 with balanced sample size ($n = m = 100$).}
    \label{fig:power_balanced_4_5}
\end{figure}

Power under alternatives is summarized in Figures~\ref{fig:power_balanced_1_3} and \ref{fig:power_balanced_4_5}. Across Sets~1--4, \texttt{TMMD(d=1)} and \texttt{TMMD-Oracle} demonstrate overwhelming superiority. In the difficult Scale-Only case ($\sigma^2=0.8$), they are the only methods with substantial power (46.3\% and 32.4\% at $q=100$). In Set~2 ($df=10$), both attain 95.3\% power at $q=1000$. In Sets~1 and 3, \texttt{TMMD(d=1)} consistently achieves near-perfect power by $q=500$. While \texttt{MMD-FUSE} occasionally yields slightly higher power in specific parameterizations (e.g., 87.8\% vs.\ 77.6\% in Set~1), \texttt{TMMD(d=1)} frequently outpaces aggregators at lower dimensions.

Notably, \texttt{TMMD(d=1)} often matches or slightly exceeds \texttt{TMMD-Oracle} in Sets~1--4. This occurs because the oracle predominantly selects $d=1$ for these alternatives (Appendix~\ref{app:freq_d}), indicating deviations are captured by a single leading direction. Fixing $d=1$ a priori bypasses the estimation noise inherent in the data-driven selection. 

\begin{figure}
    \centering
    \includegraphics[width=1.0\linewidth]{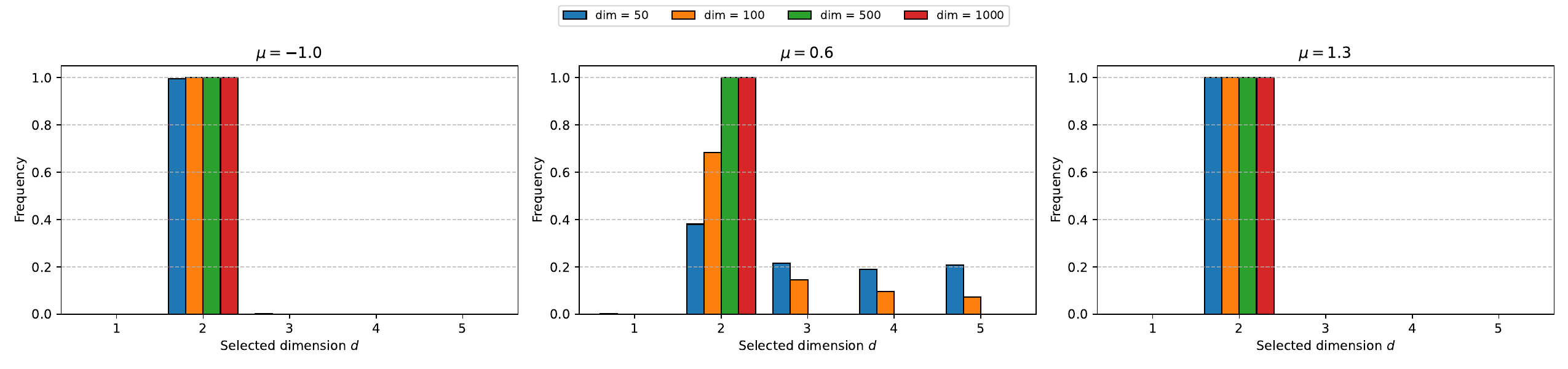}
    \caption{Frequency of truncation dimension $d$ in Set5 with balanced sample size.}
    \label{fig:d_selection_5}
\end{figure}

Set~5 (Location-Only) reveals a striking contrast: \texttt{TMMD(d=1)} exhibits trivial power (4.4\%--5.5\% for $\mu=0.6$), whereas \texttt{TMMD-Oracle} achieves 83.8\%--100.0\% by selecting $d>1$. This blind spot arises because pure mean shifts likely span multiple eigen-directions, requiring higher dimensions to capture. The failure of \texttt{TMMD(d=1)} is not a deficiency of the truncated framework, but of fixing $d=1$ universally. The oracle successfully adapts, retaining $d=1$ for Sets~1--4 while expanding $d$ for Set~5. Figure~\ref{fig:d_selection_5} reports the frequencies of selected truncation dimension $d$ for Set~5, for other Sets, see Appendix~\ref{app:freq_d}.

\paragraph{Comparisons under Unbalanced Sample Size.}
Under the unbalanced setting ($n=100, m=10$, Appendix~\ref{app:unbalanced}), all tests lose power, yet \texttt{TMMD(d=1)} and \texttt{TMMD-Oracle} remain most powerful. For DGPs 1--4, \texttt{TMMD(d=1)}'s advantage is even more pronounced because full-rank tests suffer severe power collapses when $m=10$. For instance, under DGP$_1$ ($\mu=0.05, \sigma^2=0.5$), \texttt{TMMD(d=1)} reaches 99.9\% power at $q=1000$, while \texttt{MMDAgg} drops to 91.0\% and standard \texttt{MMD} to 10.5\%. Under the difficult scale-only deviation ($\sigma^2=0.8$), \texttt{TMMD(d=1)} retains 69.2\% power, whereas \texttt{MMDAgg} falls to 11.3\%. \texttt{TMMD-Oracle} consistently underperforms relative to the fixed $d=1$ version here (e.g., 58.7\% vs.\ 69.2\% for DGP$_4$).

Under DGP$_5$ (Location-Only), a fascinating finite-sample phenomenon emerges. Unlike the balanced setting, \texttt{TMMD(d=1)} exhibits non-negligible power, reaching 100.0\% at $q=1000$ for $\mu=-1.0$. Nonetheless, \texttt{TMMD-Oracle} correctly adapts and outperforms at lower dimensions (e.g., 94.5\% vs.\ 81.0\% at $q=100$ for $\mu=1.3$). Full-rank tests perform well on DGP$_5$ but completely forfeit power under higher-moment alternatives in DGPs 1--4.

\begin{figure}[H]
    \centering
    \begin{subfigure}{\textwidth}
        \centering
        \includegraphics[width=1.0\textwidth]{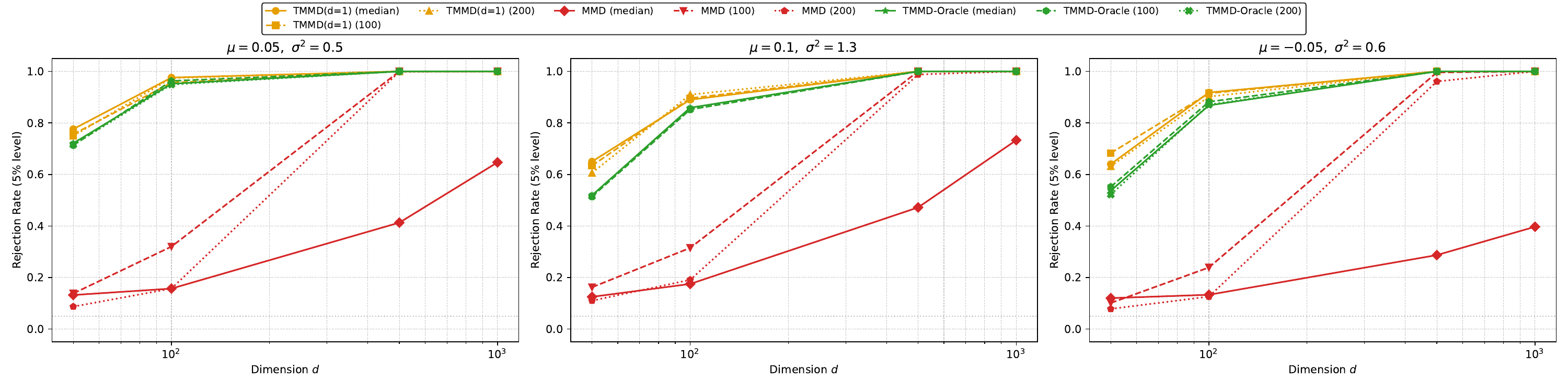}
        \caption{$\mathrm{DGP}_1$}
    \end{subfigure}
    \vspace{0.25cm}
    \begin{subfigure}{\textwidth}
        \centering
        \includegraphics[width=1.0\textwidth]{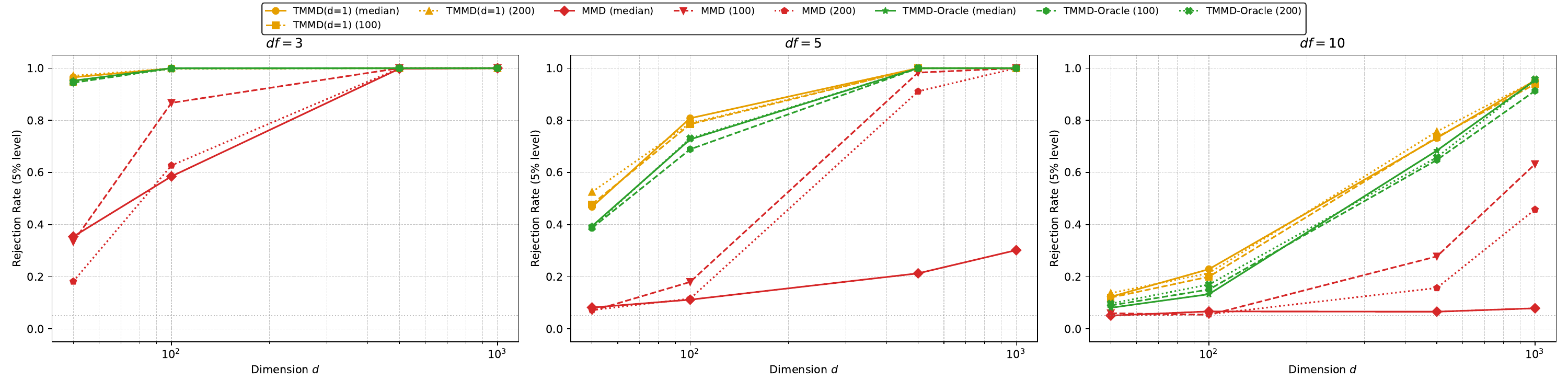}
        \caption{$\mathrm{DGP}_2$}
    \end{subfigure}
    \vspace{0.25cm}
    \begin{subfigure}{\textwidth}
        \centering
        \includegraphics[width=1.0\textwidth]{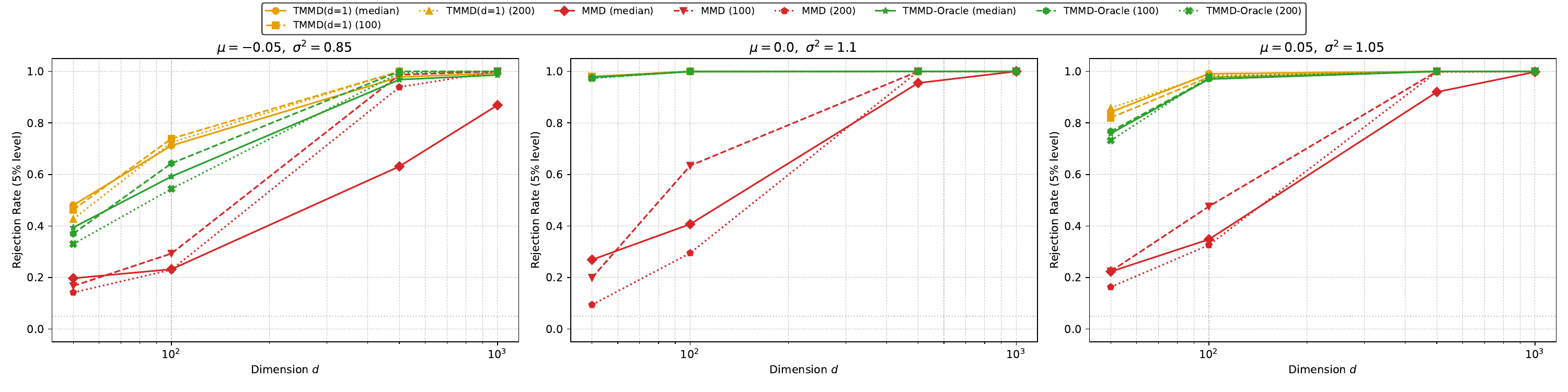}
        \caption{$\mathrm{DGP}_3$}
    \end{subfigure}
    \caption{Empirical power under different kernel choices for Set1---Set3 with balanced sample size ($n = m = 100$).}
    \label{fig:power_kernel_choice_1_3}
\end{figure}

\begin{figure}[H]
    \centering
    \begin{subfigure}{\textwidth}
        \centering
        \includegraphics[width=1.0\textwidth]{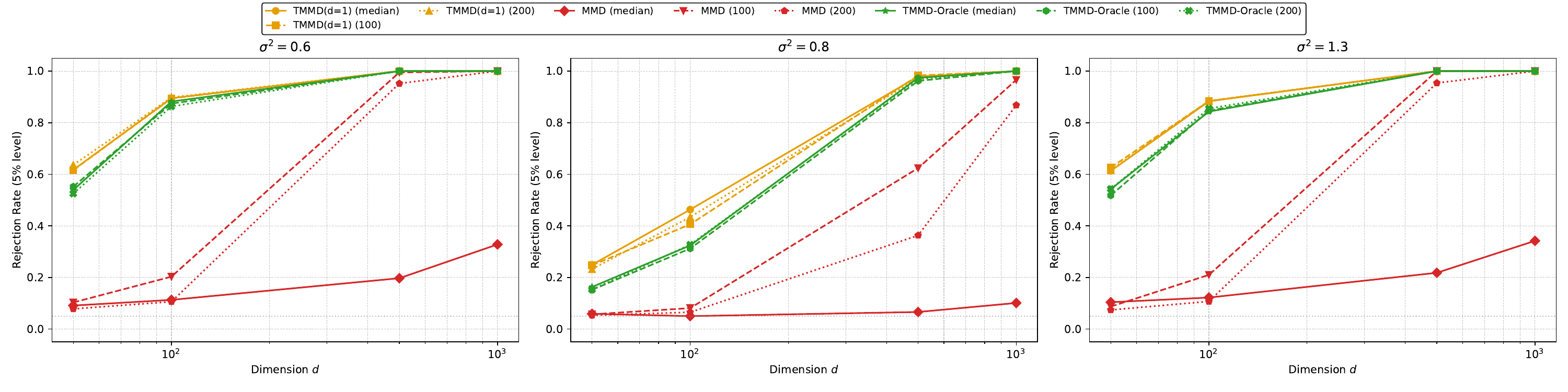}
        \caption{$\mathrm{DGP}_4$}
    \end{subfigure}
    \vspace{0.25cm}
    \begin{subfigure}{\textwidth}
        \centering
        \includegraphics[width=1.0\textwidth]{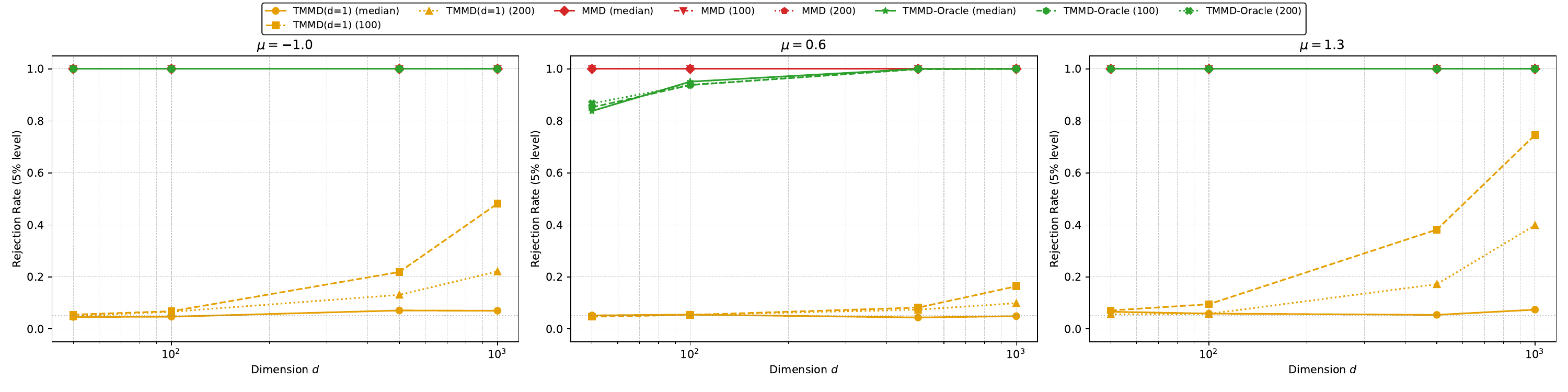}
        \caption{$\mathrm{DGP}_5$}
    \end{subfigure}
    \caption{Empirical power under different kernel choices for Set4---Set5 with balanced sample size ($n = m = 100$).}
    \label{fig:power_kernel_choice_4_5}
\end{figure}

\paragraph{Robustness to Kernel Choice.} 
We evaluate sensitivity to the Gaussian kernel bandwidth using the median heuristic and fixed bandwidths ($\sigma=100, 200$) under balanced settings (Figures~\ref{fig:power_kernel_choice_1_3} and \ref{fig:power_kernel_choice_4_5}). The truncated MMD framework exhibits inherent robustness: for DGPs 1--4, power remains remarkably stable. For instance, under the scale-only deviation ($\sigma^2=0.8$) at $q=100$, \texttt{TMMD(d=1)} achieves 46.3\%, 40.6\%, and 43.4\% across the three bandwidths, respectively, reaching 100\% at $q=1000$ for all. \texttt{TMMD-Oracle} shows parallel insensitivity (power ranging from 31.2\% to 32.7\% for the same scenario). This stability alleviates the need for delicate bandwidth tuning.

Conversely, the standard full-rank \texttt{MMD} is extremely sensitive. Under the median heuristic, its power is severely compromised (10.1\% for the scale-only deviation at $q=1000$). Inflating the bandwidth to $\sigma=100$ explosively increases power to 96.6\%. A Taylor expansion reveals that for large $\sigma$, the centered kernel approximates $2\|x-y\|_2^2/\sigma$, reducing the test to one based on squared Euclidean distances. While standard \texttt{MMD} \emph{can} detect these alternatives with an oracle bandwidth, it is practically fragile; truncated frameworks succeed by default.

Under DGP$_5$, bandwidth sensitivities diverge. With the median heuristic, \texttt{TMMD(d=1)} maintains its structural blind spot (power near 5\%). At $\sigma=100$, it mimics the linearization effect and achieves up to 74.6\% power. In contrast, \texttt{TMMD-Oracle} maintains near-perfect power across all bandwidths (100.0\% at $q=1000$ for $\mu=1.3$), confirming its data-driven selection remains highly stable even under heavy kernel linearization.

\paragraph{Time Efficiency.}
Finally, we compare computational efficiency. Benchmarked on Sets 1--3 (12 timing cases per set), sample sizes for \texttt{TMMD-Oracle} and \texttt{MMD-Oracle} are doubled to $n=m=200$ for sample-splitting, while others use $n=m=100$. In addition to the methods above, we include \texttt{LinearTimeMMD} \citep{gretton2012kernel}, a linear-time MMD estimator, which is known for its computational efficiency but often suffers from low power (hence, we did not include it in the finite sample performance comparison). The results are summarized in Table~\ref{tab:tmmd_overall_comparison}.

\texttt{TMMD(d=1)} is among the fastest, with a mean per-case runtime of 0.00320s. \texttt{TMMD-Oracle} averages 0.01079s ($3.38\times$ slower). Compared to permutation \texttt{MMD} (0.04672s), \texttt{TMMD(d=1)} offers a $14.61\times$ speedup.  Overall, \texttt{TMMD(d=1)} delivers near-minimal cost, while \texttt{TMMD-Oracle} remains markedly faster than permutation-based and spectral-regularized baselines.

\begin{table}[htbp]
\centering
\small
\caption{Overall runtime comparison: mean per-case elapsed real time (seconds) averaged over Set1--Set3}
\setlength{\tabcolsep}{10pt}
\begin{tabular}{lccccc}
\hline
Method & Set1 & Set2 & Set3 & Mean (s) & Relative to \texttt{TMMD(d=1)} \\
\hline
\texttt{BMMD}            & 0.00180 & 0.00179 & 0.00208 & 0.00189 & $0.59\times$ \\
\texttt{LinearTimeMMD}    & 0.00284 & 0.00300 & 0.00296 & 0.00294 & $0.92\times$ \\
\texttt{TMMD(d=1)}       & 0.00316 & 0.00304 & 0.00340 & 0.00320 & $1.00\times$ \\
\texttt{xMMD}            & 0.00377 & 0.00346 & 0.00378 & 0.00367 & $1.15\times$ \\
\texttt{Martingale-MMD}  & 0.00500 & 0.00467 & 0.00503 & 0.00490 & $1.53\times$ \\
\texttt{TMMD-Oracle}     & 0.01077 & 0.01020 & 0.01140 & 0.01079 & $3.38\times$ \\
\texttt{SpectrumMMD}     & 0.01885 & 0.01193 & 0.01120 & 0.01399 & $4.38\times$ \\
\texttt{Mahalanobis-MMD} & 0.01780 & 0.01748 & 0.01780 & 0.01769 & $5.53\times$ \\
\texttt{MMD-FUSE}        & 0.01902 & 0.01801 & 0.02023 & 0.01909 & $5.97\times$ \\
\texttt{MMDAgg}          & 0.03492 & 0.02725 & 0.02631 & 0.02949 & $9.22\times$ \\
\texttt{MMDAggInc}       & 0.04451 & 0.03448 & 0.03815 & 0.03905 & $12.21\times$ \\
\texttt{MMD}             & 0.04475 & 0.04672 & 0.04869 & 0.04672 & $14.61\times$ \\
\texttt{MMD-Oracle}      & 0.10665 & 0.10459 & 0.10968 & 0.10697 & $33.46\times$ \\
\texttt{SpectraRegu}     & 0.73371 & 0.73015 & 0.74042 & 0.73476 & $229.78\times$ \\
\hline
\end{tabular}
\label{tab:tmmd_overall_comparison}
\end{table}

\subsection{Empirical Studies}

We apply the proposed tests to three microarray datasets \citep{Khan:2001cz, Gordon:2002wr, Chin:2006bt} from the R package \texttt{datamicroarray}. The \cite{Khan:2001cz} dataset concerns small round blue cell tumors (63 samples, 2308 genes; testing EWS vs.\ RMS). The \cite{Gordon:2002wr} dataset contains lung cancer profiles (180 samples, 12533 genes; testing positive vs.\ negative). The \cite{Chin:2006bt} dataset involves breast cancer (118 samples, 22215 genes; testing positive vs.\ negative). We report $p$-values for \texttt{TMMD(d=1)}, \texttt{TMMD-Oracle}, and competing tests\footnote{\texttt{MMDAgg} and \texttt{MMDAggInc} are excluded as their $p$-values are difficult to compute.} in Table~\ref{tab:real_data_pvalues}.

\begin{table}[htbp]
\centering
\begin{threeparttable}
\caption{$p$-values of two-sample tests applied to microarray datasets}
\label{tab:real_data_pvalues}
\setlength{\tabcolsep}{15pt} 
\begin{tabular}{@{}lcccccc@{}}
\toprule
Dataset & T(1) & T-O & MMD & M-O & M-M & MMD-FUSE \\
\midrule
Chin et al. (2006) & 0.000 & 0.000 & 0.000 & 0.002 & 0.000 & 0.002 \\
Khan et al. (2001) & 0.028 & 0.004 & 0.000 & 0.040 & 0.000 & 0.002 \\
Gordon et al. (2002) & 0.000 & 0.000 & 0.000 & 0.084 & 0.000 & 0.002 \\
\bottomrule
\end{tabular}
\begin{tablenotes}[para,flushleft]
\item T(1): \texttt{TMMD(d=1)}; T-O: \texttt{TMMD-Oracle}; M-O: \texttt{MMD-Oracle}; M-M: \texttt{MahalanobisMMD}.
\end{tablenotes}
\end{threeparttable}
\end{table}

Across all datasets, \texttt{TMMD(d=1)} and \texttt{TMMD-Oracle} strongly reject the null hypothesis. For the Chin and Gordon datasets, both achieve $p < 0.001$. For Khan, \texttt{TMMD-Oracle} attains the smallest $p$-value among non-zero entries ($p = 0.004$). While standard \texttt{MMD} also rejects, its oracle version exhibits a larger $p$-value for Gordon ($p = 0.084$). \texttt{MMD-FUSE} and \texttt{MahalanobisMMD} produce identical outcomes across the applications.
\section{Conclusion}
\label{sec:conclusion}
In this paper, we introduced a novel kernel-based two-sample test that leverages directional components analysis to address the finite-sample inefficiencies of standard MMD tests. By recognizing that trailing directional components are often poorly estimated and introduce substantial noise, our approach truncates the spectral decomposition of the MMD to retain only the leading, well-estimated eigen-directions. This targeted aggregation of robust components significantly enhances the test's power and reliability, particularly in high-dimensional and unbalanced sample settings.

We derived the asymptotic null distribution of the test statistic, which converges to a weighted sum of independent chi-squared random variables, and analyzed its power properties under both fixed and local alternatives. To facilitate practical implementation, we developed a computationally efficient parametric bootstrap procedure for approximating critical values, which offers a significant speed-up over permutation-based methods while maintaining valid Type I error control. We also proposed a data-driven algorithm for selecting the truncation dimension $d$ and proved that this strategy preserves the test's asymptotic size.

Additionally, our perturbation analysis demonstrated that the proposed test is robust to the choice of kernel hyperparameters, provided that the leading eigenvalues exhibit sufficiently large spectral gaps. Extensive simulation studies and empirical applications confirmed our theoretical findings, showing that the proposed test maintains strict Type I error control and achieves higher statistical power than existing MMD-based tests.

\bibliography{paper-ref}

\newpage
\appendix
\begin{center}
    \Large\textbf{Online Appendix}
\end{center}


\section{Main Proofs}

\subsection{Proof of Lemma~\ref{residual}}
By the Cauchy-Schwarz inequality, we have
\[ \left|\langle \hat{\mu}_P - \hat{\mu}_Q, \hat{e}_i - e_i \rangle_{\mathcal{H}}\right| \leq \| \hat{\mu}_P - \hat{\mu}_Q \|_{\mathcal{H}} \cdot \| \hat{e}_i - e_i \|_{\mathcal{H}}. \]

Under the null hypothesis $H_0$, we have $\mu_P = \mu_Q = \mu$. By the triangle inequality and the convergence rates established in Theorem~\ref{clt},
\[ \| \hat{\mu}_P - \hat{\mu}_Q \|_{\mathcal{H}} \leq \| \hat{\mu}_P - \mu \|_{\mathcal{H}} + \| \hat{\mu}_Q - \mu \|_{\mathcal{H}} = O_p(n^{-1/2}) + O_p(m^{-1/2}) = O_p(N^{-1/2}). \]
Furthermore, assuming the positive eigengap condition $\lambda_i - \lambda_{i+1} > 0$ and aligning the sign of the empirical eigenfunction such that $\langle \hat{e}_i, e_i \rangle_{\mathcal{H}} \ge 0$, we have that $\| \hat{e}_i - e_i \|_{\mathcal{H}} = O_p(N^{-1/2})$. Combining these two bounds yields
\[ \langle \hat{\mu}_P - \hat{\mu}_Q, \hat{e}_i - e_i \rangle_{\mathcal{H}} = O_p(N^{-1}) \quad \text{under } H_0. \]

Under a fixed alternative hypothesis, $\mu_P \neq \mu_Q$. By the Strong Law of Large Numbers in Hilbert spaces, $\hat{\mu}_P \overset{a.s.}{\longrightarrow} \mu_P$ and $\hat{\mu}_Q \overset{a.s.}{\longrightarrow} \mu_Q$. Consequently, $\hat{\mu}_P - \hat{\mu}_Q \overset{a.s.}{\longrightarrow} \mu_P - \mu_Q$, which implies by the continuity of the norm that
\[ \| \hat{\mu}_P - \hat{\mu}_Q \|_{\mathcal{H}} \overset{a.s.}{\longrightarrow} \| \mu_P - \mu_Q \|_{\mathcal{H}} < \infty. \]
Since almost sure convergence to a finite constant implies boundedness in probability, we have $\| \hat{\mu}_P - \hat{\mu}_Q \|_{\mathcal{H}} = O_p(1)$. Note that the empirical eigenfunctions are constructed from the pooled sample following the mixture distribution $\rho = pP + (1-p)Q$, which remains fixed under a fixed alternative. Thus, the eigengap condition continues to ensure that $\| \hat{e}_i - e_i \|_{\mathcal{H}} = O_p(N^{-1/2})$. Multiplying these rates gives
\[ \langle \hat{\mu}_P - \hat{\mu}_Q, \hat{e}_i - e_i \rangle_{\mathcal{H}} = O_p(N^{-1/2}) \quad \text{under a fixed alternative.} \]

\subsection{Proof of Theorem \ref{null_distribution_single}}
Under $H_0$, we have $\mu_P = \mu_Q = \mu$. By Theorem \ref{clt}, the scaled difference of the empirical mean embeddings satisfies
\[ \sqrt{N}\left[(\hat{\mu}_P - \mu) - (\hat{\mu}_Q - \mu)\right] \xrightarrow{d} \frac{1}{\sqrt{p}} \xi_1 - \frac{1}{\sqrt{1-p}}\xi_2, \]
where $\xi_1$ and $\xi_2$ are independent Gaussian random elements in $\mathcal{H}$ with mean zero and covariance operator $\Sigma$. Since $\xi_1$ and $\xi_2$ are independent and identically distributed, their linear combination is equal in distribution to a single scaled Gaussian element:
\[ \frac{1}{\sqrt{p}} \xi_1 - \frac{1}{\sqrt{1-p}}\xi_2 \overset{d}{=} \sqrt{\frac{1}{p(1-p)}}\xi, \]
where $\xi \sim \mathcal{N}_{\mathcal{H}}(0, \Sigma)$, and the notation $\overset{d}{=}$ denotes equality in distribution.

We now analyze the scaled directional component $\sqrt{N}\hat{d}_i$. Recall that $\hat{d}_i = \langle \hat{\mu}_P - \hat{\mu}_Q, \hat{e}_i \rangle_{\mathcal{H}}$. Decomposing the inner product yields:
\[ \sqrt{N}\hat{d}_i = \sqrt{N} \langle \hat{\mu}_P - \hat{\mu}_Q, e_i \rangle_{\mathcal{H}} + \sqrt{N} \langle \hat{\mu}_P - \hat{\mu}_Q, \hat{e}_i - e_i \rangle_{\mathcal{H}}. \]
By Lemma \ref{residual}, the second term is $\sqrt{N} \cdot O_p(N^{-1}) = O_p(N^{-1/2}) = o_p(1)$. 

For the first term, because the inner product $\langle \cdot, e_i \rangle_{\mathcal{H}}$ is a continuous linear functional on $\mathcal{H}$, the continuous mapping theorem implies
\[ \sqrt{N} \langle (\hat{\mu}_P - \mu) - (\hat{\mu}_Q - \mu), e_i \rangle_{\mathcal{H}} \xrightarrow{d} \sqrt{\frac{1}{p(1-p)}} \langle \xi, e_i \rangle_{\mathcal{H}}. \]

Combining the limits of the two terms via Slutsky's theorem, we conclude that
\[ \sqrt{N} \hat{d}_i \xrightarrow{d} \sqrt{\frac{1}{p(1-p)}} \langle \xi, e_i \rangle_{\mathcal{H}}. \]
Since $\xi$ is a centered Gaussian element with covariance operator $\Sigma$, the random variable $\langle \xi, e_i \rangle_{\mathcal{H}}$ is a centered real-valued Gaussian with variance $\langle \Sigma e_i, e_i \rangle_{\mathcal{H}}$. Therefore, the limiting distribution is $\mathcal{N}\left(0, \frac{1}{p(1-p)} \langle \Sigma e_i, e_i \rangle_{\mathcal{H}}\right)$.

\subsection{Proof of Theorem \ref{null_distribution_multiple}}
Under $H_0$, we have $\mu_P = \mu_Q = \mu$. By Theorem \ref{clt}, the scaled difference of the empirical mean embeddings satisfies
\[ \sqrt{N}\left[(\hat{\mu}_P - \mu) - (\hat{\mu}_Q - \mu)\right] \xrightarrow{d} \sqrt{\frac{1}{p(1-p)}}\xi, \]
where $\xi$ is a centered Gaussian random element in $\mathcal{H}$ with covariance operator $\Sigma$.

By the argument of Appendix~\ref{orthogonal transformation}, there exists a sequence of random orthogonal matrices $O_N \in \mathbb{R}^{d \times d}$ such that the empirical eigenfunctions $\{\hat{e}_j\}_{j=1}^d$ closely approximate a rotated population eigenbasis. Specifically, defining the approximation error $\delta_j = \hat{e}_j - \sum_{k=1}^d (O_N)_{jk} e_k$, we have
\[ \sum_{j=1}^d \|\delta_j\|_{\mathcal{H}}^2 = O_p(N^{-1}). \]
Define the rotated empirical eigenfunctions as $\tilde{e}_i = \sum_{j=1}^d (O_N)_{ji} \hat{e}_j$ for $i = 1, \dots, d$. Using the orthonormality of the columns of $O_N$, we can bound their deviation from the population eigenfunctions $\{e_i\}_{i=1}^d$:
\[ \sum_{i=1}^d \|\tilde{e}_i - e_i\|_{\mathcal{H}}^2 = \sum_{i=1}^d \left\| \sum_{j=1}^d (O_N)_{ji} \delta_j \right\|_{\mathcal{H}}^2 = \sum_{i=1}^d \|\delta_i\|_{\mathcal{H}}^2 = O_p(N^{-1}). \]
Consequently, for each $i = 1, \dots, d$, we have $\|\tilde{e}_i - e_i\|_{\mathcal{H}} = O_p(N^{-1/2})$.

We can rewrite $\tilde{d}_{N,i}$ as the projection of the empirical mean difference onto $\tilde{e}_i$:
\[ \tilde{d}_{N,i} = \sum_{j=1}^d (O_N)_{ji} \langle \hat{\mu}_P - \hat{\mu}_Q, \hat{e}_j \rangle_{\mathcal{H}} = \langle \hat{\mu}_P - \hat{\mu}_Q, \tilde{e}_i \rangle_{\mathcal{H}}. \]
Scaling by $\sqrt{N}$ and applying the Cauchy-Schwarz inequality alongside the fact that $\|\hat{\mu}_P - \hat{\mu}_Q\|_{\mathcal{H}} = O_p(N^{-1/2})$ under $H_0$ (Lemma~\ref{residual}), we obtain:
\[ \sqrt{N}\,\tilde{d}_{N,i} = \sqrt{N}\langle \hat{\mu}_P - \hat{\mu}_Q, e_i \rangle_{\mathcal{H}} + \sqrt{N}\langle \hat{\mu}_P - \hat{\mu}_Q, \tilde{e}_i - e_i \rangle_{\mathcal{H}} = \sqrt{N}\langle \hat{\mu}_P - \hat{\mu}_Q, e_i \rangle_{\mathcal{H}} + o_p(1). \]
Since the linear functionals $f \mapsto \langle f, e_i \rangle_{\mathcal{H}}$ are continuous, the continuous mapping theorem applied to the joint vector and Theorem~\ref{clt} yield:
\[ \sqrt{N}(\tilde{d}_{N,1},\dots,\tilde{d}_{N,d}) \xrightarrow{d} \sqrt{\frac{1}{p(1-p)}}(\langle \xi,e_1\rangle_{\mathcal{H}},\dots,\langle \xi,e_d\rangle_{\mathcal{H}}). \]

Next, observe that $\{\hat{e}_i\}_{i=1}^d$ and $\{\tilde{e}_i\}_{i=1}^d$ are related by the orthogonal transformation $O_N$. Since orthogonal transformations preserve the Euclidean norm, the test statistic can be equivalently written as:
\[ D_N^d = N \sum_{i=1}^d \langle \hat{\mu}_P - \hat{\mu}_Q, \hat{e}_i \rangle_{\mathcal{H}}^2 = N \sum_{i=1}^d \langle \hat{\mu}_P - \hat{\mu}_Q, \tilde{e}_i \rangle_{\mathcal{H}}^2 = N \sum_{i=1}^d \tilde{d}_{N,i}^2. \]
Applying the continuous mapping theorem to the function $x \mapsto \|x\|_2^2$, we conclude that
\[ D_N^d \xrightarrow{d} \frac{1}{p(1-p)} \sum_{i=1}^d \langle \xi, e_i \rangle_{\mathcal{H}}^2. \]

Let $\mathbf{Z} = (\langle \xi, e_1 \rangle_{\mathcal{H}}, \dots, \langle \xi, e_d \rangle_{\mathcal{H}})^\top$. Since $\xi$ is a centered Gaussian random element in $\mathcal{H}$ with covariance operator $\Sigma$, the vector $\mathbf{Z}$ follows a $d$-dimensional multivariate normal distribution, $\mathbf{Z} \sim \mathcal{N}(0, \Sigma_d)$, where $(\Sigma_d)_{ij} = \langle \Sigma e_i, e_j \rangle_{\mathcal{H}}$. 

The limiting quadratic form can be expressed as $\sum_{i=1}^d \langle \xi, e_i \rangle_{\mathcal{H}}^2 = \mathbf{Z}^\top \mathbf{Z}$. Because $\Sigma_d$ is a real symmetric positive semi-definite matrix, it admits the spectral decomposition $\Sigma_d = U \Lambda U^\top$, where $U \in \mathbb{R}^{d \times d}$ is an orthogonal matrix and $\Lambda = \operatorname{diag}(\sigma_1, \dots, \sigma_d)$ contains the eigenvalues of $\Sigma_d$. Consequently,
\[ \mathbf{Z}^\top \mathbf{Z} = \mathbf{Z}^\top U U^\top \mathbf{Z} = (U^\top \mathbf{Z})^\top (U^\top \mathbf{Z}). \]
Since $U^\top \mathbf{Z} \sim \mathcal{N}(0, U^\top \Sigma_d U) = \mathcal{N}(0, \Lambda)$, the components of $U^\top \mathbf{Z}$ are independent. Therefore, the quadratic form is equal in distribution to a weighted sum of independent chi-squared random variables:
\[ \mathbf{Z}^\top \mathbf{Z} \overset{d}{=} \sum_{i=1}^d \sigma_i \chi_{1,i}^2, \]
where $\chi_{1,i}^2$ are independent chi-squared variables with one degree of freedom. Substituting this back yields the final result:
\[ D_N^d \xrightarrow{d} \sum_{i=1}^d \frac{\sigma_i}{p(1-p)} \chi_{1,i}^2. \]

\subsection{Proof of Lemma \ref{lm:consistency_cov_matrix}}
By the construction of orthogonal matrices in Appendix~\ref{orthogonal transformation}, we have 
\[ \tilde{e}_i = \sum_{j=1}^{d} (O_N)_{ji} \hat{e}_j, \quad i=1,\ldots,d, \]
and satisfy $ \|\tilde{e}_i -e_i\|_{\mathcal{H}}=O_p(N^{-1/2}) $. Note that $ \mathcal{H}- $convergence implies pointwise convergence:
\[ |\tilde{e}_i(z_\alpha) - e_i(z_\alpha)| = |\langle \tilde{e}_i - e_i, k(z_\alpha, \cdot) \rangle_{\mathcal{H}}| \le \|k(z_\alpha, \cdot)\|_{\mathcal{H}} \| \tilde{e}_i - e_i \|_{\mathcal{H}} \le \sqrt{\bar{k}} \| \tilde{e}_i - e_i \|_{\mathcal{H}} = O_p(N^{-1/2}). \]

Let $ \tilde{U}_i^\alpha = \tilde{e}_i(z_\alpha) - \sum_{\gamma=1}^N \tilde{e}_i(z_\gamma) /N$ and $ U_i^\alpha = e_i(z_\alpha) - \mathbb{E}[e_i(Z)] $. Then, we have
\begin{align*}
&\left| \tilde{U}_i^\alpha \tilde{U}_j^\alpha -  U_i^\alpha U_j^\alpha \right| \\
& = \left| (\tilde{e}_i^\alpha \tilde{e}_j^\alpha - e_i^\alpha e_j^\alpha) + (\tilde{e}_i^\alpha \bar{\tilde{e}}_j - e_i^\alpha \mathbb{E}[e_j(Z)]) + (\tilde{e}_j^\alpha \bar{\tilde{e}}_i - e_j^\alpha \mathbb{E}[e_i(Z)]) + (\bar{\tilde{e}}_i \bar{\tilde{e}}_j - \mathbb{E}[e_i(Z)]\mathbb{E}[e_j(Z)]) \right| \\
& \le \left| \tilde{e}_i^\alpha\right| \left| \tilde{e}_j^\alpha - e_j^\alpha \right| + \left| e_j^\alpha \right| \left| \tilde{e}_i^\alpha - e_i^\alpha \right| + \left| \tilde{e}_i^\alpha\right| \left|\bar{\tilde{e}}_j - \mathbb{E}[e_j(Z)]\right| + \left| e_i^\alpha\right| \left|\bar{\tilde{e}}_j - \mathbb{E}[e_j(Z)]\right|\\ 
& \quad + \left| \tilde{e}_j^\alpha\right| \left|\bar{\tilde{e}}_i - \mathbb{E}[e_i(Z)]\right| + \left| e_j^\alpha\right| \left|\bar{\tilde{e}}_i - \mathbb{E}[e_i(Z)]\right| + \left|\bar{\tilde{e}}_i\right| \left|\bar{\tilde{e}}_j - \mathbb{E}[e_j(Z)]\right|+ \left|\mathbb{E}[e_j(Z)]\right| \left|\bar{\tilde{e}}_i-\mathbb{E}[e_i(Z)]\right|,
\end{align*} 
where we have used the equality of $ ab-cd = a(b-d)+d(a-c) $. Since $ |\tilde{e}_i^\alpha| = \langle k(z_\alpha, \cdot), \tilde{e}_i \rangle_{\mathcal{H}} \le \|\tilde{e}_i\|_{\mathcal{H}} \|k(z_\alpha, \cdot)\|_{\mathcal{H}} = \|k(z_\alpha, \cdot)\|_{\mathcal{H}} \le \sqrt{\bar{k}} $, $ |e_i^\alpha| \le \sqrt{\bar{k}} $ for the same reason, both $ |\tilde{e}_i^\alpha - e_i^\alpha| $ and $ |\tilde{e}_j^\alpha - e_j^\alpha| $ are of order $ O_p(N^{-1/2}) $ and $ |\bar{\tilde{e}}_i - \mathbb{E}[e_i(Z)]| $ and $ |\bar{\tilde{e}}_j - \mathbb{E}[e_j(Z)]| $ are also of order $ O_p(N^{-1/2}) $ by the triangle inequality and the weak law of large numbers. Therefore, we have $ \left| \tilde{U}_i^\alpha \tilde{U}_j^\alpha -  U_i^\alpha U_j^\alpha \right| = O_p(N^{-1/2}) $. By the weak law of large numbers, we have
\[ (\tilde{\Sigma}_d)_{ij} = \frac{N}{N-1} \frac{1}{N} \sum_{\alpha=1}^N (\tilde{U}_i^\alpha \tilde{U}_j^\alpha) \xrightarrow{p} \mathbb{E}[U_i^\alpha U_j^\alpha] = (\Sigma_d)_{ij}. \]     

Thus, $ \tilde{\Sigma}_d \xrightarrow{p} \Sigma_d $ in any matrix norm due to the entry-wise convergence. Since eigenvalues of $ \tilde{\Sigma}_d $ are exactly the same as those of $ \hat{\Sigma}_d $, and because the mapping from a covariance matrix to its eigenvalues is continuous, we have $ \hat{\sigma}_i \xrightarrow{p} \sigma_i $ for each $ i=1,\ldots,d $.   

\subsection{Proof of Lemma~\ref{lm:parametric_bootstrap_consistency}}
Construct $ \tilde{\Gamma}_d = \tilde{\Sigma}_d / (\hat{p}(1-\hat{p})) $. By Lemma~\ref{lm:consistency_cov_matrix}, we have $ \tilde{\Sigma}_d \xrightarrow{p} \Sigma_d $ and $ \hat{p} \to p$, which implies $ \tilde{\Gamma}_d \xrightarrow{p} \Sigma_d / (p(1-p)) = \Gamma_d $. 

Let $ W^{(b)} \sim \mathcal{N}(\boldsymbol{0},\hat{\Gamma}_d) $, and define $ \tilde{W}^{(b)} = O_N W^{(b)} $. Since $ O_N $ is an orthogonal matrix, we have $ \tilde{W}^{(b)} \sim \mathcal{N}(\boldsymbol{0}, O_N \hat{\Gamma}_d O_N^\top) = \mathcal{N}(\boldsymbol{0}, \tilde{\Gamma}_d) $, and $ \|W^{(b)}\|^2 = \| \tilde{W}^{(b)} \|^2 $. Thus, the distribution of $ \|W^{(b)}\|^2 $ is the same as that of $ \|\tilde{W}^{(b)}\|^2 $, the distribution of the latter is 
\[ \| \tilde{W}^{(b)} \|^2 \sim \sum_{i=1}^d \frac{\hat{\sigma}_i}{\hat{p}(1-\hat{p})} \chi_{1,i}^2, \]
where $ \hat{\sigma}_i $ are the eigenvalues of both $ \hat{\Sigma}_d $ and $ \tilde{\Sigma}_d $, and $ \chi_{1,i}^2 $ are independent chi-squared random variables with one degree of freedom. By Lemma~\ref{lm:consistency_cov_matrix}, we have $ \hat{\sigma}_i \xrightarrow{p} \sigma_i $ for each $ i=1,\ldots,d $, and since $ \hat{p} \to p $, we have $ \hat{\sigma}_i / (\hat{p}(1-\hat{p})) \xrightarrow{p} \sigma_i / (p(1-p)) $. By the continuous mapping theorem, we conclude that
\[ \sum_{i=1}^d \frac{\hat{\sigma}_i}{\hat{p}(1-\hat{p})} \chi_{1,i}^2 \xrightarrow{d} \sum_{i=1}^d \frac{\sigma_i}{p(1-p)} \chi_{1,i}^2. \]
Thus, the conditional distribution of $ \|W^{(b)}\|^2 $ converges to the null distribution $ F $  of the test statistic $ D_N^d $. Since $ F $ is continuous, weak convergence is equivalent to uniform convergence of the cumulative distribution functions. Therefore, we have $ \sup_{x\in \mathbb{R}} |\hat{F}_N(x) - F(x)| \to 0 $ in probability, where $ \hat{F}_N $ is the conditional cumulative distribution function of $ \|W^{(b)}\|^2 $ given the data.

\subsection{Proof of Lemma \ref{alternative_expression}}
Define the linear functional $T$ on a block matrix $M \in \mathbb{R}^{N \times N}$ (partitioned conformably with $K^{(d)}$ into $M_{XX}, M_{YY}, M_{XY}$) as:
\[
T(M)=\frac{N}{n(n-1)}\bigl(\mathbf{1}_n^{\top}M_{XX}\mathbf{1}_n-\operatorname{Tr}(M_{XX})\bigr)
+\frac{N}{m(m-1)}\bigl(\mathbf{1}_m^{\top}M_{YY}\mathbf{1}_m-\operatorname{Tr}(M_{YY})\bigr)
-\frac{2N}{nm}\mathbf{1}_n^{\top}M_{XY}\mathbf{1}_m .
\]
By definition, the unbiased test statistic is $\tilde{D}_N^d = T(K^{(d)})$. Because $T$ is a linear functional, establishing $T(K^{(d)}) = T(\bar{K}^{(d)})$ is equivalent to showing that $T(K^{(d)} - \bar{K}^{(d)}) = 0$.

Let $J = \mathbf{1}_N\mathbf{1}_N^{\top}$. Since $H_N = I_N - \frac{1}{N}J$, we can expand the difference as:
\[
K^{(d)} - \bar{K}^{(d)} = K^{(d)} - H_N K^{(d)} H_N = \frac{1}{N}K^{(d)}J + \frac{1}{N}JK^{(d)} - \frac{1}{N^2}JK^{(d)}J.
\]
Because $K^{(d)}$ is symmetric, the vector $u = K^{(d)}\mathbf{1}_N$ satisfies $K^{(d)}J = u\mathbf{1}_N^{\top}$ and $JK^{(d)} = \mathbf{1}_N u^{\top}$. Furthermore, $JK^{(d)}J = (\mathbf{1}_N^{\top} K^{(d)} \mathbf{1}_N) \mathbf{1}_N\mathbf{1}_N^{\top} = c J$, where $c = \mathbf{1}_N^{\top} K^{(d)} \mathbf{1}_N$. Substituting these expressions yields:
\[
K^{(d)} - \bar{K}^{(d)} = \frac{1}{N}(\mathbf{1}_N u^{\top} + u\mathbf{1}_N^{\top}) - \frac{c}{N^2} J.
\]
By the linearity of $T$, it suffices to prove that $T(L) = 0$ for matrices $L$ of the following two types:
\begin{itemize}
\item[(i)] $L = \mathbf{1}_N u^{\top} + u\mathbf{1}_N^{\top}$ (a symmetric, rank-2 matrix),
\item[(ii)] $L = c\,\mathbf{1}_N\mathbf{1}_N^{\top}$ (a constant matrix).
\end{itemize}

\textbf{Case (i):} Let $L = \mathbf{1}_N u^{\top} + u\mathbf{1}_N^{\top}$. Partition $u = (u_X^{\top}, u_Y^{\top})^{\top}$ with $u_X\in\mathbb{R}^n$ and $u_Y\in\mathbb{R}^m$. The corresponding submatrices of $L$ are:
\[
L_{XX}= \mathbf{1}_n u_X^{\top}+u_X\mathbf{1}_n^{\top},\quad
L_{YY}= \mathbf{1}_m u_Y^{\top}+u_Y\mathbf{1}_m^{\top},\quad
L_{XY}= \mathbf{1}_n u_Y^{\top}+u_X\mathbf{1}_m^{\top}.
\]
Let $S_X=\sum_{i=1}^n (u_X)_i$ and $S_Y=\sum_{j=1}^m (u_Y)_j$. For the first term of $T(L)$, we compute:
\[
\mathbf{1}_n^{\top}L_{XX}\mathbf{1}_n = \mathbf{1}_n^{\top}(\mathbf{1}_n u_X^{\top}+u_X\mathbf{1}_n^{\top})\mathbf{1}_n = n S_X + n S_X = 2nS_X,
\]
\[
\operatorname{Tr}(L_{XX}) = \operatorname{Tr}(\mathbf{1}_n u_X^{\top})+\operatorname{Tr}(u_X\mathbf{1}_n^{\top}) = S_X + S_X = 2S_X.
\]
Thus, the $XX$ component evaluates to:
\[
\frac{N}{n(n-1)}\bigl(\mathbf{1}_n^{\top}L_{XX}\mathbf{1}_n-\operatorname{Tr}(L_{XX})\bigr)
= \frac{N}{n(n-1)}(2nS_X-2S_X)= \frac{2N}{n}S_X.
\]
By an identical argument, the $YY$ component evaluates to $\frac{2N}{m}S_Y$. For the cross term, we have:
\[
\mathbf{1}_n^{\top}L_{XY}\mathbf{1}_m = \mathbf{1}_n^{\top}(\mathbf{1}_n u_Y^{\top}+u_X\mathbf{1}_m^{\top})\mathbf{1}_m = nS_Y + mS_X,
\]
which implies:
\[
-\frac{2N}{nm}\mathbf{1}_n^{\top}L_{XY}\mathbf{1}_m = -\frac{2N}{nm}(nS_Y+mS_X)= -\frac{2N}{m}S_Y-\frac{2N}{n}S_X.
\]
Summing the three components yields $T(L) = \frac{2N}{n}S_X + \frac{2N}{m}S_Y - \frac{2N}{m}S_Y - \frac{2N}{n}S_X = 0$.

\textbf{Case (ii):} Let $L = c\,\mathbf{1}_N\mathbf{1}_N^{\top} = cJ$. The submatrices are $L_{XX}=c\mathbf{1}_n\mathbf{1}_n^{\top}$, $L_{YY}=c\mathbf{1}_m\mathbf{1}_m^{\top}$, and $L_{XY}=c\mathbf{1}_n\mathbf{1}_m^{\top}$. We compute:
\[
\mathbf{1}_n^{\top}L_{XX}\mathbf{1}_n = c\,n^2, \quad \operatorname{Tr}(L_{XX})=c\,n.
\]
The first term of $T(L)$ is therefore $\frac{N}{n(n-1)}(c n^2 - c n) = Nc$. Similarly, the second term evaluates to $Nc$. For the cross term:
\[
\mathbf{1}_n^{\top}L_{XY}\mathbf{1}_m = c\,nm \implies -\frac{2N}{nm}\mathbf{1}_n^{\top}L_{XY}\mathbf{1}_m = -2Nc.
\]
Summing the components gives $T(L) = Nc + Nc - 2Nc = 0$.

Since $T(L) = 0$ for both basis cases, linearity guarantees that $T(K^{(d)} - \bar{K}^{(d)}) = 0$. Therefore, $T(K^{(d)}) = T(\bar{K}^{(d)})$, which concludes the proof.

\subsection{Proof of Theorem \ref{null_distribution_test_statistic}}
Recall from Theorem \ref{null_distribution_multiple} that the biased test statistic converges as 
\[ D_N^d \xrightarrow{d} \sum_{i=1}^d \frac{\sigma_i}{p(1-p)} \chi_{1,i}^2, \]
which has an expected value of $\frac{1}{p(1-p)} \sum_{i=1}^d \sigma_i = \frac{1}{p(1-p)} \operatorname{Tr}(\Sigma_d) = \frac{1}{p(1-p)} \sum_{i=1}^d \langle \Sigma e_i, e_i \rangle_{\mathcal{H}}$.

From the definitions of $D_N^d$ and $\tilde{D}_N^d$ in Lemma \ref{alternative_expression} and Remark 6, their difference is given by:
\[ D_N^d - \tilde{D}_N^d = \frac{N}{n^2} \operatorname{Tr}(\bar{K}_{XX}^{(d)}) + \frac{N}{m^2} \operatorname{Tr}(\bar{K}_{YY}^{(d)}) + o_p(1), \]
where the $o_p(1)$ term arises from the minor scaling differences between $1/n^2$ and $1/(n(n-1))$ in the quadratic forms, which vanish asymptotically. By Slutsky's theorem, it suffices to show that the trace correction terms converge in probability to the expected value of the limiting distribution of $D_N^d$:
\[ \frac{N}{n^2} \operatorname{Tr}(\bar{K}_{XX}^{(d)}) + \frac{N}{m^2} \operatorname{Tr}(\bar{K}_{YY}^{(d)}) \xrightarrow{p} \frac{1}{p(1-p)} \sum_{i=1}^d \langle \Sigma e_i, e_i \rangle_{\mathcal{H}}. \]

To evaluate the trace terms, let $\hat{\mathbf{e}}_i = (\hat{e}_i(z_1), \dots, \hat{e}_i(z_N))^\top \in \mathbb{R}^N$ be the vector of empirical eigenfunction evaluations. By Lemma \ref{svd_empirical}, we have $\hat{\mathbf{e}}_i = \sqrt{\hat{\lambda}_i N} \mathbf{u}_i$, where $\mathbf{u}_i$ is the $i$-th eigenvector of the Gram matrix $K$. Consequently, the scaled and truncated Gram matrix can be expressed as:
\[ K^{(d)} = N \sum_{i=1}^d \hat{\lambda}_i \mathbf{u}_i \mathbf{u}_i^\top = \sum_{i=1}^d \hat{\mathbf{e}}_i \hat{\mathbf{e}}_i^\top. \]
Applying the centering matrix $H_N = I_N - \frac{1}{N}\mathbf{1}_N\mathbf{1}_N^\top$ yields:
\[ \bar{K}^{(d)} = H_N K^{(d)} H_N = \sum_{i=1}^d (H_N \hat{\mathbf{e}}_i)(H_N \hat{\mathbf{e}}_i)^\top. \]
The $XX$ submatrix is therefore $\bar{K}_{XX}^{(d)} = \sum_{i=1}^d (H_N \hat{\mathbf{e}}_i)_X (H_N \hat{\mathbf{e}}_i)_X^\top$, where the subscript $X$ denotes the restriction to the first $n$ entries corresponding to the sample from $P$. The trace of this submatrix is:
\[ \operatorname{Tr}(\bar{K}_{XX}^{(d)}) = \sum_{i=1}^d \| (H_N \hat{\mathbf{e}}_i)_X \|^2 = \sum_{i=1}^d \left( (H_N \hat{\mathbf{e}}_i)_X \right)^\top \left( (H_N \hat{\mathbf{e}}_i)_X \right). \]
Scaling by $N/n^2$, we obtain:
\begin{align*}
    \frac{N}{n^2} \operatorname{Tr}(\bar{K}_{XX}^{(d)}) &= \frac{N}{n} \sum_{i=1}^d \frac{1}{n} \left( (H_N \hat{\mathbf{e}}_i)_X \right)^\top \left( (H_N \hat{\mathbf{e}}_i)_X \right).
\end{align*}
Under $H_0$, the empirical eigenfunctions $\{\hat{e}_i\}_{i=1}^d$ consistently estimate the population eigenfunctions $\{e_i\}_{i=1}^d$ up to an orthogonal rotation within eigenspaces of multiplicity greater than one. By Lemma \ref{lm:trace_convergence}, which explicitly accounts for this rotational invariance, the scaled sum converges in probability to the population variance:
\[ \sum_{i=1}^d \frac{1}{n} \left( (H_N \hat{\mathbf{e}}_i)_X \right)^\top \left( (H_N \hat{\mathbf{e}}_i)_X \right) \xrightarrow{p} \sum_{i=1}^d \operatorname{Var}(e_i(Z)) = \sum_{i=1}^d \langle \Sigma e_i, e_i \rangle_{\mathcal{H}}. \]
Since $N/n \longrightarrow 1/p$, we have:
\[ \frac{N}{n^2} \operatorname{Tr}(\bar{K}_{XX}^{(d)}) \xrightarrow{p} \frac{1}{p} \sum_{i=1}^d \langle \Sigma e_i, e_i \rangle_{\mathcal{H}}. \]
By an identical argument for the $YY$ block, utilizing $N/m \longrightarrow 1/(1-p)$, we have:
\[ \frac{N}{m^2} \operatorname{Tr}(\bar{K}_{YY}^{(d)}) \xrightarrow{p} \frac{1}{1-p} \sum_{i=1}^d \langle \Sigma e_i, e_i \rangle_{\mathcal{H}}. \]

Combining these two limits, the trace correction converges in probability to:
\[ \frac{1}{p} \sum_{i=1}^d \langle \Sigma e_i, e_i \rangle_{\mathcal{H}} + \frac{1}{1-p} \sum_{i=1}^d \langle \Sigma e_i, e_i \rangle_{\mathcal{H}} = \frac{1}{p(1-p)} \sum_{i=1}^d \langle \Sigma e_i, e_i \rangle_{\mathcal{H}} = \frac{1}{p(1-p)} \sum_{i=1}^d \sigma_i. \]

Finally, the limiting distribution of the unbiased test statistic is the limiting distribution of $D_N^d$ minus its asymptotic expectation:
\[ \tilde{D}_N^d \xrightarrow{d} \sum_{i=1}^d \frac{\sigma_i}{p(1-p)} \chi_{1,i}^2 - \frac{1}{p(1-p)} \sum_{i=1}^d \sigma_i = \sum_{i=1}^d \frac{\sigma_i}{p(1-p)} (\chi_{1,i}^2-1). \]

\subsection{Proof of Theorem \ref{th:fixed-alternative}}

We begin by isolating the dominant terms in the expansion of the test statistic and verifying that the remainder terms are of sufficiently small order to be neglected after scaling by $\sqrt{N}$. 

Let $\hat{P}_k$ denote the empirical projection onto the subspace spanned by the eigenfunctions corresponding to the $k$-th distinct empirical eigenvalue. The perturbation of the population operator is $\delta T_N = T_N - T_{\mathcal{H}}$.  The analytic perturbation theory summarised in Theorem~\ref{th:perturbation_space} guarantees that with probability tending to one,
\[
\hat{P}_k - P_k = \mathcal{S}_k(\delta T_N) P_k + P_k (\delta T_N) \mathcal{S}_k + R_k,
\]
where $\mathcal{S}_k = \sum_{l \neq k} \frac{P_l}{\lambda_l - \lambda_k}$ is the reduced resolvent and the remainder satisfies $\|R_k\|_{op} =O_p(N^{-1})$. Expanding the symmetric combination yields the more convenient form
\[
\hat{P}_k - P_k = \sum_{l \neq k} \frac{P_l \delta T_N P_k + P_k \delta T_N P_l}{\lambda_l - \lambda_k} + R_k. \tag{1}
\]
We will denote $\delta P_k = \hat{P}_k - P_k$ throughout.

The unbiased statistic differs from the biased one by a trace correction:
\[
\tilde{D}_N^d = D_N^d - \frac{N}{n^2}\operatorname{Tr}(\bar{K}_{XX}^{(d)}) - \frac{N}{m^2}\operatorname{Tr}(\bar{K}_{YY}^{(d)}) + o_p(1).
\]
Under the fixed alternative $H_1$, the diagonal entries of $\bar{K}_{XX}^{(d)}$ are $O_p(1)$, implying $\operatorname{Tr}(\bar{K}_{XX}^{(d)}) = O_p(n)$. Thus, the trace correction terms are $O_p(1)$. Consequently, their contribution to $\frac{1}{N}\tilde{D}_N^d$ is $O_p(N^{-1})$, and after multiplication by $\sqrt{N}$ they contribute $O_p(N^{-1/2}) = o_p(1)$. It thus suffices to establish the asymptotic distribution of the biased statistic $D_N^d = N \sum_{k=1}^v \|\hat{P}_k (\hat{\mu}_P - \hat{\mu}_Q)\|_{\mathcal{H}}^2$.

Set $h = \hat{\mu}_P - \hat{\mu}_Q$. Under $H_1$ we have $h = g + \Delta$ with $\Delta = (\hat{\mu}_P - \mu_P) - (\hat{\mu}_Q - \mu_Q)$. Theorem~\ref{clt} implies $\|\Delta\|_{\mathcal{H}} = O_p(N^{-1/2})$. Expanding the squared norm gives
\[
\|\hat{P}_k h\|_{\mathcal{H}}^2 = \|(P_k + \delta P_k)(g + \Delta)\|_{\mathcal{H}}^2.
\]
Distributing the terms leads to
\begin{align*}
\|\hat{P}_k h\|_{\mathcal{H}}^2 &= \|P_k g\|_{\mathcal{H}}^2 + 2\langle P_k g, P_k \Delta \rangle_{\mathcal{H}} + 2\langle P_k g, \delta P_k g \rangle_{\mathcal{H}} \\
&\quad + 2\langle P_k g, \delta P_k \Delta \rangle_{\mathcal{H}} + \|P_k \Delta + \delta P_k g + \delta P_k \Delta\|_{\mathcal{H}}^2.
\end{align*}
Because $\|P_k\|_{\mathrm{op}} \le 1$, $\|\delta P_k\|_{\mathrm{HS}} = O_p(N^{-1/2})$ (Lemma~\ref{bound_projections}), and $\|\Delta\|_{\mathcal{H}} = O_p(N^{-1/2})$, all terms in the second line are of order $O_p(N^{-1})$. Summing over $k = 1,\dots,v$, we obtain
\[
\frac{1}{N} D_N^d = \sum_{k=1}^v \|P_k g\|_{\mathcal{H}}^2 + 2\sum_{k=1}^v \langle P_k g, \Delta \rangle_{\mathcal{H}} + 2\sum_{k=1}^v \langle P_k g, \delta P_k g \rangle_{\mathcal{H}} + R_N,
\]
where $R_N = O_p(N^{-1})$. Note that since $\sum_{k=1}^v P_k = P_d$, the term involving $\Delta$ simplifies to $2\langle P_d g, \Delta \rangle_{\mathcal{H}}$, which by the reproducing property yields
\[ 2 \langle P_{d} g, \Delta \rangle_{\mathcal{H}} = \frac{2}{n}\sum_{\alpha=1}^n P_{d} g(x_\alpha) - 2\mathbb{E}_P[P_{d} g(X)] - \frac{2}{m}\sum_{\beta=1}^m P_{d} g(y_\beta) + 2\mathbb{E}_Q[P_{d} g(Y)]. \]

We now analyse the perturbation sum $\sum_{k=1}^v \langle P_k g, \delta P_k g \rangle_{\mathcal{H}}$. Substituting the expansion for $\delta P_k$ (Equaution (1)) and using the orthogonality $P_l P_k = 0$ for $l \neq k$, we find
\[
\langle P_k g, \delta P_k g \rangle_{\mathcal{H}} = \sum_{l \neq k} \frac{\langle P_k g, P_l \delta T_N P_k g + P_k \delta T_N P_l g \rangle_{\mathcal{H}}}{\lambda_l - \lambda_k} + \langle P_k g, R_k g \rangle_{\mathcal{H}}.
\]
Since $P_l P_k = 0$ for $l \neq k$, the first term in the numerator vanishes. The remainder $\langle P_k g, R_k g \rangle_{\mathcal{H}}$ is $O_p(N^{-1})$. Summing over $k$ and separating the sums over $l \le v$ and $l > v$ yields
\begin{align*}
\sum_{k=1}^v \langle P_k g, \delta P_k g \rangle_{\mathcal{H}}
&= \sum_{k=1}^v \sum_{\substack{ l \le v \\ l \neq k }} \frac{\langle P_k g, \delta T_N P_l g \rangle_{\mathcal{H}}}{\lambda_l - \lambda_k} 
+ \sum_{k=1}^v \sum_{l > v} \frac{\langle P_k g, \delta T_N P_l g \rangle_{\mathcal{H}}}{\lambda_l - \lambda_k} + O_p(N^{-1}).
\end{align*}
Because $\delta T_N$ is self-adjoint, $\langle P_k g, \delta T_N P_l g \rangle_{\mathcal{H}} = \langle P_l g, \delta T_N P_k g \rangle_{\mathcal{H}}$, while the denominators satisfy $\lambda_l - \lambda_k = -(\lambda_k - \lambda_l)$. Consequently, the contributions of each pair $(k,l)$ and $(l,k)$ in the first double sum cancel exactly. The entire sum over $k,l \le v$ therefore vanishes, leaving only the terms with $l > v$:
\[
\sum_{k=1}^v \langle P_k g, \delta P_k g \rangle_{\mathcal{H}} = \sum_{k=1}^v \sum_{l > v} \frac{\langle P_k g, \delta T_N P_l g \rangle_{\mathcal{H}}}{\lambda_l - \lambda_k}.
\]
Define $w_k = \sum_{l > v} \frac{P_l g}{\lambda_l - \lambda_k}$. The series converges in $\mathcal{H}$ because the eigenvalues $\lambda_l$ decay and the eigengap $\lambda_k - \lambda_{v+1}$ is bounded away from zero. The right-hand side then becomes $\sum_{k=1}^v \langle P_k g, \delta T_N w_k \rangle_{\mathcal{H}}$.

For each $k$, the empirical operator perturbation evaluates to $\langle P_k g, \delta T_N w_k \rangle_{\mathcal{H}} = \frac{1}{N}\sum_{\gamma=1}^N P_k g(z_\gamma) w_k(z_\gamma)$, since $\mathbb{E}_\rho[P_k g w_k] = \langle P_k g, T_H w_k \rangle_{\mathcal{H}} = 0$. 

Splitting the pooled sample sum into the $X$ and $Y$ samples and grouping with the $\Delta$ terms, we obtain
\begin{align*}
2\langle P_{d} g, \Delta \rangle_{\mathcal{H}} + 2\sum_{k=1}^v \langle P_k g, \delta T_N w_k \rangle_{\mathcal{H}} 
&= \frac{2}{n}\sum_{\alpha=1}^n \left[ P_{d} g(x_\alpha) + \frac{n}{N} \sum_{k=1}^v P_k g(x_\alpha)w_k(x_\alpha) \right] - 2\mathbb{E}_P[P_{d} g(X)] \\
&\quad - \frac{2}{m}\sum_{\beta=1}^m \left[ P_{d} g(y_\beta) - \frac{m}{N} \sum_{k=1}^v P_k g(y_\beta)w_k(y_\beta) \right] + 2\mathbb{E}_Q[P_{d} g(Y)].
\end{align*}
Because $n$ and $m$ are deterministic sequences satisfying $n/N \to p$ and $m/N \to 1-p$, we can substitute $\frac{n}{N} = p + O(N^{-1})$ and $\frac{m}{N} = 1-p + O(N^{-1})$. The remainder terms induced by this substitution are of order $O_p(N^{-1})$. For instance, 
 $ \frac{2}{n}\sum_{\alpha=1}^n O(N^{-1}) \sum_{k=1}^v P_k g(x_\alpha)w_k(x_\alpha) = O_p(N^{-1}) $.
Grouping the sample averages precisely identifies the influence functions $\xi_P$ and $\xi_Q$:
\begin{align*}
\frac{1}{N} D_N^d - \|P_{d} g\|_{\mathcal{H}}^2 = \frac{1}{n}\sum_{\alpha=1}^n \xi_P(x_\alpha) + \frac{1}{m}\sum_{\beta=1}^m \xi_Q(y_\beta) - 2\mathbb{E}_P[P_{d} g(X)] + 2\mathbb{E}_Q[P_{d} g(Y)] + O_p(N^{-1}).
\end{align*}

To characterize the limiting distribution, we introduce the centered influence functions $\bar{\xi}_P(x) = \xi_P(x) - \mathbb{E}_P[\xi_P(X)]$ and $\bar{\xi}_Q(y) = \xi_Q(y) - \mathbb{E}_Q[\xi_Q(Y)]$. Substituting these into the expansion, we obtain
\begin{align*}
\frac{1}{N} D_N^d - \|P_{d} g\|_{\mathcal{H}}^2 &= \frac{1}{n}\sum_{\alpha=1}^n \bar{\xi}_P(x_\alpha) + \frac{1}{m}\sum_{\beta=1}^m \bar{\xi}_Q(y_\beta) \\
&\quad + \mathbb{E}_P[\xi_P(X)] + \mathbb{E}_Q[\xi_Q(Y)] - 2\mathbb{E}_P[P_{d} g(X)] + 2\mathbb{E}_Q[P_{d} g(Y)] + O_p(N^{-1}).
\end{align*}
We now show that the deterministic centering constants exactly cancel. Evaluating the expectations of the influence functions gives
\begin{align*}
\mathbb{E}_P[\xi_P(X)] + \mathbb{E}_Q[\xi_Q(Y)] &= 2\mathbb{E}_P[P_{d} g(X)] - 2\mathbb{E}_Q[P_{d} g(Y)] \\
&\quad + 2p\sum_{k=1}^v \mathbb{E}_P[P_k g w_k] + 2(1-p)\sum_{k=1}^v \mathbb{E}_Q[P_k g w_k].
\end{align*}
By the orthogonality of distinct eigenspaces, $\mathbb{E}_\rho[P_k g w_k] = p\mathbb{E}_P[P_k g w_k] + (1-p)\mathbb{E}_Q[P_k g w_k] = \langle P_k g, T_H w_k \rangle_{\mathcal{H}} = 0$. Hence, 
\begin{align*}
\mathbb{E}_P[\xi_P(X)] + \mathbb{E}_Q[\xi_Q(Y)] &= 2\mathbb{E}_P[P_{d} g(X)] - 2\mathbb{E}_Q[P_{d} g(Y)] \\
&= 2\langle \mu_P - \mu_Q, P_{d} g \rangle_{\mathcal{H}} = 2\langle g, P_{d} g \rangle_{\mathcal{H}} = 2\|P_{d} g\|_{\mathcal{H}}^2.
\end{align*}
Substituting this back, we note that $2\mathbb{E}_P[P_{d} g(X)] - 2\mathbb{E}_Q[P_{d} g(Y)] = 2\|P_{d} g\|_{\mathcal{H}}^2$, which perfectly offsets the $-2\mathbb{E}_P[P_{d} g(X)] + 2\mathbb{E}_Q[P_{d} g(Y)]$ term. This leaves
\begin{align*}
\frac{1}{N} D_N^d - \|P_{d} g\|_{\mathcal{H}}^2 = \frac{1}{n}\sum_{\alpha=1}^n \bar{\xi}_P(x_\alpha) + \frac{1}{m}\sum_{\beta=1}^m \bar{\xi}_Q(y_\beta) + O_p(N^{-1}).
\end{align*}

Multiplying by $\sqrt{N}$, and noting that $N/n \to 1/p$ and $N/m \to 1/(1-p)$, the variances of the scaled sample means converge as follows:
\begin{align*}
\operatorname{Var}\left(\frac{\sqrt{N}}{n} \sum_{\alpha=1}^n \bar{\xi}_P(x_\alpha)\right) = \frac{N}{n} \operatorname{Var}_P(\xi_P(X)) \to \frac{1}{p} \operatorname{Var}_P(\xi_P(X)),
\end{align*}
and similarly for the $Y$ sample. Because the two samples are independent and the centered influence functions possess finite variance, the classical Central Limit Theorem ensures joint asymptotic normality. Finally, we conclude that
\begin{align*}
\sqrt{N} \left( \frac{1}{N} \tilde{D}_N^d - \sum_{k=1}^v \|P_k g\|_{\mathcal{H}}^2 \right) \xrightarrow{d} \mathcal{N}\left(0, \frac{1}{p} \operatorname{Var}_P(\xi_P(X)) + \frac{1}{1-p} \operatorname{Var}_Q(\xi_Q(Y))\right),
\end{align*}
which completes the proof.

\subsection{Proof of Theorem \ref{th:local_alternative}}

The density path assumption guarantees that the total variation distance \(\|P_N-P\|_{\mathrm{TV}} = O(N^{-1/2})\); the same rate holds for the pooled mixture \(\rho_N = pP_N + (1-p)Q\) relative to the null mixture \(\rho = pP + (1-p)Q\). This measure perturbation is the source of all the technical adjustments required under the local alternative.

For any probability measure \(\nu\) the population integral operator can be written as a Bochner integral of rank-one operators,
\[
T_{\mathcal H}(\nu) = \int_{\mathcal Z} \bigl(k(\cdot,z)\otimes_{\mathcal H}k(\cdot,z)\bigr)\,d\nu(z).
\]
For a fixed \(z\) and any \(f\in\mathcal H\) with \(\|f\|_{\mathcal H}\le 1\),
\[
\|(k(\cdot,z)\otimes_{\mathcal H}k(\cdot,z))f\|_{\mathcal H}
   = |\langle f,k(\cdot,z)\rangle_{\mathcal H}|\,\|k(\cdot,z)\|_{\mathcal H}
   \le \|f\|_{\mathcal H}\|k(\cdot,z)\|_{\mathcal H}^{2}
   \le k(z,z) \le \bar k,
\]
so the operator norm of the rank-one kernel satisfies \(\|k(\cdot,z)\otimes_{\mathcal H}k(\cdot,z)\|_{\mathrm{op}} \le \bar k\).  Hence
\[
\|T_{\mathcal H}(\nu_1)-T_{\mathcal H}(\nu_2)\|_{\mathrm{op}}
   \le \int_{\mathcal Z} \|k(\cdot,z)\otimes_{\mathcal H}k(\cdot,z)\|_{\mathrm{op}}\,d|\nu_1-\nu_2|(z)
   \le \bar k \int_{\mathcal Z} d|\nu_1-\nu_2|(z)
   = 2\bar k\,\|\nu_1-\nu_2\|_{\mathrm{TV}},
\]
where $\| \cdot\|_{TV}$ denotes the total variation distance of probability measures. Applying this to \(\rho_N\) and \(\rho\) yields \(\|T_{\mathcal H}(\rho_N)-T_{\mathcal H}(\rho)\|_{\mathrm{op}} = O(N^{-1/2})\).

Let \(P_d\) be the orthogonal projection onto the span of the leading \(d\) eigenfunctions of \(T_{\mathcal H}(\rho)\) and \(P_d^{(N)}\) the corresponding projection for \(T_{\mathcal H}(\rho_N)\).  By assumption, the spectral gap \(\eta_d = \frac12(\lambda_d-\lambda_{d+1})\) is positive; for large \(N\) the operator perturbation is smaller than \(\eta_d\), so Theorem~\ref{th:perturbation_space} in Appendix~\ref{app:additional_tech_results} gives
\[
\|P_d^{(N)}-P_d\|_{\mathrm{op}} = O(N^{-1/2}).
\]
The orthogonal alignment of Appendix~\ref{orthogonal transformation} provides a deterministic orthogonal matrix \(O_N^{(1)}\) such that the rotated population eigenfunctions \(e_i^{(N)} = \sum_{j=1}^d (O_N^{(1)})_{ji}\,e_j(\rho_N)\) satisfy
\[
\|e_i^{(N)}-e_i\|_{\mathcal H} = O(N^{-1/2}),\qquad i=1,\dots,d.
\]

The empirical eigenfunctions \(\hat e_i\) are computed from the pooled sample \(\{Z_\alpha\}_{\alpha=1}^N\sim\rho_N\).  
Conditionally on \(\rho_N\), Lemma~\ref{bound_projections} gives \(\|\hat P_d - P_d^{(N)}\|_{\mathrm{HS}} = O_p(N^{-1/2})\), where \(\hat P_d\) projects onto \(\operatorname{span}\{\hat e_1,\dots,\hat e_d\}\).  
The spectral gaps of \(\rho_N\) approach those of \(\rho\), so the rate is uniform in \(N\).  
Applying the alignment construction again yields a random orthogonal matrix \(O_N^{(2)}\) such that the rotated empirical eigenfunctions \(\tilde e_i = \sum_{j=1}^d (O_N^{(2)})_{ji}\hat e_j\) satisfy \(\|\tilde e_i - e_i^{(N)}\|_{\mathcal H}=O_p(N^{-1/2})\).  
By the triangle inequality,
\[
\|\tilde e_i - e_i\|_{\mathcal H} = O_p(N^{-1/2}),\qquad i=1,\dots,d.
\tag{1}
\]
Since squared norms are invariant under orthogonal transformations, the biased statistic can be written as
\[
D_N^d = N\sum_{i=1}^d\langle \hat\mu_P - \hat\mu_Q,\tilde e_i\rangle_{\mathcal H}^2.
\]

We now examine the asymptotic distribution of the scaled mean difference.  Under the density path condition the distributions \(P_N\) are contiguous to \(P\); moreover the total variation bound implies that the covariance operator \(\Sigma_{P_N} = \mathbb E_{X\sim P_N}[(k(X,\cdot)-\mu_{P_N})\otimes (k(X,\cdot)-\mu_{P_N})]\) converges to \(\Sigma\) in trace norm at rate \(O(N^{-1/2})\).  Consider the triangular array of independent random elements \(X_{N,\alpha} = k(x_\alpha,\cdot) - \mu_{P_N}\) with \(\alpha = 1,\dots,n\) and \(x_\alpha \stackrel{\mathrm{i.i.d.}}{\sim} P_N\).  Because \(\|k(z,\cdot)\|_{\mathcal H} \le \sqrt{\bar k}\) and \(\mu_{P_N} \to \mu\), there exists a constant \(M\) independent of \(N\) such that \(\|X_{N,\alpha}\|_{\mathcal H} \le M\) almost surely.  For any \(\varepsilon > 0\) and sufficiently large \(N\), \(\varepsilon\sqrt{n} > M\) so that the indicator \(\mathbf 1_{\{\|X_{N,\alpha}\|_{\mathcal H} > \varepsilon\sqrt{n}\}}\) is identically zero; the Lindeberg condition for Hilbert-space valued triangular arrays is therefore trivially satisfied.  By Theorem~1.1 of \citet{kundu2000central}, this condition together with the convergence of the covariance operators implies
\[
\sqrt{n}(\hat\mu_P - \mu_{P_N}) \xrightarrow{d} \frac{1}{\sqrt{p}}\,\xi_1.
\]
The analogous limit for the \(Q\)-sample,\(\sqrt{m}(\hat\mu_Q - \mu) \xrightarrow{d} \frac{1}{\sqrt{1-p}}\,\xi_2\) with \(\xi_2\) independent of \(\xi_1\), follows directly from Theorem~\ref{clt} because \(Q=\mu\) is unchanged.  Since \(\mu_{P_N} = \mu + g/\sqrt{N}\), we obtain
\[
\sqrt{N}(\hat\mu_P - \hat\mu_Q) = \sqrt{N}\bigl((\hat\mu_P - \mu_{P_N}) - (\hat\mu_Q - \mu)\bigr) + g \xrightarrow{d} \frac{1}{\sqrt{p(1-p)}}\,\xi + g,
\]
where \(\xi = \sqrt{1-p}\,\xi_1 - \sqrt{p}\,\xi_2\).

Expand the inner product using the rotated basis:
\[
\langle \hat\mu_P - \hat\mu_Q,\tilde e_i\rangle_{\mathcal H}
= \langle \hat\mu_P - \hat\mu_Q, e_i\rangle_{\mathcal H}
+ \langle \hat\mu_P - \hat\mu_Q, \tilde e_i - e_i\rangle_{\mathcal H}.
\]
From the limit above and the local alternative, \(\|\hat\mu_P - \hat\mu_Q\|_{\mathcal H} = O_p(N^{-1/2})\).  Using \((1)\) and the reproducing property we obtain the uniform bound \(|\tilde e_i(z)-e_i(z)| = O_p(N^{-1/2})\).  Hence the second inner product is \(O_p(N^{-1})\), and
\[
\sqrt{N}\langle \hat\mu_P - \hat\mu_Q,\tilde e_i\rangle_{\mathcal H}
= \sqrt{N}\langle \hat\mu_P - \hat\mu_Q, e_i\rangle_{\mathcal H} + o_p(1).
\]
Applying the continuous mapping theorem to the joint limit of \(\sqrt{N}(\hat\mu_P - \hat\mu_Q)\) yields
\[
\sqrt{N}\bigl(\langle \hat\mu_P - \hat\mu_Q,\tilde e_i\rangle_{\mathcal H}\bigr)_{i=1}^d
\xrightarrow{d}
\Bigl( \frac{1}{\sqrt{p(1-p)}}\langle \xi,e_i\rangle_{\mathcal H} + \langle g,e_i\rangle_{\mathcal H} \Bigr)_{i=1}^d,
\]
and therefore
\[
D_N^d \xrightarrow{d}
\sum_{i=1}^d \Bigl( \frac{1}{\sqrt{p(1-p)}}\langle \xi,e_i\rangle_{\mathcal H} + \langle g,e_i\rangle_{\mathcal H} \Bigr)^2.
\tag{2}
\]

It remains to verify that the trace corrections that convert \(D_N^d\) to the unbiased statistic \(\tilde D_N^d\) converge to the same deterministic limit as under the null.  

First, note that 
\[
\bar{K}^{(d)} = H_N \left( \sum_{i=1}^d v_i v_i^\top \right) H_N = \sum_{i=1}^d (H_N v_i)(H_N v_i)^\top = \sum_{i=1}^d c_i c_i^\top,
\]
where $v_i = \sqrt{N \hat{\lambda}_i} u_i$ and $u_i$ is the i-th eigenvector of $K^{(d)}$, and the centered vector $c_i = H_N v_i$ has entries $c_{i,\alpha} = \tilde{e}_i(z_\alpha) - \bar{e}_i$ with \(\bar e_i = \frac{1}{N}\sum_{\gamma=1}^N \tilde e_i(z_\gamma)\).

The matrix $\bar{K}^{(d)}$ is $N \times N$, and its top-left $n \times n$ block is $\bar{K}^{(d)}_{XX}$. The trace of this block is simply the sum of its first $n$ diagonal entries:
\[
\operatorname{Tr}(\bar{K}^{(d)}_{XX}) = \sum_{\alpha=1}^n (\bar{K}^{(d)})_{\alpha\alpha}.
\]
Since $(\bar{K}^{(d)})_{\alpha\alpha} = \sum_{i=1}^d c_{i,\alpha} c_{i,\alpha} = \sum_{i=1}^d c_{i,\alpha}^2$, we have:
\begin{align*}
    \operatorname{Tr}(\bar{K}^{(d)}_{XX}) &= \sum_{\alpha=1}^n \sum_{i=1}^d (\tilde{e}_i(x_\alpha) - \bar{e}_i)^2 \\
    &= \sum_{i=1}^d \sum_{\alpha=1}^n (\tilde{e}_i(x_\alpha) - \bar{e}_i)^2,
\end{align*}
leading to 
\[
\frac{N}{n^2}\operatorname{Tr}(\bar{K}^{(d)}_{XX}) = \frac{N}{n} \sum_{i=1}^d \left[ \frac{1}{n} \sum_{\alpha=1}^n (\tilde{e}_i(x_\alpha) - \bar{e}_i)^2 \right].
\]

Now, look at the term inside the bracket: $\frac{1}{n} \sum_{\alpha=1}^n (\tilde{e}_i(x_\alpha) - \bar{e}_i)^2$. This is the sample second moment of $\tilde{e}_i(X)$ centered at the \textit{pooled} mean $\bar{e}_i$. 

Let $\hat{\mu}_{i,X} = \frac{1}{n} \sum_{\alpha=1}^n \tilde{e}_i(x_\alpha)$ be the sample mean of $\tilde{e}_i(X)$ for the $X$ group. We can decompose the deviation from the pooled mean into a deviation from the group mean plus the difference between the group mean and the pooled mean:
\[\tilde{e}_i(x_\alpha) - \bar{e}_i = (\tilde{e}_i(x_\alpha) - \hat{\mu}_{i,X}) + (\hat{\mu}_{i,X} - \bar{e}_i)\]

Squaring and averaging over $\alpha$ gives:
\begin{align*}
    \frac{1}{n} \sum_{\alpha=1}^n (\tilde{e}_i(x_\alpha) - \bar{e}_i)^2 &= \frac{1}{n} \sum_{\alpha=1}^n \Big[ (\tilde{e}_i(x_\alpha) - \hat{\mu}_{i,X})^2 \\
    &\quad + 2(\tilde{e}_i(x_\alpha) - \hat{\mu}_{i,X})(\hat{\mu}_{i,X} - \bar{e}_i) \\
    &\quad + (\hat{\mu}_{i,X} - \bar{e}_i)^2 \Big]
\end{align*}
Because $\sum_{\alpha=1}^n (\tilde{e}_i(x_\alpha) - \hat{\mu}_{i,X}) = 0$, the cross-term vanishes, leaving the standard bias-variance decomposition:
\begin{equation*}
    \frac{1}{n} \sum_{\alpha=1}^n (\tilde{e}_i(x_\alpha) - \bar{e}_i)^2 = \frac{1}{n} \sum_{\alpha=1}^n (\tilde{e}_i(x_\alpha) - \hat{\mu}_{i,X})^2 + (\hat{\mu}_{i,X} - \bar{e}_i)^2.
\end{equation*}

Consequently, we have
\[
\frac{N}{n^2}\operatorname{Tr}(\bar K_{XX}^{(d)}) = \frac{N}{n}\Bigl(
\sum_{i=1}^d\frac{1}{n}\sum_{\alpha=1}^n (\tilde e_i(x_\alpha)-\hat\mu_{i,X})^2
+ \sum_{i=1}^d (\hat\mu_{i,X}-\bar e_i)^2\Bigr).
\]
 
From the uniform pointwise bound just established, \(|\tilde e_i(z)-e_i(z)| = O_p(N^{-1/2})\).  Because \(\|P_N-P\|_{\mathrm{TV}} = O(N^{-1/2})\), the expectations of \(e_i\) and \(e_i^2\) under \(P_N\) differ from those under \(P\) by \(O(N^{-1/2})\).  A standard law of large numbers argument then gives
\[
\sum_{i=1}^d\frac{1}{n}\sum_{\alpha=1}^n (\tilde e_i(x_\alpha)-\hat\mu_{i,X})^2 \xrightarrow{p} \sum_{i=1}^d \operatorname{Var}_P(e_i(X)) = \sum_{i=1}^d \langle\Sigma e_i,e_i\rangle_{\mathcal H}.
\]
For the second term we employ the identity \(\hat\mu_{i,X} - \bar e_i = \frac{m}{N}(\hat\mu_{i,X} - \hat\mu_{i,Y})\).  Under \(H_{1N}\),
\[
\hat\mu_{i,X} - \hat\mu_{i,Y} = \langle \mu_{P_N} - \mu_Q, e_i\rangle_{\mathcal H} + O_p(N^{-1/2})
= \frac{1}{\sqrt{N}}\langle g,e_i\rangle_{\mathcal H} + O_p(N^{-1/2}) = O_p(N^{-1/2}),
\]
so this squared term is \(O_p(N^{-1})\) and its scaled contribution vanishes.  The \(Y\)-sample yields the symmetric limit, and together we obtain
\[
\frac{N}{n^2}\operatorname{Tr}(\bar K_{XX}^{(d)}) + \frac{N}{m^2}\operatorname{Tr}(\bar K_{YY}^{(d)}) \xrightarrow{p}
\frac{1}{p(1-p)}\sum_{i=1}^d \langle \Sigma e_i,e_i\rangle_{\mathcal H}.
\tag{3}
\]

Finally, Lemma~\ref{alternative_expression} and Remark~\ref{rk:biased_and_unbiased} state that \(\tilde D_N^d = D_N^d - \frac{N}{n^2}\operatorname{Tr}(\bar K_{XX}^{(d)}) - \frac{N}{m^2}\operatorname{Tr}(\bar K_{YY}^{(d)}) + o_p(1)\).  Subtracting the deterministic limit \((3)\) from the stochastic limit \((2)\) via Slutsky's theorem gives the claimed distribution:
\[
\tilde D_N^d \xrightarrow{d}
\sum_{i=1}^d \Bigl( \frac{1}{\sqrt{p(1-p)}}\langle \xi,e_i\rangle_{\mathcal H} + \langle g,e_i\rangle_{\mathcal H} \Bigr)^2
- \frac{1}{p(1-p)}\sum_{i=1}^d \langle \Sigma e_i,e_i\rangle_{\mathcal H},
\]
which completes the proof.

\subsection{Proof of Lemma \ref{lm:integral_perturb}}
Define the perturbation operator $\delta L = L_{\theta'} - L_\theta$. This is an integral operator on $L^2(\rho)$ with the kernel $\delta k(z, z') = k_{\theta'}(z, z') - k_\theta(z, z')$. 

Recall that the operator norm of an integral operator is bounded by its Hilbert-Schmidt norm. Therefore, we have:
\[
    \|\delta L\|_{\mathrm{op}} \le \|\delta L\|_{\mathrm{HS}}.
\]
The squared Hilbert-Schmidt norm is given by the double integral of the squared absolute value of the kernel:
\[
    \|\delta L\|_{\mathrm{HS}}^2 = \iint_{\mathcal{Z} \times \mathcal{Z}} |\delta k(z, z')|^2 \, d\rho(z) \, d\rho(z').
\]
By Fubini's Theorem, which applies since the boundedness of the kernel ensures integrability, we can express this double integral as an iterated integral:
\[
    \|\delta L\|_{\mathrm{HS}}^2 = \int_{\mathcal{Z}} \left( \int_{\mathcal{Z}} |k_{\theta'}(z, z') - k_\theta(z, z')|^2 \, d\rho(z') \right) d\rho(z).
\]
Observe that the inner integral constitutes the squared $L^2(\rho)$ norm of the function $k_{\theta'}(z, \cdot) - k_\theta(z, \cdot)$ for a fixed $z$. By the uniform Lipschitz assumption, we can bound this inner integral uniformly in $z$:
\[
    \int_{\mathcal{Z}} |k_{\theta'}(z, z') - k_\theta(z, z')|^2 \, d\rho(z') = \|k_{\theta'}(z, \cdot) - k_\theta(z, \cdot)\|_{\rho}^2 \le L^2 \|\theta' - \theta\|^2.
\]
Substituting this bound into the outer integral yields:
\[
    \|\delta L\|_{\mathrm{HS}}^2 \le \int_{\mathcal{Z}} L^2 \|\theta' - \theta\|^2 \, d\rho(z) = L^2 \|\theta' - \theta\|^2 \int_{\mathcal{Z}} d\rho(z).
\]
Since $\rho$ is a probability measure, $\int_{\mathcal{Z}} d\rho(z) = 1$. Thus, we have:
\[
    \|\delta L\|_{\mathrm{HS}}^2 \le L^2 \|\theta' - \theta\|^2.
\]
Taking the square root of both sides gives $\|\delta L\|_{\mathrm{HS}} \le L \|\theta' - \theta\|$. Finally, combining this with the relation between the operator and Hilbert-Schmidt norms, we conclude:
\[
    \|L_{\theta'} - L_\theta\|_{\mathrm{op}} \le \|L_{\theta'} - L_\theta\|_{\mathrm{HS}} \le L \|\theta' - \theta\| = O(\|\theta' - \theta\|).
\]

\subsection{Proof of Theorem \ref{th:space_bound}}
We prove the three statements in sequence.
    
\textbf{Proof of (1).} By assumption, the perturbation size satisfies $\|\delta L\|_{\mathrm{op}} < \eta_i$. Applying standard operator perturbation theory (Theorem~\ref{th:perturbation_space}) directly yields the exact bound:
\[
    \| P_i(\theta') - P_i(\theta)\|_{\mathrm{op}} \le \frac{\| \delta L \|_{\mathrm{op}}}{\eta_i - \|\delta L\|_{\mathrm{op}}}.
\]
Since $\|\delta L\|_{\mathrm{op}} = O(\|\theta' - \theta\|)$, we have $\|\delta L\|_{\mathrm{op}} \to 0$ as $\theta' \to \theta$. Therefore, for sufficiently small $\|\theta' - \theta\|$, we are guaranteed that $\|\delta L\|_{\mathrm{op}} \le c\eta_i$ for some $c \in (0,1)$. This implies the denominator is bounded below by $\eta_i - \|\delta L\|_{\mathrm{op}} \ge (1-c)\eta_i$. Substituting this into the exact bound gives:
\[
    \| P_i(\theta') - P_i(\theta)\|_{\mathrm{op}} \le \frac{\|\delta L\|_{\mathrm{op}}}{(1-c)\eta_i}.
\]
Finally, substituting $\|\delta L\|_{\mathrm{op}} = O(\|\theta' - \theta\|)$ yields the leading-order behavior:
\[
    \| P_i(\theta') - P_i(\theta)\|_{\mathrm{op}} = O\left(\frac{\|\theta' - \theta\|}{\eta_i}\right).
\]

\textbf{Proof of (2).} For any perturbed eigenfunction $\phi_{i,j}(\theta')$ corresponding to the perturbed eigenvalue $\lambda_{i,j}(\theta')$, Theorem~\ref{th:perturbation_eigenvalue} provides the exact eigenvalue decomposition:
\begin{align*}
    \lambda_{i,j}(\theta')-\lambda_i &= \langle (\delta L) \phi_{i,j}(\theta'),\phi_{i,j}(\theta')\rangle_{\rho} \\
    &\quad + \langle (P_i(\theta')-P_i(\theta)) (L_{\theta}-\lambda_i I) (P_i(\theta')-P_i(\theta))\phi_{i,j}(\theta'),\phi_{i,j}(\theta') \rangle_{\rho}.
\end{align*}
We bound the two terms separately. The first term is bounded by the operator norm:
\[
    \left| \langle (\delta L) \phi_{i,j}(\theta'),\phi_{i,j}(\theta')\rangle_{\rho} \right| \le \|\delta L\|_{\mathrm{op}} \|\phi_{i,j}(\theta')\|_{\rho}^2 = O(\|\theta' - \theta\|).
\]
For the second term, applying the operator norm bound gives:
\[
    \left| \langle (P_i(\theta')-P_i(\theta)) (L_{\theta}-\lambda_i I) (P_i(\theta')-P_i(\theta))\phi_{i,j}(\theta'),\phi_{i,j}(\theta') \rangle_{\rho} \right| \le \|P_i(\theta')-P_i(\theta)\|_{\mathrm{op}}^2 \| L_{\theta}-\lambda_i I\|_{\mathrm{op}} \|\phi_{i,j}(\theta')\|_{\rho}^2 .
\]
Since $\|L_\theta - \lambda_i I\|_{\mathrm{op}}$ and $\|\phi_{i,j}(\theta')\|_{\rho}^2$ are bounded, the second term is of order $\|P_i(\theta')-P_i(\theta)\|_{\mathrm{op}}^2=O\left(\frac{\|\theta' - \theta\|^2}{\eta_i^2}\right)$ by result (1). Because $\|\delta L\|_{\mathrm{op}} < \eta_i $, the first term dominates, yielding:
\[
    |\lambda_{i,j}(\theta')-\lambda_i| = O(\|\theta' - \theta\|).
\]

\textbf{Proof of (3).} We quantify the change in the contribution of the $i$-th block by expanding the difference:
\[
    C_i(\theta') - C_i(\theta) = (\lambda_i(\theta')-\lambda_i)\|P_i(\theta') \boldsymbol{h}\|_{\rho}^2 + \lambda_i \left(\|P_i(\theta') \boldsymbol{h}\|_{\rho}^2 - \|P_i(\theta) \boldsymbol{h}\|_{\rho}^2\right).
\]
The first term is bounded using result (2) and the fact that $\|\boldsymbol{h}\|_\rho < \infty$:
\[
    |\lambda_i(\theta')-\lambda_i| \|P_i(\theta') \boldsymbol{h}\|_{\rho}^2 \le |\lambda_i(\theta')-\lambda_i| \| \boldsymbol{h} \|_{\rho}^2 = O(\|\theta' - \theta\|).
\]
For the second term, we apply the polarization identity $\|a\|^2 - \|b\|^2 = \langle a-b, a+b \rangle$ to the inner products:
\begin{align*}
    \lambda_i \bigl| \|P_i(\theta') \boldsymbol{h}\|_{\rho}^2 - \|P_i(\theta) \boldsymbol{h}\|_{\rho}^2 \bigr| 
    &= \lambda_i \bigl| \langle (P_i(\theta')-P_i(\theta)) \boldsymbol{h}, (P_i(\theta')+P_i(\theta)) \boldsymbol{h} \rangle_{\rho} \bigr| \\
    &\le \lambda_i \|P_i(\theta')-P_i(\theta)\|_{\mathrm{op}} \|P_i(\theta')+P_i(\theta)\|_{\mathrm{op}} \|\boldsymbol{h}\|_\rho^2.
\end{align*}
Since $\|P_i(\theta')+P_i(\theta)\|_{\mathrm{op}} \le 2$, applying result (1) yields:
\[
    \lambda_i \bigl| \|P_i(\theta') \boldsymbol{h}\|_{\rho}^2 - \|P_i(\theta) \boldsymbol{h}\|_{\rho}^2 \bigr| = O\left(\frac{\lambda_i\|\theta' - \theta\|}{\eta_i}\right).
\]
Combining the bounds for both terms, we conclude that:
\[
    |C_i(\theta') - C_i(\theta)| = O\!\left(\frac{\lambda_i \|\theta' - \theta\|}{\eta_i}\right) + O(\|\theta' - \theta\|),
\]
which completes the proof.

\subsection{Proof of Lemma \ref{lm:gram_perturb}}
The operator norm is bounded by the Frobenius norm:
\begin{align*}
    \| K(\theta') - K(\theta)\|_{op} & \le \| K(\theta') - K(\theta)\|_{F} = \sqrt{\sum_{\alpha,\beta=1}^N \left( \frac{k_{\theta'}(z_\alpha, z_\beta) - k_{\theta}(z_\alpha, z_\beta)}{N} \right)^2} \\
    & \le \sqrt{\sum_{\alpha,\beta=1}^N \frac{L^2 \|\theta' - \theta\|^2}{N^2}} = L \|\theta' - \theta\|,
\end{align*}
for some finite $0<L<\infty$.

\subsection{Proof of Theorem \ref{th:statistic_bound}}
We first prove the finite-sample conditional bound. We condition on the fixed dataset $\{z_\alpha\}_{\alpha=1}^N$. In this setting, $N, n, m$ are fixed constants, and the Gram matrix $K(\theta)$ is a deterministic matrix-valued function of $\theta$. 

We analyze the biased statistic, $D_N^d(\theta) = N\delta^\top \Pi_d(\theta) (NK(\theta)) \Pi_d(\theta) \delta$. The difference is given by:
\begin{align*}
    |D_N^d(\theta) - D_N^d(\theta')| &= N \bigl| \delta^\top \bigl[ \Pi_d(\theta) NK(\theta) \Pi_d(\theta) - \Pi_d(\theta') NK(\theta') \Pi_d(\theta') \bigr] \delta \bigr| \\
    &\le N \bigl| \delta^\top \Pi_d(\theta) NK(\theta) (\Pi_d(\theta) - \Pi_d(\theta')) \delta \bigr| \\
    &\quad + N \bigl| \delta^\top (\Pi_d(\theta) - \Pi_d(\theta')) NK(\theta') \Pi_d(\theta') \delta \bigr| \\
    &\quad + N \bigl| \delta^\top \Pi_d(\theta) N(K(\theta) - K(\theta')) \Pi_d(\theta') \delta \bigr|,
\end{align*}
where we used the algebraic identity $A_1 B_1 C_1 - A_2 B_2 C_2 = A_1 B_1 (C_1 - C_2) + (A_1 - A_2) B_2 C_2 + A_1 (B_1 - B_2) C_2$.

Applying the Cauchy-Schwarz inequality to the first term, we get:
\begin{align*}
    N \bigl| \delta^\top \Pi_d(\theta) NK(\theta) (\Pi_d(\theta) - \Pi_d(\theta')) \delta \bigr| &\le N \| NK(\theta) \Pi_d(\theta) \delta \|_2 \| (\Pi_d(\theta) - \Pi_d(\theta')) \delta \|_2 \\
    &\le N \| NK(\theta) \|_{op} \| \Pi_d(\theta) \delta \|_2 \| \Pi_d(\theta) - \Pi_d(\theta') \|_{op} \| \delta \|_2.
\end{align*}
For the fixed dataset, $\|\delta\|_2^2 = 1/n + 1/m$ and $\| \Pi_d(\theta) \delta \|_2 \le \| \delta \|_2$ are constants. Furthermore, $\| NK(\theta) \|_{op}$ is bounded deterministically by $N \bar{k}$. 

By Theorem~\ref{th:perturbation_space} applied to the empirical matrix $K(\theta)$, if the perturbation satisfies $\|K(\theta') - K(\theta)\|_{op} < \hat{\eta}_d(\theta)$, the projection perturbation is bounded by the empirical spectral gap:
\[ \|\Pi_d(\theta) - \Pi_d(\theta')\|_{op} \le C \frac{\|K(\theta') - K(\theta)\|_{op}}{\hat{\eta}_d(\theta)}. \]
By Lemma \ref{lm:gram_perturb}, $\|K(\theta') - K(\theta)\|_{op} \le L \|\theta' - \theta\|$. Assuming $L\|\theta' - \theta\| < \hat{\eta}_d(\theta)$, we obtain:
\[ \|\Pi_d(\theta) - \Pi_d(\theta')\|_{op} \le C \frac{L \|\theta' - \theta\|}{\hat{\eta}_d(\theta)}. \]
Substituting these constants yields a bound of $C_1 \frac{\|\theta' - \theta\|}{\hat{\eta}_d(\theta)}$ for the first term. By symmetry, the second term satisfies the same bound.

For the third term, Cauchy-Schwarz yields:
\begin{align*}
    N \bigl| \delta^\top \Pi_d(\theta) N(K(\theta) - K(\theta')) \Pi_d(\theta') \delta \bigr| &\le N \| \Pi_d(\theta) \delta \|_2 \| N(K(\theta) - K(\theta')) \|_{op} \| \Pi_d(\theta') \delta \|_2 \\
    &\le N \|\delta\|_2^2 N L \|\theta' - \theta\| = O(\|\theta' - \theta\|),
\end{align*}
which is dominated by the first two terms. 

For the unbiased statistic, we must also bound the difference of the trace correction terms, $|\mathrm{Tr}_1(\theta') - \mathrm{Tr}_1(\theta)| + |\mathrm{Tr}_2(\theta') - \mathrm{Tr}_2(\theta)|$. Recall from Lemma~\ref{alternative_expression} that $\mathrm{Tr}_1(\theta) = \frac{N}{n(n-1)} \mathrm{Tr}(\bar{K}_{XX}^{(d)}(\theta))$ and $\mathrm{Tr}_2(\theta) = \frac{N}{m(m-1)} \mathrm{Tr}(\bar{K}_{YY}^{(d)}(\theta))$. 
Let $P_n = \begin{pmatrix} I_n & 0 \\ 0 & 0 \end{pmatrix}$ and $P_m = \begin{pmatrix} 0 & 0 \\ 0 & I_m \end{pmatrix}$ be the projection matrices that select the $XX$ and $YY$ blocks, respectively, so that $\bar{K}_{XX}^{(d)}(\theta) = P_n \bar{K}^{(d)}(\theta) P_n$ and $\bar{K}_{YY}^{(d)}(\theta) = P_m \bar{K}^{(d)}(\theta) P_m$. 
Define the difference matrix $\Delta \bar{K}^{(d)} = \bar{K}^{(d)}(\theta') - \bar{K}^{(d)}(\theta)$. Since $K^{(d)}(\theta)$ and $K^{(d)}(\theta')$ have rank at most $d$, $\Delta \bar{K}^{(d)}$ has rank at most $2d$. Consequently, the projected difference $P_n \Delta \bar{K}^{(d)} P_n$ also has rank at most $2d$. Applying the trace inequality $|\mathrm{Tr}(A)| \le \mathrm{rank}(A) \|A\|_{\mathrm{op}}$ yields:
\begin{align*}
    \bigl| \mathrm{Tr}(\bar{K}_{XX}^{(d)}(\theta')) - \mathrm{Tr}(\bar{K}_{XX}^{(d)}(\theta)) \bigr| &= \bigl| \mathrm{Tr}(P_n \Delta \bar{K}^{(d)} P_n) \bigr| \\
    &\le 2d \| P_n \Delta \bar{K}^{(d)} P_n \|_{\mathrm{op}} \\
    &\le 2d \| \Delta \bar{K}^{(d)} \|_{\mathrm{op}} \\
    &\le 2d \| K^{(d)}(\theta') - K^{(d)}(\theta) \|_{\mathrm{op}} \\
    &\le 2d \| K(\theta') - K(\theta) \|_{\mathrm{op}} \\
    &\le 2d L \|\theta' - \theta\| \quad \text{(by Lemma~\ref{lm:gram_perturb})}.
\end{align*}
An identical bound holds for the $YY$ block trace difference. Since $N, n, m, d$, and $L$ are fixed constants, the sum of the trace correction term differences is bounded by $O(\|\theta' - \theta\|)$. This establishes the conditional bound.

To prove the unconditional bound, we note that by Lemma~\ref{bound_eigenvalues}, $\hat{\lambda}_d(\theta) \xrightarrow{p} \lambda_d(\theta)$ and $\hat{\lambda}_{d+1}(\theta) \xrightarrow{p} \lambda_{d+1}(\theta)$. Define the event $\mathcal{E}_N = \{ \hat{\eta}_d(\theta) \ge \frac{1}{2}\eta_d(\theta) \}$. Since $\eta_d(\theta) > 0$, it follows that $\mathbb{P}(\mathcal{E}_N) \to 1$ as $N \to \infty$. On $\mathcal{E}_N$, we have:
\[ \frac{1}{\hat{\eta}_d(\theta)} \le \frac{2}{\eta_d(\theta)}. \]
Thus, on the event $\mathcal{E}_N$, the conditional bound translates to:
\[ |\tilde{D}_N^d(\theta) - \tilde{D}_N^d(\theta')| \le C' \frac{\|\theta' - \theta\|}{\eta_d(\theta)} + O(\|\theta' - \theta\|). \]
Because $\mathcal{E}_N$ holds with probability tending to 1, this deterministic sequence bound (with respect to the deterministic denominator $\eta_d(\theta)$) holds unconditionally, yielding $ |\tilde{D}_N^d(\theta) - \tilde{D}_N^d(\theta')| = O_p\left(\frac{\| \theta - \theta'\|}{\eta_d(\theta)} \right) +O(\|\theta' - \theta\|)$.

\subsection{Proof of Proposition \ref{prop: l_2_covariance}}
We first verify that $\boldsymbol{C}_\theta = \mathbb{E} \left[ (\Phi_\theta(Z) - \nu_\theta) \otimes_{\rho} (\Phi_\theta(Z) - \nu_\theta) \right].$ is positive semi-definite. For any test function $f \in L^2(\rho)$, consider the quadratic form $\langle f, \boldsymbol{C}_{\theta} f \rangle_{\rho}$. By the definition of $\boldsymbol{C}_{\theta}$ and the bilinearity of the inner product, we have:
    \begin{align*}
        \langle f, \boldsymbol{C}_{\theta} f \rangle_{\rho} &= \left\langle f, \mathbb{E} \left[ \langle f, \Phi_\theta(Z) - \nu_\theta \rangle_{\rho} (\Phi_\theta(Z) - \nu_\theta) \right] \right\rangle_{\rho} \\
        &= \mathbb{E} \left[ \langle f, \Phi_\theta(Z) - \nu_\theta \rangle_{\rho} \langle f, \Phi_\theta(Z) - \nu_\theta \rangle_{\rho} \right] \\
        &= \mathbb{E} \left[ \langle f, \Phi_\theta(Z) - \nu_\theta \rangle_{\rho}^2 \right].
    \end{align*}
    Define the real-valued random variable $X_Z = \langle f, \Phi_\theta(Z) \rangle_{\rho}$. Its expectation is:
    \[
        \mathbb{E}[X_Z] = \mathbb{E} \left[ \langle f, \Phi_\theta(Z) \rangle_{\rho} \right] = \left\langle f, \mathbb{E}[\Phi_\theta(Z)] \right\rangle_{\rho} = \langle f, \nu_\theta \rangle_{\rho}.
    \]
    Therefore, the quadratic form can be rewritten as:
    \[
        \langle f, \boldsymbol{C}_{\theta} f \rangle_{\rho} = \mathbb{E} \left[ \left( X_Z - \mathbb{E}[X_Z] \right)^2 \right] = \text{Var}_{Z \sim \rho} \left( \langle f, \Phi_\theta(Z) \rangle_{\rho} \right) \ge 0.
        \tag{1}
    \]
    Since the variance is intrinsically non-negative, this establishes that $\boldsymbol{C}_{\theta}$ is positive semi-definite. The self-adjointness of $\boldsymbol{C}_{\theta}$ follows immediately because $\langle f, \boldsymbol{C}_{\theta} g \rangle_{\rho} = \text{Cov}(\langle f, \Phi_\theta(Z) \rangle, \langle g, \Phi_\theta(Z) \rangle)$ is symmetric in $f$ and $g$.

    Next, we derive the explicit operator identity by evaluating the action of $\boldsymbol{C}_{\theta}$ on an arbitrary $f \in L^2(\rho)$ pointwise at $z' \in \mathcal{Z}$. Let $c = \langle f, \nu_\theta \rangle_{\rho}$. Expanding the definition yields:
    \begin{align*}
        (\boldsymbol{C}_{\theta} f)(z') &=\left(\mathbb{E} \left[ (\Phi_\theta(Z) - \nu_\theta) \otimes_{\rho} (\Phi_\theta(Z) - \nu_\theta) \right]\right)f(z')\\
        & = \mathbb{E} \left[ \langle \Phi_\theta(Z) - \nu_\theta, f \rangle_{\rho} (\Phi_\theta(Z) - \nu_\theta)(z') \right]\\
        &= \mathbb{E} \left[ \left( L_\theta f(Z) - c \right) \left( k_\theta(Z, z') - \nu_\theta(z') \right) \right] \\
        &= \mathbb{E} \Big[ L_\theta f(Z) k_\theta(Z, z') - L_\theta f(Z) \nu_\theta(z') - c k_\theta(Z, z') + c \nu_\theta(z') \Big].
    \end{align*}
    By linearity of expectation, we separate this into four integrals. Notice that $c$ and $\nu_\theta(z')$ are constants with respect to the integration variable $Z$:
    \begin{enumerate}
        \item $\mathbb{E} \left[ L_\theta f(Z) k_\theta(Z, z') \right] = \int_{\mathcal{Z}} (L_\theta f)(z) k_\theta(z, z') d\rho(z) = L_\theta(L_\theta f)(z') = L_\theta^2 f(z')$.
        \item $\mathbb{E} \left[ L_\theta f(Z) \nu_\theta(z') \right] = \nu_\theta(z') \mathbb{E} \left[ L_\theta f(Z) \right] = \nu_\theta(z') \langle L_\theta f, \mathbf{1} \rangle_{\rho}$.
        \item $\mathbb{E} \left[ c k_\theta(Z, z') \right] = c \mathbb{E} \left[ k_\theta(Z, z') \right] = c \nu_\theta(z')$.
        \item $\mathbb{E} \left[ c \nu_\theta(z') \right] = c \nu_\theta(z')$.
    \end{enumerate}
    Substituting these four terms back into the expectation:
    \begin{align*}
        (\boldsymbol{C}_{\theta} f)(z') &= L_\theta^2 f(z') - \nu_\theta(z') \langle L_\theta f, \mathbf{1} \rangle_{\rho} - c \nu_\theta(z') + c \nu_\theta(z') \\
        &= L_\theta^2 f(z') - \nu_\theta(z') \langle L_\theta f, \mathbf{1} \rangle_{\rho}.
    \end{align*}
    Because $L_\theta$ is self-adjoint on $L^2(\rho)$, we have $\langle L_\theta f, \mathbf{1} \rangle_{\rho} = \langle f, L_\theta \mathbf{1} \rangle_{\rho} = \langle f, \nu_\theta \rangle_{\rho} = c$. Substituting this back yields:
    \begin{align*}
        (\boldsymbol{C}_{\theta} f)(z') = L_\theta^2 f(z') - c \nu_\theta(z') = L_\theta^2 f(z') - \langle f, \nu_\theta \rangle_{\rho} \nu_\theta(z'),
    \end{align*}
    which proves the operator identity $\boldsymbol{C}_{\theta} = L_\theta^2 - \nu_\theta \otimes_{\rho} \nu_\theta$.

    The trace of $\boldsymbol{C}_{\theta}$ is bounded by:
    \[
        \text{Tr}(\boldsymbol{C}_{\theta}) = \mathbb{E} \left[ \|\Phi_\theta(Z) - \nu_\theta\|_{\rho}^2 \right] \le \mathbb{E} \left[ \|\Phi_\theta(Z)\|_{\rho}^2 \right] = \int_{\mathcal{Z}} k_\theta(z,z) d\rho(z) \le \bar{k} < \infty.
    \]
    Since $\boldsymbol{C}_{\theta}$ is positive, self-adjoint, and has finite trace, it is a trace-class operator.

    Finally, we establish the relation between $\boldsymbol{C}_{\theta}$ and the covariance matrix $\Sigma_d(\theta)$. Let $M(\theta)$ be the matrix representation of the restricted operator $P_d(\theta) \boldsymbol{C}_{\theta} P_d(\theta)$ in the orthonormal basis $\mathcal{B} = \{\phi_i(\theta)\}_{i=1}^d$. By definition, its entries are:
    \[
        M_{ij}(\theta) = \langle \phi_i(\theta), P_d \boldsymbol{C}_{\theta} P_d \phi_j(\theta) \rangle_{\rho} = \langle \phi_i(\theta), \boldsymbol{C}_{\theta} \phi_j(\theta) \rangle_{\rho}.
    \]
    Substituting the operator identity $\boldsymbol{C}_{\theta} = L_\theta^2 - \nu_\theta \otimes_{\rho} \nu_\theta$, we split this into two terms:
    \[
        M_{ij}(\theta) = \langle \phi_i(\theta), L_\theta^2 \phi_j(\theta) \rangle_{\rho} - \langle \phi_i(\theta), (\nu_\theta \otimes_{\rho} \nu_\theta) \phi_j(\theta) \rangle_{\rho}.
    \]
    For the first term, since $L_\theta \phi_j(\theta) = \lambda_j \phi_j(\theta)$, we have $L_\theta^2 \phi_j(\theta) = \lambda_j^2 \phi_j(\theta)$, and thus $\langle \phi_i(\theta), L_\theta^2 \phi_j(\theta) \rangle_{\rho} = \lambda_j^2 \delta_{ij}$. Since $\delta_{ij}$ is non-zero only when $i=j$, we can write this symmetrically as $\lambda_i \lambda_j \delta_{ij}$. For the second term, we evaluate:
    \[
        \langle \phi_i(\theta), (\nu_\theta \otimes_{\rho} \nu_\theta) \phi_j(\theta) \rangle_{\rho} = \langle \phi_j(\theta), \nu_\theta \rangle_{\rho} \langle \phi_i(\theta), \nu_\theta \rangle_{\rho}.
    \]
    Recall that $\nu_\theta = L_\theta \mathbf{1}$. Thus, $\langle \phi_j(\theta), \nu_\theta \rangle_{\rho} = \langle \phi_j(\theta), L_\theta \mathbf{1} \rangle_{\rho} = \langle L_\theta \phi_j(\theta), \mathbf{1} \rangle_{\rho} = \lambda_j \langle \phi_j(\theta), \mathbf{1} \rangle_{\rho}$. Substituting this back, the second term becomes $\lambda_i \lambda_j \langle \phi_i(\theta), \mathbf{1} \rangle_{\rho} \langle \phi_j(\theta), \mathbf{1} \rangle_{\rho}$. Combining the two terms yields the exact matrix entries for $M(\theta)$:
    \[
        M_{ij}(\theta) = \lambda_i \lambda_j \left( \delta_{ij} - \langle \phi_i(\theta), \mathbf{1} \rangle_{\rho} \langle \phi_j(\theta), \mathbf{1} \rangle_{\rho} \right).
    \]
    Now, recall $e_i(\theta) = \sqrt{\lambda_i}\phi_i(\theta)$ for $\rho-$almost surely by Lemma~\ref{svd_population}, and the entries of the covariance matrix $\Sigma_d(\theta)$:
    \begin{align*}
        (\Sigma_d(\theta))_{ij} &= \text{Cov}(e_i(\theta), e_j(\theta)) = \sqrt{\lambda_i \lambda_j} \left( \mathbb{E}[\phi_i(\theta) \phi_j(\theta)] - \mathbb{E}[\phi_i(\theta)]\mathbb{E}[\phi_j(\theta)] \right)\\ 
        &= \sqrt{\lambda_i \lambda_j} \left( \delta_{ij} - \langle \phi_i(\theta), \mathbf{1} \rangle_{\rho} \langle \phi_j(\theta), \mathbf{1} \rangle_{\rho} \right).
    \end{align*}
    Comparing the two matrices entry-wise, we find that $M_{ij}(\theta) = \sqrt{\lambda_i \lambda_j} (\Sigma_d(\theta))_{ij}$. In matrix form, this yields the algebraic relation:
    \[
        M(\theta) = [\Lambda_d]_{\mathcal{B}}^{1/2}(\theta) \Sigma_d(\theta) [\Lambda_d]_{\mathcal{B}}^{1/2}(\theta),
    \]
    where $ [\Lambda_d]_{\mathcal{B}}^{1/2}(\theta) $ is the matrix representation of $\Lambda_d^{1/2}(\theta)$ in terms of the basis $\mathcal{B}$.

    Rearranging this equation by pre- and post-multiplying by $[\Lambda_d]_{\mathcal{B}}^{-1/2}(\theta)$ gives:
    \[
        \Sigma_d(\theta) = [\Lambda_d]_{\mathcal{B}}^{-1/2}(\theta) M(\theta) [\Lambda_d]_{\mathcal{B}}^{-1/2}(\theta).
    \]
    Recall that $M(\theta)$ is the matrix representation of $P_d(\theta) \boldsymbol{C}_{\theta} P_d(\theta)$. The matrix representation of the composite operator $T_\theta = \Lambda_d^{-1/2}(\theta) P_d(\theta) \boldsymbol{C}_{\theta} P_d(\theta) \Lambda_d^{-1/2}(\theta)$ is precisely the product of the individual matrix representations:
    \[
        [T_\theta]_{\mathcal{B}} = [\Lambda_d]_{\mathcal{B}}^{-1/2}(\theta) M(\theta) [\Lambda_d]_{\mathcal{B}}^{-1/2}(\theta) = \Sigma_d(\theta).
    \]
    By the Theorem~\ref{th:spectral_equivalence}, an operator and its matrix representation share the same eigenvalues. Therefore, the eigenvalues of $\Sigma_d(\theta)$, which are the asymptotic variance weights $\sigma_i(\theta)$, are exactly the eigenvalues of the operator $T_\theta$.

\subsection{Proof of Lemma \ref{lm:l2_cov_bound}}
For any $f \in L^2(\rho)$ with $\|f\|_{\rho} \le 1$, we bound the difference of the quadratic forms. Introducing the shorthand notation $A(Z) = \langle f, \Phi_{\theta'}(Z) \rangle_{\rho}$ and $B(Z) = \langle f, \Phi_{\theta}(Z) \rangle_{\rho}$. By the definition of the covariance operator and the result (1) from the proof of Proposition~\ref{prop: l_2_covariance}, we have:
\begin{align*}
    |\langle f, (\boldsymbol{C}_{\theta'} - \boldsymbol{C}_{\theta}) f \rangle_{\rho}| &= \left| \text{Var}(A) - \text{Var}(B) \right| \\
    &= \left| \left( \mathbb{E}[A^2] - (\mathbb{E}A)^2 \right) - \left( \mathbb{E}[B^2] - (\mathbb{E}B)^2 \right) \right| \\
    &= \left| \bigl( \mathbb{E}[A^2] - \mathbb{E}[B^2] \bigr) - \bigl( (\mathbb{E}A)^2 - (\mathbb{E}B)^2 \bigr) \right|.
\end{align*}
Applying the triangle inequality yields:
\begin{equation*}
    \left| \text{Var}(A) - \text{Var}(B) \right| \le \bigl| \mathbb{E}[A^2] - \mathbb{E}[B^2] \bigr| + \bigl| (\mathbb{E}A)^2 - (\mathbb{E}B)^2 \bigr|.
\end{equation*}

We bound the first term by first invoking Jensen's inequality to move the absolute value inside the expectation, and then applying the algebraic identity $|a^2 - b^2| = |a-b||a+b|$:
\begin{align*}
    \bigl| \mathbb{E}[A^2] - \mathbb{E}[B^2] \bigr| &\le \mathbb{E}\bigl[ |A^2 - B^2| \bigr] \\
    &= \mathbb{E}\bigl[ |A-B||A+B| \bigr].
\end{align*}
Since the kernel is uniformly bounded, we have $|A| = |\langle f, \Phi_{\theta'}(Z) \rangle_{\rho}| \le \|f\|_{\rho}\|\Phi_{\theta'}(Z)\|_{\rho} \le \bar{k}$, and similarly $|B| \le \bar{k}$. Consequently, $|A+B| \le |A| + |B| \le 2\bar{k}$, which allows us to bound the expectation:
\begin{equation*}
    \mathbb{E}\bigl[ |A-B||A+B| \bigr] \le 2\bar{k} \, \mathbb{E} |A - B|.
\end{equation*}

For the second term, we apply the same algebraic identity directly to the squared expectations:
\begin{equation*}
    \bigl| (\mathbb{E}A)^2 - (\mathbb{E}B)^2 \bigr| = \bigl| \mathbb{E}A - \mathbb{E}B \bigr| \cdot \bigl| \mathbb{E}A + \mathbb{E}B \bigr|.
\end{equation*}
By Jensen's inequality, $|\mathbb{E}A| \le \mathbb{E}|A| \le \bar{k}$ and $|\mathbb{E}B| \le \bar{k}$, implying $|\mathbb{E}A + \mathbb{E}B| \le 2\bar{k}$. Furthermore, noting that $\mathbb{E}A - \mathbb{E}B = \mathbb{E}[A - B]$, we obtain:
\begin{equation*}
    \bigl| (\mathbb{E}A)^2 - (\mathbb{E}B)^2 \bigr| \le 2\bar{k} \bigl| \mathbb{E}[A - B] \bigr|.
\end{equation*}

Combining the bounds for both terms, we find:
\begin{equation*}
    |\langle f, (\boldsymbol{C}_{\theta'} - \boldsymbol{C}_{\theta}) f \rangle_{\rho}| \le 2\bar{k} \, \mathbb{E} |A - B| + 2\bar{k} \bigl| \mathbb{E}[A - B] \bigr|.
\end{equation*}
Finally, applying Jensen's inequality once more ensures that the absolute value of the expectation is dominated by the expectation of the absolute value, i.e., $\bigl| \mathbb{E}[A - B] \bigr| \le \mathbb{E}|A - B|$. Substituting this relation yields:
\begin{align*}
    |\langle f, (\boldsymbol{C}_{\theta'} - \boldsymbol{C}_{\theta}) f \rangle_{\rho}| &\le 4\bar{k} \, \mathbb{E} |A - B| \\
    &= 4\bar{k} \, \mathbb{E} \left| \langle f, \Phi_{\theta'}(Z) - \Phi_{\theta}(Z) \rangle_{\rho} \right|.
\end{align*}
    By the Cauchy-Schwarz inequality and the uniform Lipschitz assumption:
    \begin{align*}
        \mathbb{E} \left| \langle f, \Phi_{\theta'}(Z) - \Phi_{\theta}(Z) \rangle_{\rho} \right| &\le \mathbb{E} \left[ \|f\|_{\rho} \|\Phi_{\theta'}(Z) - \Phi_{\theta}(Z)\|_{\rho} \right] \\
        &\le \sup_{z \in \mathcal{Z}} \|k_{\theta'}(z, \cdot) - k_{\theta}(z, \cdot)\|_{\rho} \\
        &\le L \|\theta' - \theta\|.
    \end{align*}
    Since this holds uniformly for all $\|f\|_{\rho} \le 1$, we conclude that $\|\boldsymbol{C}_{\theta'} - \boldsymbol{C}_{\theta}\|_{op} \le 4\bar{k} L \|\theta' - \theta\| = O(\|\theta' - \theta\|)$.

\subsection{Proof of Theorem \ref{th:perturbation_distribution}}

Note the Wasserstein-1 distance is bounded by:
\begin{align*}
    W_1(F_\theta, F_{\theta'}) &\le \frac{1}{p(1-p)} \mathbb{E} \left| \sum_{i=1}^d \left(\sigma_i(\theta) - \sigma_i(\theta')\right) (\chi^2_{1,i} - 1) \right| \\
    &\le \frac{1}{p(1-p)} \sqrt{ \mathbb{E} \left[ \sum_{i=1}^d \left(\sigma_i(\theta) - \sigma_i(\theta')\right) (\chi^2_{1,i} - 1) \right]^2 }.
\end{align*}
The first inequality follows from the synchronous coupling of the chi-squared random variables, i.e., we couple $\chi^2_{1,i}$ in $F_\theta$ and $F_{\theta'}$ to be the same random variable for each $i$. The second inequality is an application of Jensen's inequality.

Expanding the squared sum and using the independence and variance of $(\chi^2_{1,i} - 1)$, the off-diagonal terms vanish, yielding:
\[
    W_1(F_\theta, F_{\theta'}) \le \frac{\sqrt{2}}{p(1-p)} \|\boldsymbol{\sigma}(\theta) - \boldsymbol{\sigma}(\theta')\|_2,
\]
where $\boldsymbol{\sigma}(\theta) = (\sigma_1(\theta),\ldots,\sigma_d(\theta))^\top$, and $\|\cdot\|_2$ is the usual Euclidean norm.  From Proposition~\ref{prop: l_2_covariance}, the weights $\sigma_i(\theta)$ are precisely the eigenvalues of the operator $T_\theta$. Applying Theorem~\ref{th:perturbation_eigenvalue} to the operators $T_{\theta}$ and $T_{\theta'}$, for any perturbed eigenvalue $\sigma_i(\theta')$ with corresponding perturbed eigenvector $v_i(\theta')$, we obtain the exact decomposition:
\begin{align*}
    \sigma_i(\theta') - \sigma_i(\theta) &= \langle (T_{\theta'} - T_\theta) v_i(\theta'), v_i(\theta') \rangle_{\rho} \\
    &\quad + \langle (P_i(\theta') - P_i(\theta)) (T_\theta - \sigma_i(\theta) I) (P_i(\theta') - P_i(\theta)) v_i(\theta'), v_i(\theta') \rangle_{\rho},
\end{align*}
where $P_i(\theta)$ and $P_i(\theta')$ are the spectral projections onto the corresponding eigenspaces, and $I$ is the identity operator over the space $L^2(\rho)$. Taking absolute values and applying standard operator norm bounds yields:
\[
    |\sigma_i(\theta') - \sigma_i(\theta)| \le \|T_{\theta'} - T_\theta\|_{op} + \|P_i(\theta') - P_i(\theta)\|_{op}^2 \|T_\theta - \sigma_i(\theta) I\|_{op}.
\]
By applying Theorem~\ref{th:perturbation_space} to the operators $T_\theta$ and $T_{\theta'}$, the projection perturbation is bounded by $\|P_i(\theta') - P_i(\theta)\|_{op} \le \|T_{\theta'} - T_\theta\|_{op} / (\eta_i - \|T_{\theta'} - T_\theta\|_{op})< 2 \|T_{\theta'} - T_\theta\|_{op} / \eta_i$ for some sufficiently small $\|T_{\theta'} - T_\theta\|_{op} $. Consequently, the second term is of order $O(\|T_{\theta'} - T_\theta\|_{op}^2 / \eta_i^2)$ and is asymptotically negligible compared to the first term by the assumption that $\|T_{\theta'} - T_\theta\|_{op} < \eta_i$. Therefore, $|\sigma_i(\theta') - \sigma_i(\theta)| = O(\|T_{\theta'} - T_\theta\|_{op})$, which implies that up to a constant factor, we have:
\[
    \|\boldsymbol{\sigma}(\theta) - \boldsymbol{\sigma}(\theta')\|_2 \le \sqrt{d} O(\|T_{\theta'} - T_\theta\|_{op}).
\]

We now bound $\|T_\theta - T_{\theta'}\|_{op}$. To simplify notation, let $S_\theta = \Lambda_d^{-1/2}(\theta)$ and $A_\theta = P_d(\theta) \boldsymbol{C}_{\theta} P_d(\theta)$, so $T_\theta = S_\theta A_\theta S_\theta$. Adding and subtracting the cross term $S_\theta A_\theta S_{\theta'}$ yields:
\begin{align*}
    T_\theta - T_{\theta'} &= S_\theta A_\theta S_\theta - S_{\theta'} A_{\theta'} S_{\theta'} \\
    &= S_\theta A_\theta (S_\theta - S_{\theta'}) + (S_\theta A_\theta - S_{\theta'} A_{\theta'}) S_{\theta'}.
\end{align*}
For the second term, we add and subtract $S_\theta A_{\theta'} S_{\theta'}$:
\[
    S_\theta A_\theta - S_{\theta'} A_{\theta'} = S_\theta (A_\theta - A_{\theta'}) + (S_\theta - S_{\theta'}) A_{\theta'}.
\]
Combining these and taking the operator norm, we obtain:
\begin{align*}
    \|T_\theta - T_{\theta'}\|_{op} &\le \|S_\theta\|_{op} \|A_\theta\|_{op} \|S_\theta - S_{\theta'}\|_{op} + \|S_\theta\|_{op} \|S_{\theta'}\|_{op} \|A_\theta - A_{\theta'}\|_{op} \\
    &\quad + \|S_\theta - S_{\theta'}\|_{op} \|A_{\theta'}\|_{op} \|S_{\theta'}\|_{op}.
\end{align*}
We evaluate the norms of these components:
\begin{enumerate}
    \item Since $\eta_d > 0$ ensures $\lambda_d > 0$, the operators $S_\theta = \Lambda_d^{-1/2}(\theta)$ are uniformly bounded: $\|S_\theta\|_{op} = \lambda_d^{-1/2} = O(1)$.
    \item The operators $A_\theta = P_d(\theta) \boldsymbol{C}_{\theta} P_d(\theta)$ are uniformly bounded by $\|\boldsymbol{C}_{\theta}\|_{op} \le \bar{k}^2$.
    \item Since the eigenvalues $\lambda_i(\theta)$ are Lipschitz continuous with respect to $\theta$ (by Theorem~\ref{th:space_bound}), the operator $S_\theta$ is also Lipschitz: $\|S_\theta - S_{\theta'}\|_{op} = O(\|\theta - \theta'\|)$.
    \item For the term $\|A_\theta - A_{\theta'}\|_{op}$, we bound it using Lemma~\ref{lm:l2_cov_bound} and Theorem~\ref{th:space_bound}:
    \begin{align*}
        \|A_\theta - A_{\theta'}\|_{op} &\le \|P_d(\theta) \boldsymbol{C}_{\theta} P_d(\theta) - P_d(\theta') \boldsymbol{C}_{\theta} P_d(\theta')\|_{op} + \|P_d(\theta') (\boldsymbol{C}_{\theta} - \boldsymbol{C}_{\theta'}) P_d(\theta')\|_{op} \\
        &\le O(\|P_d(\theta) - P_d(\theta')\|_{op}) + O(\|\boldsymbol{C}_{\theta} - \boldsymbol{C}_{\theta'}\|_{op}) \\
        &= O\left(\frac{\|\theta - \theta'\|}{\eta_d}\right) + O(\|\theta - \theta'\|).
    \end{align*}
\end{enumerate}
Substituting these bounds back:
\[
    \|T_\theta - T_{\theta'}\|_{op} = O\left( \frac{\|\theta - \theta'\|}{\eta_d} \right) + O(\|\theta - \theta'\|),
\]
which subsequently yields $\|\boldsymbol{\sigma}(\theta) - \boldsymbol{\sigma}(\theta')\|_2 = O\left( \frac{\|\theta - \theta'\|}{\eta_d} \right) + O(\|\theta - \theta'\|)$ and completes the proof.

\section{Results on Truncation Dimension Selection}
\label{app:truncation_dim_selection}
\subsection{Pseudocode for Truncation Dimension Selection}
\label{app:algorithm}
Algorithm~\ref{alg:select_d} outlines the procedure for data-driven selection of the truncation dimension $d$ via signal-to-noise ratio (SNR) maximization. The algorithm takes as input the training samples, the maximum candidate dimension $\bar{d}$, the kernel function, and the sample proportion $\hat{p}$. It computes the Gram matrix, extracts the top eigenpairs, and iteratively evaluates the SNR for each candidate dimension to identify the optimal $d^*$ that maximizes the SNR.

\begin{algorithm}[htbp]
\caption{Data-Driven Selection of $d$ via SNR}
\label{alg:select_d}
\begin{algorithmic}[1]
\Require Training samples $X' = \{x_1,\ldots,x_{n'}\}$ and $Y' = \{y_1,\ldots,y_{m'}\}$ (jointly $Z'$), maximum truncation level $\bar{d}$, kernel $k$, proportion $\hat{p}$
\State Set $N' \gets n' + m'$
\State Compute the full Gram matrix $K \in \mathbb{R}^{N' \times N'}$ with entries $k(z_\alpha, z_\beta)$
\State Compute the top $\bar{d}$ eigenvalues $\hat{\lambda}_1 \geq \dots \geq \hat{\lambda}_{\bar{d}}$ and eigenvectors $\boldsymbol{u}_1,\dots,\boldsymbol{u}_{\bar{d}}$ of $K$
\State Construct $\boldsymbol{U}_{\bar{d}} \gets [\boldsymbol{u}_1 \cdots \boldsymbol{u}_{\bar{d}}]$ and $\boldsymbol{\Lambda}_{\bar{d}} \gets \operatorname{diag}(\hat{\lambda}_1,\dots,\hat{\lambda}_{\bar{d}})$
\State Compute the score matrix $\boldsymbol{S} \gets \boldsymbol{U}_{\bar{d}} \boldsymbol{\Lambda}_{\bar{d}}^{1/2} \in \mathbb{R}^{N' \times \bar{d}}$
\State Center the scores: $\tilde{\boldsymbol{S}} \gets \left(\boldsymbol{I}_{N'} - \frac{1}{N'}\boldsymbol{1}_{N'}\boldsymbol{1}_{N'}^\top\right)\boldsymbol{S}$
\State Compute the empirical covariance: $\hat{\boldsymbol{\Sigma}}_{\bar{d}} \gets \frac{1}{N'-1} \tilde{\boldsymbol{S}}^\top\tilde{\boldsymbol{S}}$
\State Compute the asymptotic covariance scaling: $\hat{\boldsymbol{\Gamma}}_{\bar{d}} \gets \frac{\hat{\boldsymbol{\Sigma}}_{\bar{d}}}{\hat{p}(1-\hat{p})}$
\State Initialize $\text{SNR}_{\max} \gets -\infty$, $d^* \gets 1$
\For{$d = 1, \dots, \bar{d}$}
    \State Extract $\boldsymbol{U}_d \gets \boldsymbol{U}_{\bar{d}}[:, 1:d]$ and $\hat{\boldsymbol{\lambda}}_d \gets \hat{\boldsymbol{\lambda}}_{1:d}$
    \State Compute $\tilde{D}_{N'}^d$ using the unbiased statistic formula with $K^{(d)} = \boldsymbol{U}_d \operatorname{diag}(\hat{\boldsymbol{\lambda}}_d) \boldsymbol{U}_d^\top$
    \State Compute $\boldsymbol{\Gamma}_d \gets \hat{\boldsymbol{\Gamma}}_{\bar{d}}[1:d, 1:d]$
    \State Compute $\text{variance}_d \gets \operatorname{tr}(\boldsymbol{\Gamma}_d \boldsymbol{\Gamma}_d)$
    \State $\text{noise}_d \gets \sqrt{\text{variance}_d}$ 
    \State $\text{SNR}(d) \gets \tilde{D}_{N'}^d / \text{noise}_d$
    \State \textbf{if} $\text{SNR}(d) > \text{SNR}_{\max}$ \textbf{then}
        \State $\text{SNR}_{\max} \gets \text{SNR}(d)$, $\hat{d} \gets d$
    \State \textbf{end if}
\EndFor
\State \Return $\hat{d}$
\end{algorithmic}
\end{algorithm}

\subsection{Theoretical Results for Data-Driven $d$ Selection}
\label{app:theory_d_selection}
In this section, we show that under mild conditions, the data-driven selection procedure for $d$ via SNR maximization preserves the nominal Type I error and is consistent in the sense that it asymptotically selects the optimal truncation dimension that maximizes the population SNR.

\begin{lemma}
\label{lm:datadriven}
Assume the null hypothesis $H_0: P = Q$ holds. Let $\mathcal{D}_{\text{train}}$ and $\mathcal{D}_{\text{test}}$ be independent random partitions of the pooled sample, with sizes $N'$ and $N$ respectively. Suppose $\hat{d}$ is chosen according to any measurable function of $\mathcal{D}_{\text{train}}$ alone (e.g., $\hat{d} = \arg\max_{d \in \{1,\dots,\bar{d}\}} \widehat{\text{SNR}}(d)$). Let $P_{\text{final}}$ be the parametric bootstrap $p$-value obtained by applying the test with truncation level $\hat{d}$ to the test set $\mathcal{D}_{\text{test}}$. Then, for any significance level $\alpha \in (0,1)$,
\[ \lim_{N', N \to \infty} \mathbb{P}_{H_0}\bigl( P_{\text{final}} \le \alpha \bigr) = \alpha. \]
\end{lemma}

Denote the population SNR as a function of $ d $ by 
\[ Q(d) = \frac{\sum_{i=1}^d d_i^2}{\operatorname{tr}(\Gamma_d)}, \] 
and the scaled empirical SNR by
\[ Q_{N'}(d) =\frac{1}{N'}\frac{\tilde{D}_{N'}^d}{\operatorname{tr}(\hat{\Gamma}_d)}. \]
The selected dimension is $\hat{d} = \arg\max_{1 \le d \le \bar{d}} Q_{N'}(d)$, and the optimal dimension is $d^* = \arg\max_{1 \le d \le \bar{d}} Q(d)$. The following theorem shows that $\hat{d} \xrightarrow{P} d^*$ as $N' \to \infty$.

\begin{lemma}
    \label{lm:d_consistency}
    Under a fixed alternative hypothesis, assume the function $ Q(d) $ has an unique maximizer $ d^* $ in the candidate set $ \mathcal{D} = \{1,2,\ldots,\bar{d}\} $. Then the selected truncation dimension $\hat{d}$ satisfies $\hat{d} \xrightarrow{P} d^*$ as $N' \to \infty$. 
\end{lemma}

The proofs of these results are provided in Appendix~\ref{app:other_proofs}.

\section{Additional Simulation Results}
\label{app:simulations}
\subsection{Empirical Size Reports}
\label{app:size}
\begin{figure}
    \centering
    \includegraphics[width=0.9\textwidth]{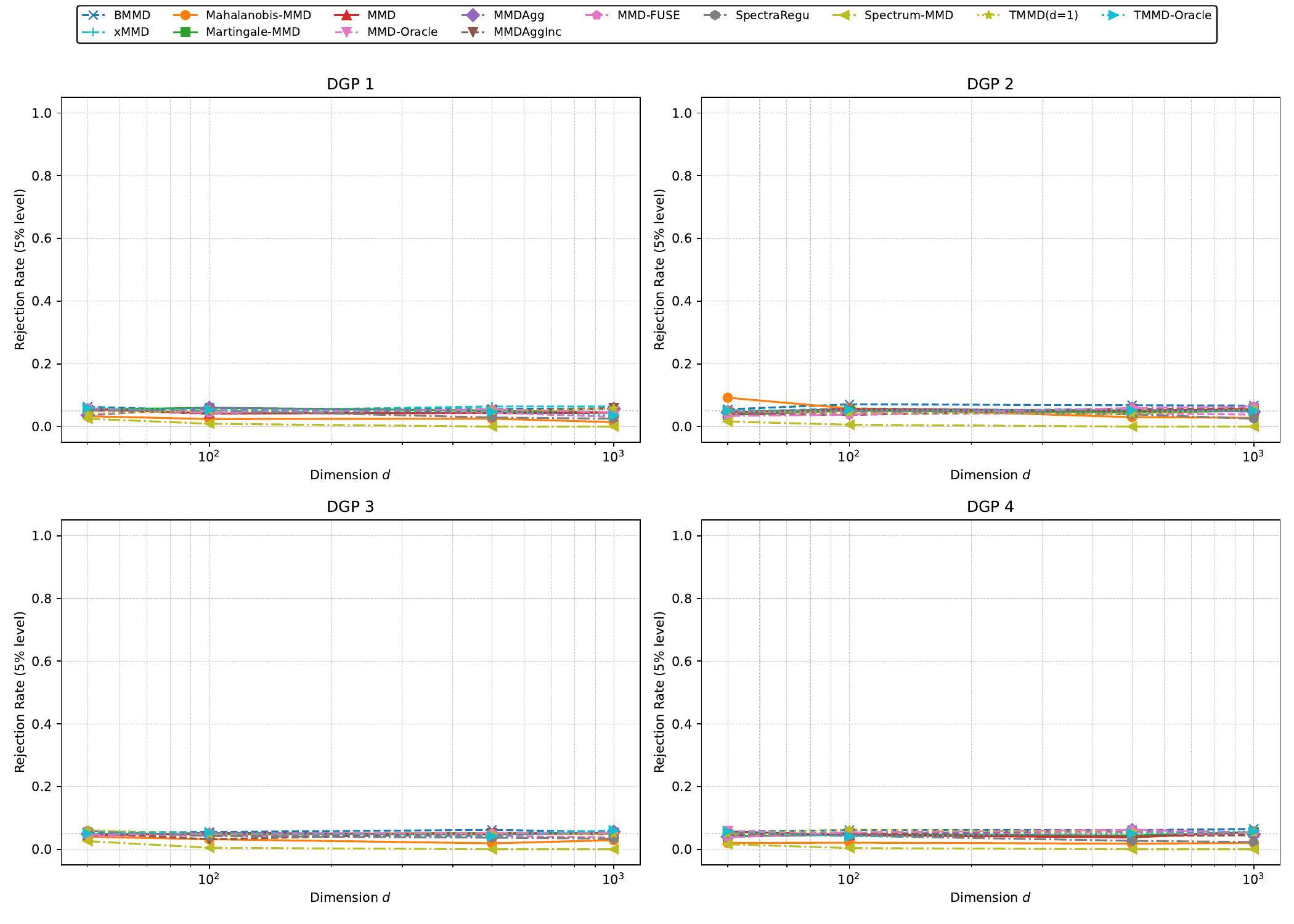}
    \caption{Empirical size of the tests under the null scenarios with balanced sample size ($n = m = 100$). The nominal level is $\alpha = 0.05$.}
    \label{fig:size_balanced}
\end{figure}

\begin{figure}
    \centering
    \includegraphics[width=0.9\textwidth]{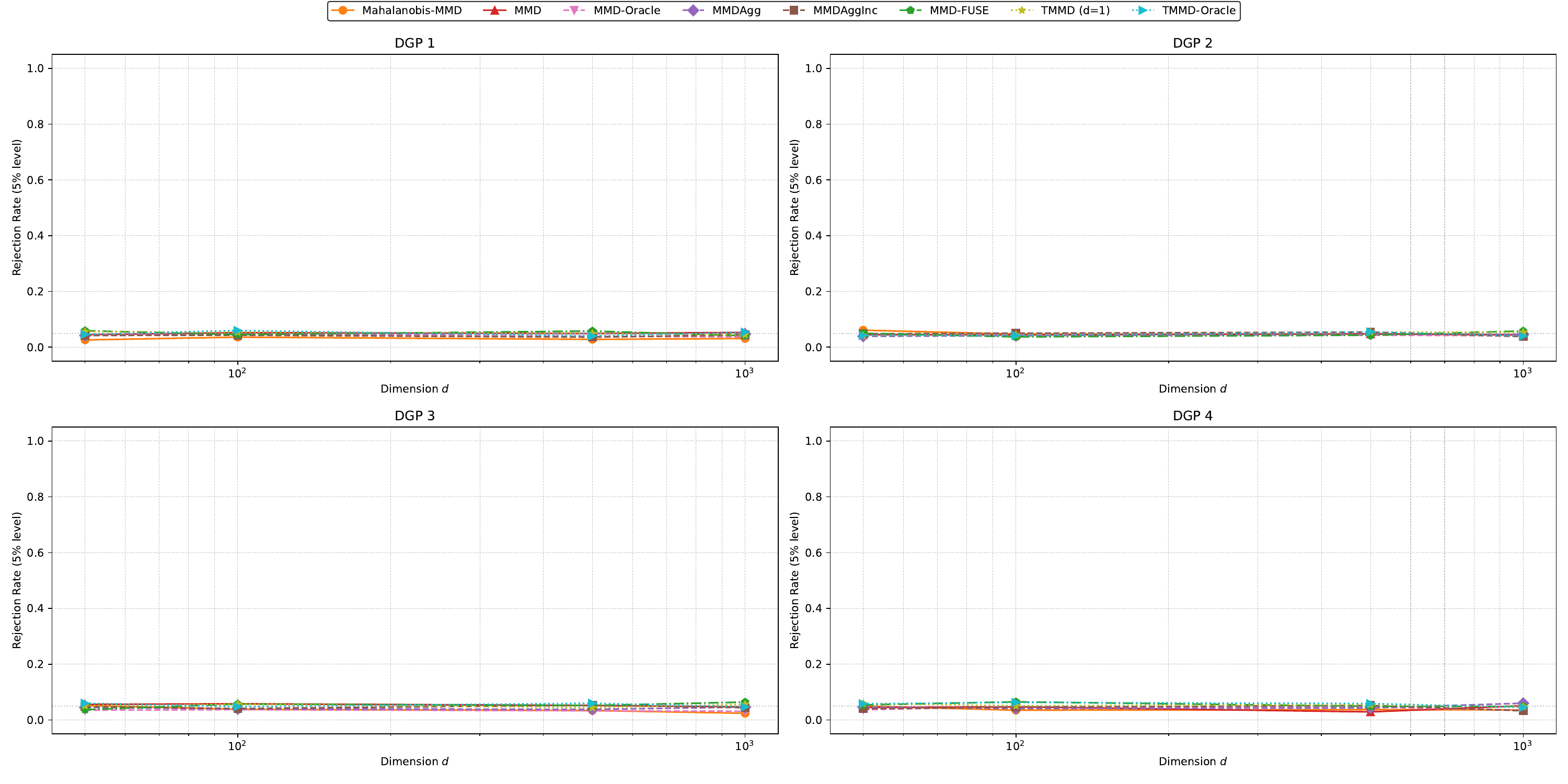}
    \caption{Empirical size of the tests under the null scenarios with unbalanced sample size ($n = 100$, $m = 10$). The nominal level is $\alpha = 0.05$.}
    \label{fig:size_unbalanced}
\end{figure}

\begin{figure}
    \centering
    \includegraphics[width=0.9\textwidth]{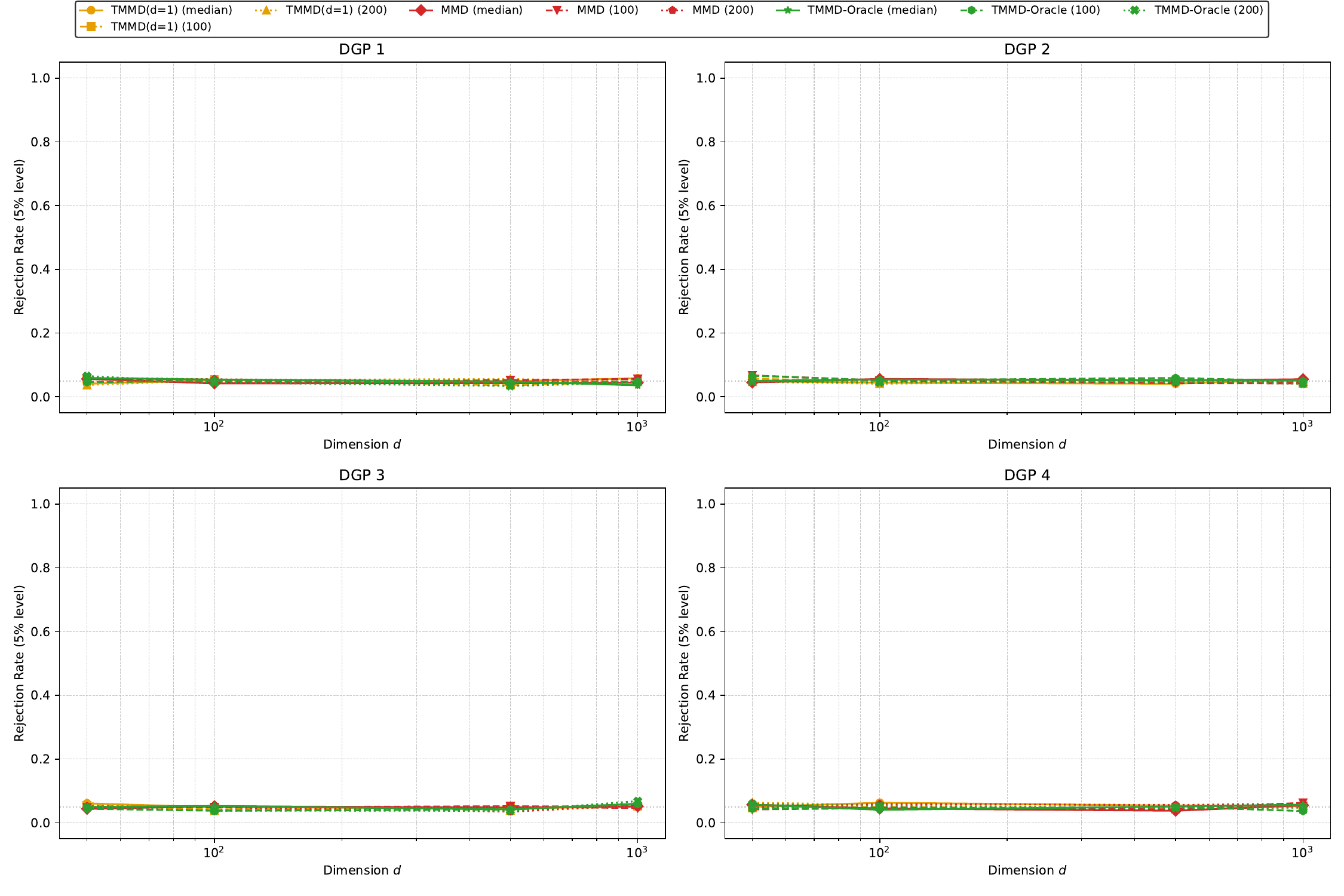}
    \caption{Empirical size of the tests under the null scenarios with different kernels (sample size $n = 100$, $m = 100$). The nominal level is $\alpha = 0.05$.}
    \label{fig:size_insensitive}
\end{figure}

Figure \ref{fig:size_balanced} reports the empirical size of the tests under the null scenarios with balanced sample size ($n = m = 100$). Figure~\ref{fig:size_unbalanced} reports the empirical size of the tests under the null scenarios with unbalanced sample size ($n = 100$, $m = 10$). Figure~\ref{fig:size_insensitive} reports the empirical size of the tests under the null scenarios with different kernels (sample size $n = 100$, $m = 100$). In all cases, the nominal level is $\alpha = 0.05$. 

\subsection{Simulation Results for Unbalanced Sample Size}
\label{app:unbalanced}
\begin{figure}
    \centering
    \begin{subfigure}{\textwidth}
        \centering
        \includegraphics[width=1.0\textwidth]{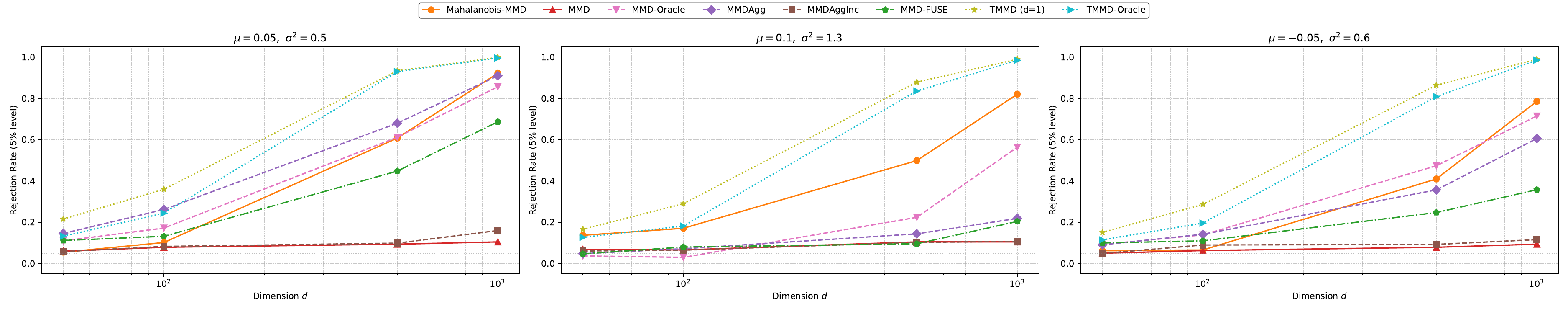}
        \caption{$\mathrm{DGP}_1$}
    \end{subfigure}
    \vspace{0.5cm}
    \begin{subfigure}{\textwidth}
        \centering
        \includegraphics[width=1.0\textwidth]{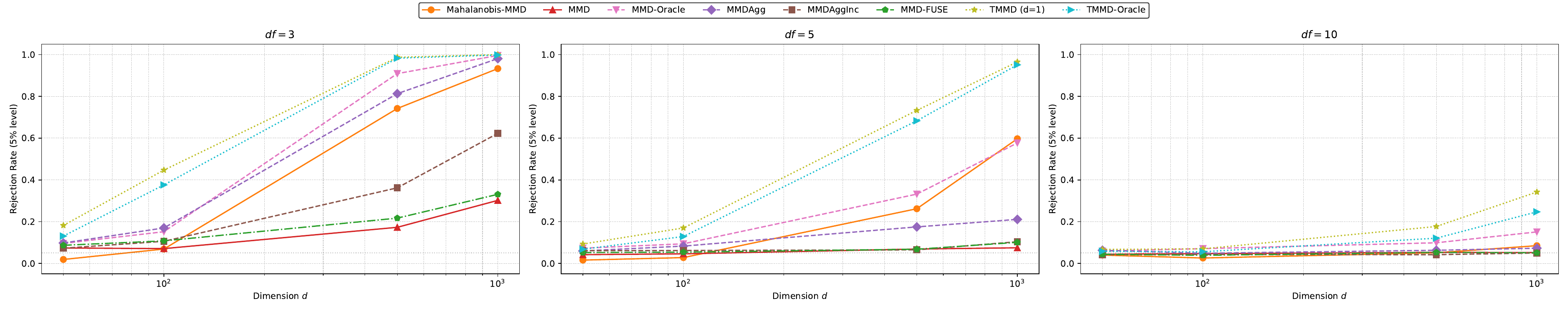}
        \caption{$\mathrm{DGP}_2$}
    \end{subfigure}
    \vspace{0.5cm}
    \begin{subfigure}{\textwidth}
        \centering
        \includegraphics[width=1.0\textwidth]{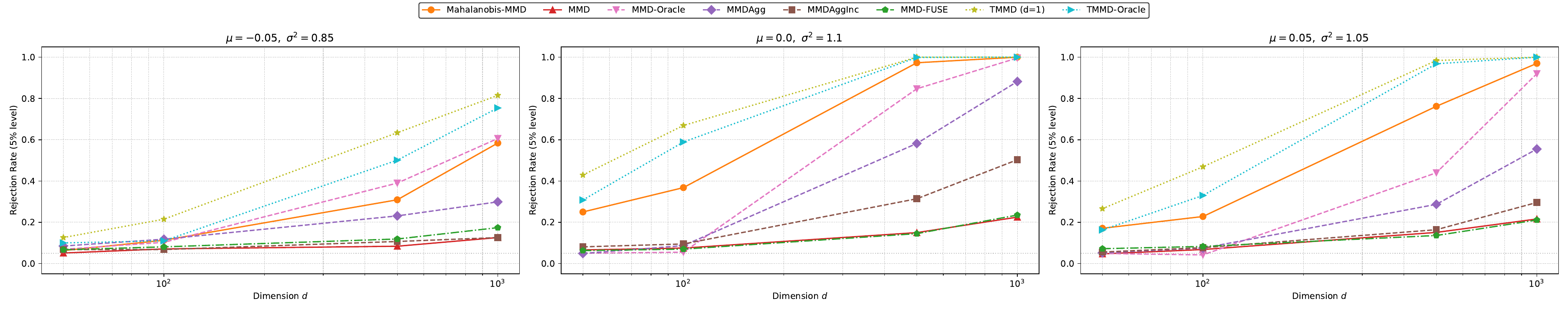}
        \caption{$\mathrm{DGP}_3$}
    \end{subfigure}
    \vspace{0.5cm}
    \begin{subfigure}{\textwidth}
        \centering
        \includegraphics[width=1.0\textwidth]{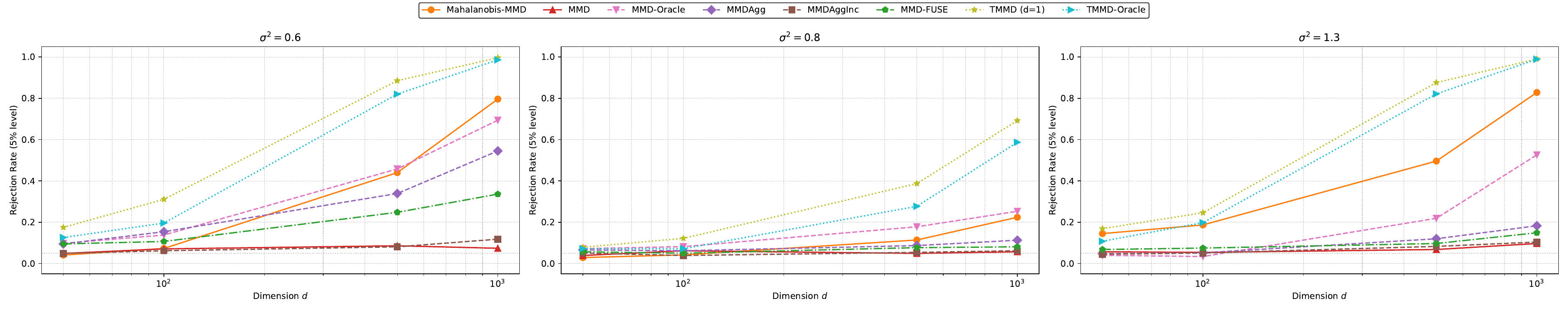}
        \caption{$\mathrm{DGP}_4$}
    \end{subfigure}
    \vspace{0.5cm}
    \begin{subfigure}{\textwidth}
        \centering
        \includegraphics[width=1.0\textwidth]{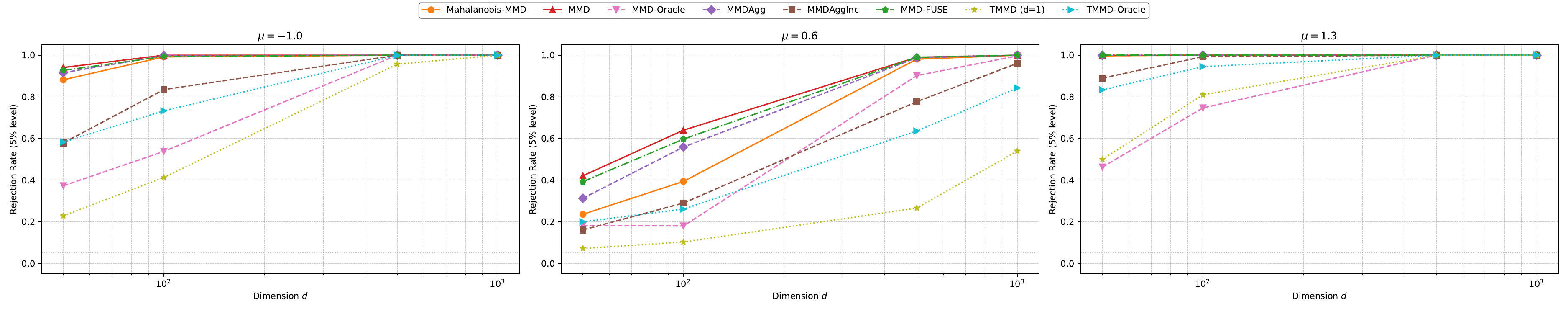}
        \caption{$\mathrm{DGP}_5$}
    \end{subfigure}
    \caption{Empirical power of the tests under alternative scenarios with unbalanced sample size ($n = 100$, $m = 10$).}
    \label{fig:power_unbalanced}
\end{figure}

In the highly unbalanced sample size regime, we restrict our comparison to a subset of the test statistics evaluated earlier. Test statistics are omitted for two reasons: either they lack theoretical justification under unbalanced sample sizes, or they exhibited comparatively poor finite--sample power in the balanced setting. Figure \ref{fig:power_unbalanced} reports the results. For discussions, please refer to Section~\ref{sec:simulation} in the main text.

\subsection{Frequency of $ d $ Selection}
\label{app:freq_d}
\begin{figure}
    \centering
    \includegraphics[width=0.9\textwidth]{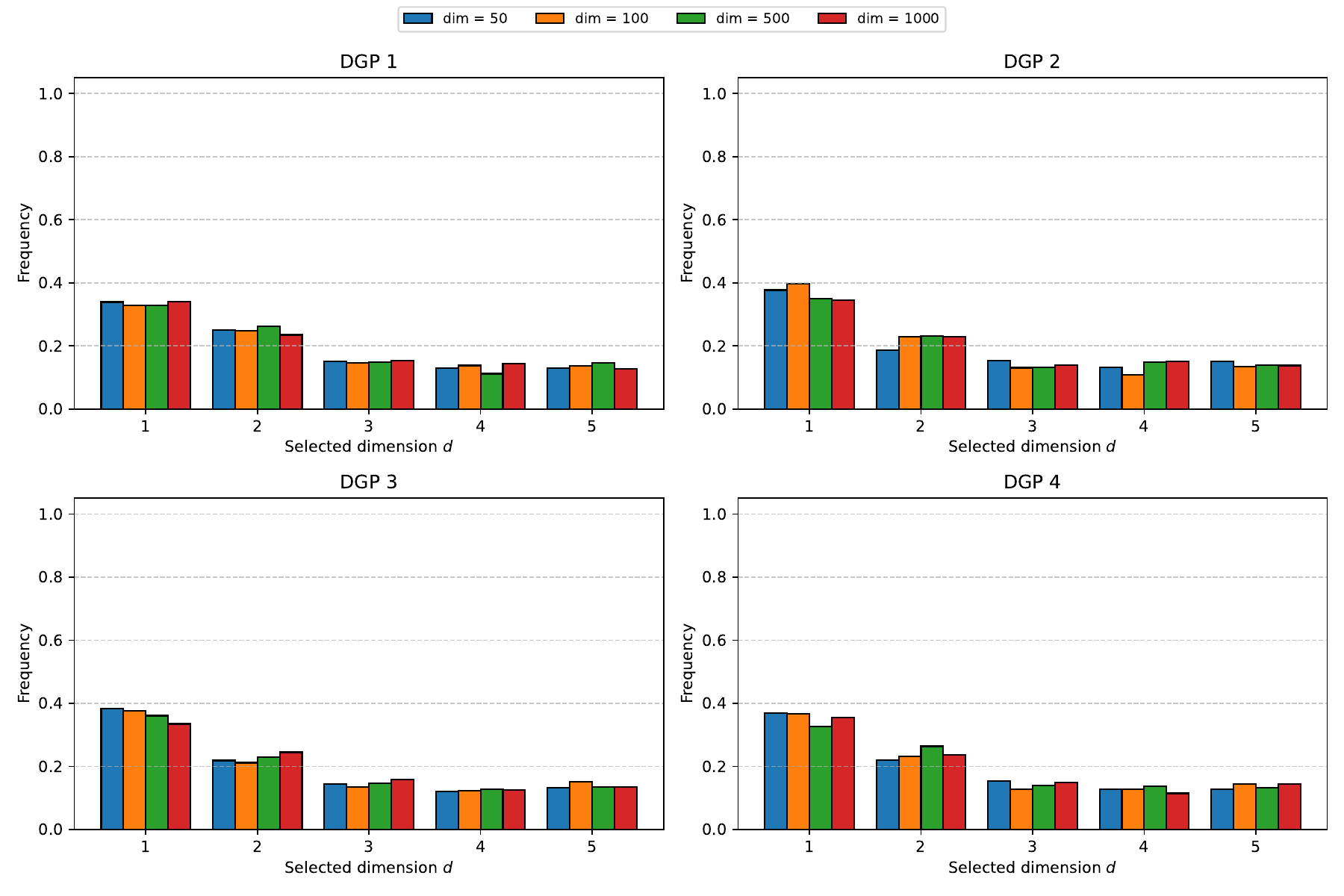}
    \caption{Frequency of $d$ selection under the null scenarios with balanced sample size ($n = m = 200$).}
    \label{fig:d_freq_size}
\end{figure}

\begin{figure}
    \centering
    \begin{subfigure}{0.8\textwidth}
        \centering
        \includegraphics[width=1.0\textwidth]{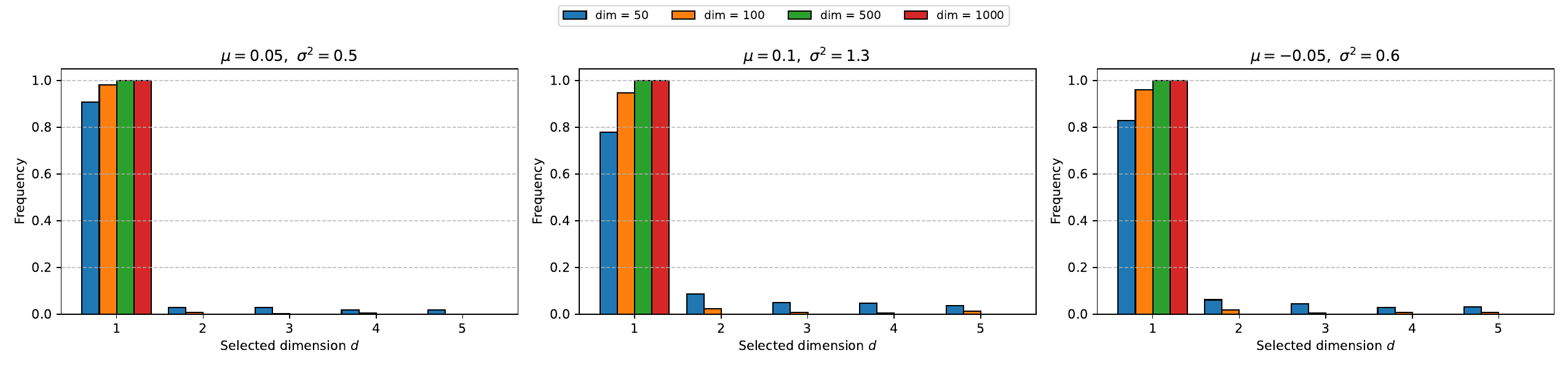}
        \caption{$\mathrm{DGP}_1$}
    \end{subfigure}
    \vspace{0.5cm}
    \begin{subfigure}{0.8\textwidth}
        \centering
        \includegraphics[width=1.0\textwidth]{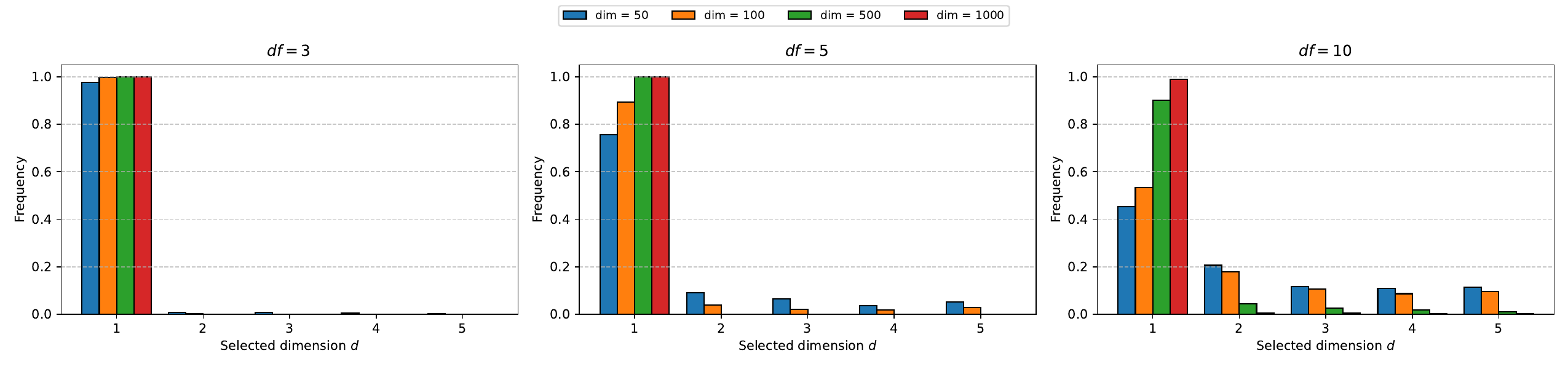}
        \caption{$\mathrm{DGP}_2$}
    \end{subfigure}
    \vspace{0.5cm}
    \begin{subfigure}{0.8\textwidth}
        \centering
        \includegraphics[width=1.0\textwidth]{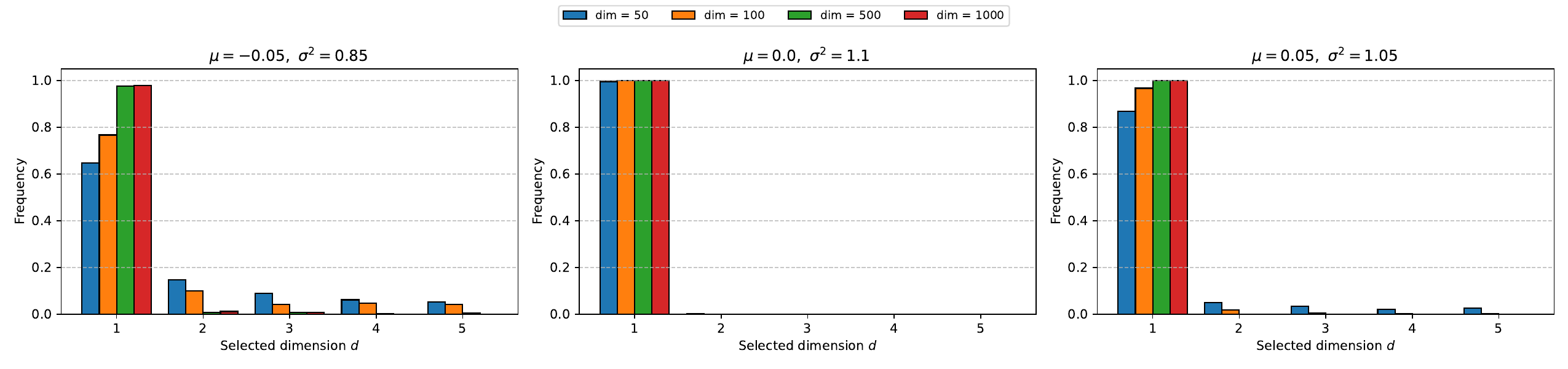}
        \caption{$\mathrm{DGP}_3$}
    \end{subfigure}
    \vspace{0.5cm}
    \begin{subfigure}{0.8\textwidth}
        \centering
        \includegraphics[width=1.0\textwidth]{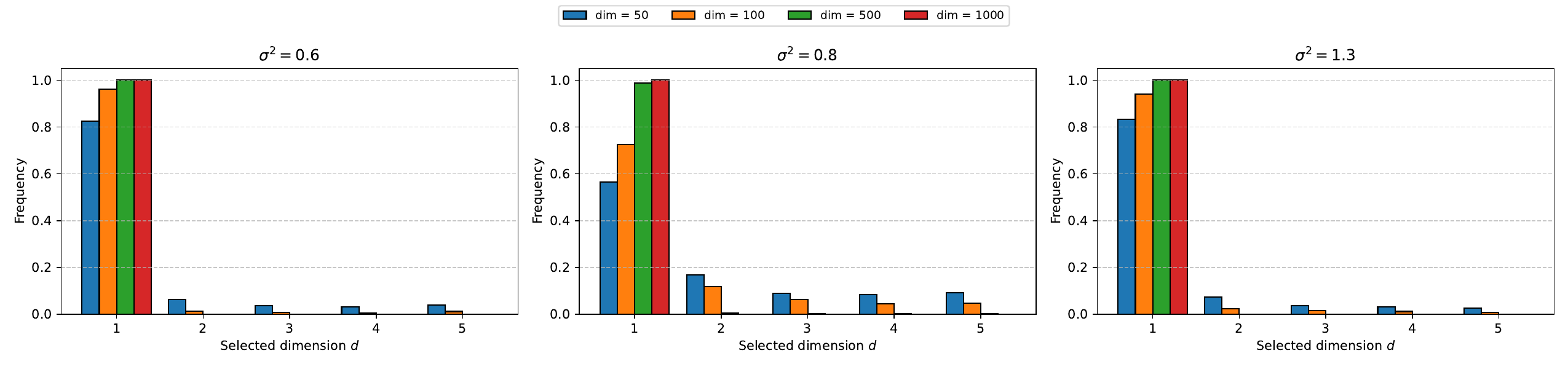}
        \caption{$\mathrm{DGP}_4$}
    \end{subfigure}
    \vspace{0.5cm}
    \begin{subfigure}{0.8\textwidth}
        \centering
        \includegraphics[width=1.0\textwidth]{art/freq_d/Set_5_freq_d.pdf}
        \caption{$\mathrm{DGP}_5$}
    \end{subfigure}
    \caption{Frequency of $d$ selection under alternative scenarios with balanced sample size ($n =  m =200$).}
    \label{fig:d_freq_power}
\end{figure}

In this section, we analyze the empirical frequencies of selected truncation dimensions across the simulation DGPs. Using a maximum search space of $\bar{d}=5$, Figures~\ref{fig:d_freq_size}--\ref{fig:d_freq_power} summarizes the selection distributions under the different scenarios.

Under the null distribution (Set 0), the selection procedure exhibits no strong preference, distributing choices relatively uniformly across the candidate dimensions $1$ through $5$ (e.g., frequencies of $0.339$, $0.250$, and $0.152$ for $d \in \{1, 2, 3\}$, respectively, under DGP 1 at $d=50$). This confirms that no single principal component spuriously dominates when the null hypothesis is true. 

In contrast, under Sets 1--4, the data-driven mechanism overwhelmingly concentrates on the smallest possible dimension, $d=1$. For instance, under the location--scale deviation (Set 1, $\mu=0.05, \sigma^2=0.5$), the procedure selects $d=1$ exactly $100\%$ of the time at $d=1000$. This identical behavior is observed across $t$--distribution, mixture, and scale--only deviations. The spectral geometry of these alternatives ensures that the leading eigencomponent of the centered kernel matrix entirely captures the higher-moment signal in high dimensions, perfectly justifying the use of \texttt{TMMD(d=1)} for these specific alternatives.

The most revealing finding emerges under the location-only deviations (Set 5). Here, the oracle selection mechanism systematically rejects $d=1$ and strongly concentrates on $d=2$. For the setting $\mu=1.3$ at $d=1000$, $d=2$ is selected in $100\%$ of the trials, while $d=1$ is never selected. This empirically isolates the exact cause of the structural blind spot: fixing the truncation dimension at $d=1$ blindly discards the exact spectral subspace where the location signal resides, whereas the data-driven oracle successfully adapts to $d=2$ to detect it.

\subsection{The Choice of Ceiling Parameter $ \bar{d} $}
\label{app:d_bar_selection}
\begin{figure}
    \centering
    \includegraphics[width=0.9\textwidth]{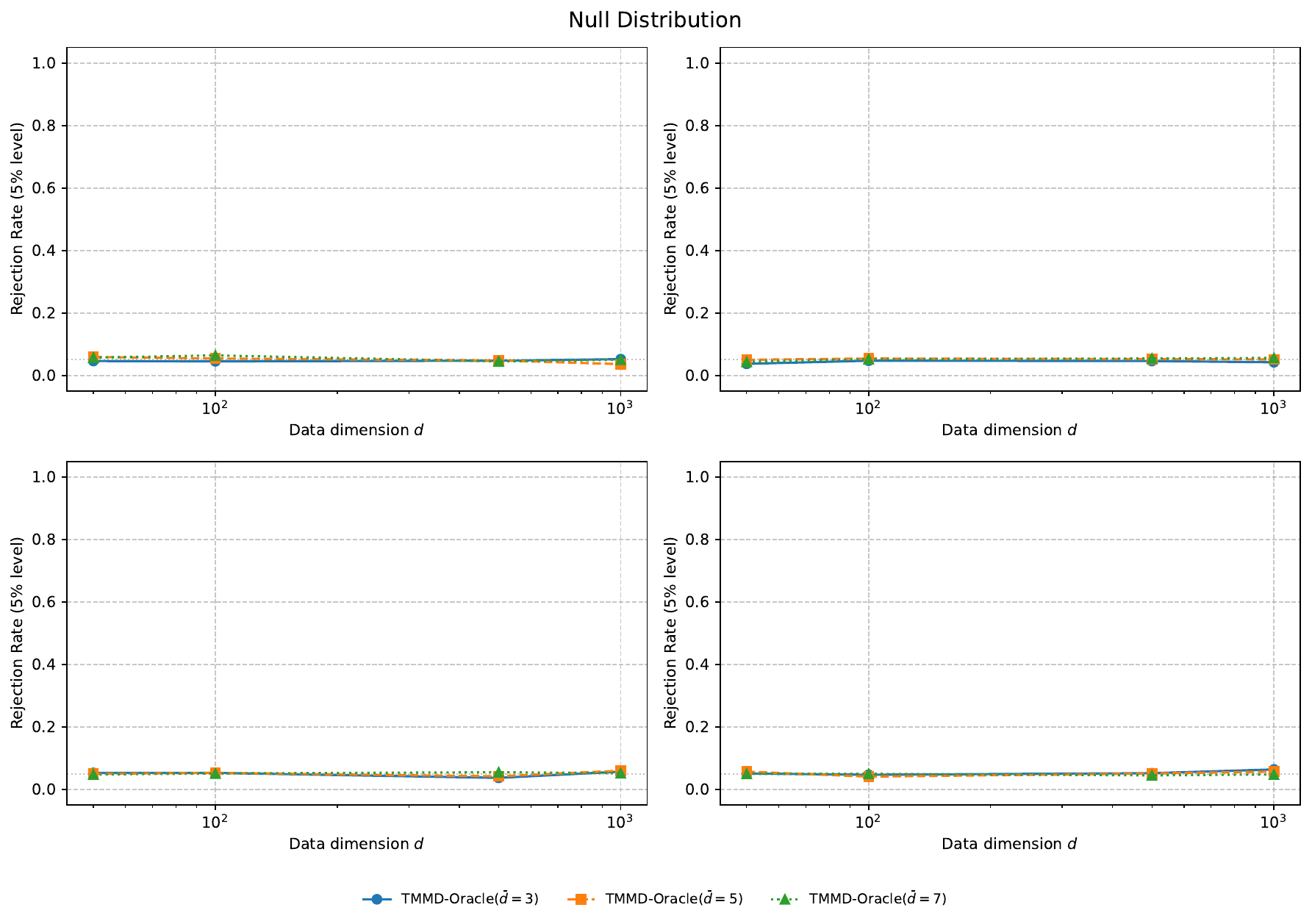}
    \caption{Empirical size control under the null scenarios with balanced sample size ($n = m = 200$).}
    \label{fig:d_bar_sensitivity_size}
\end{figure}

\begin{figure}
    \centering
    \begin{subfigure}{0.8\textwidth}
        \centering
        \includegraphics[width=1.0\textwidth]{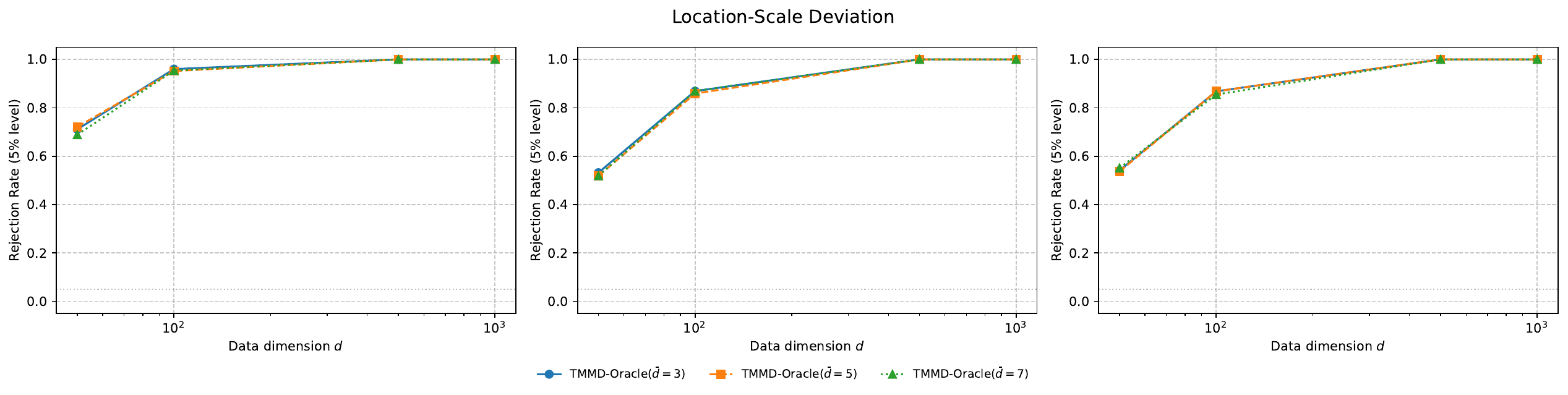}
        \caption{$\mathrm{DGP}_1$}
    \end{subfigure}
    \vspace{0.5cm}
    \begin{subfigure}{0.8\textwidth}
        \centering
        \includegraphics[width=1.0\textwidth]{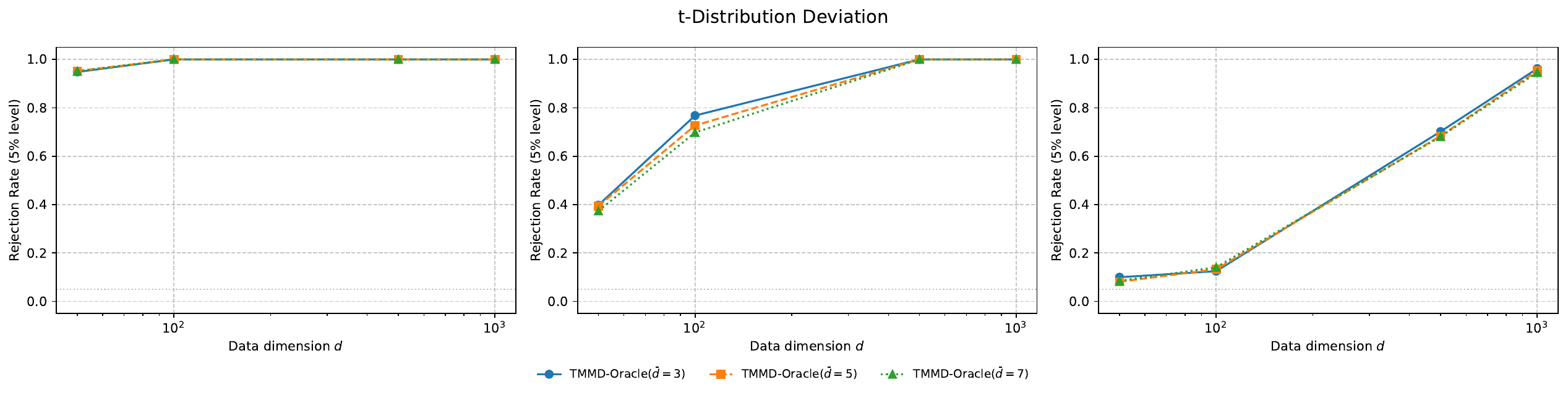}
        \caption{$\mathrm{DGP}_2$}
    \end{subfigure}
    \vspace{0.5cm}
    \begin{subfigure}{0.8\textwidth}
        \centering
        \includegraphics[width=1.0\textwidth]{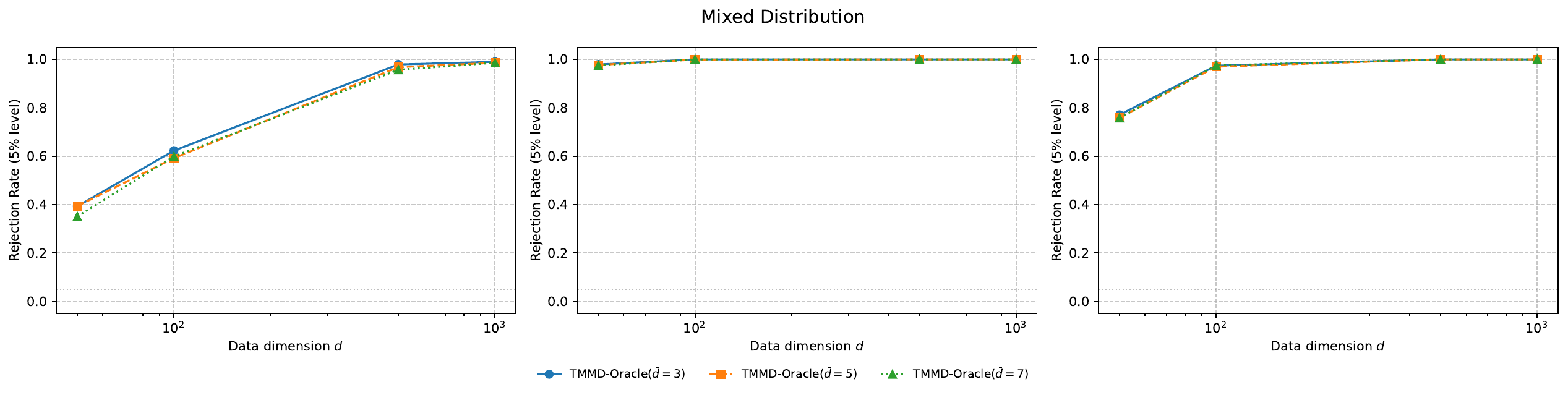}
        \caption{$\mathrm{DGP}_3$}
    \end{subfigure}
    \vspace{0.5cm}
    \begin{subfigure}{0.8\textwidth}
        \centering
        \includegraphics[width=1.0\textwidth]{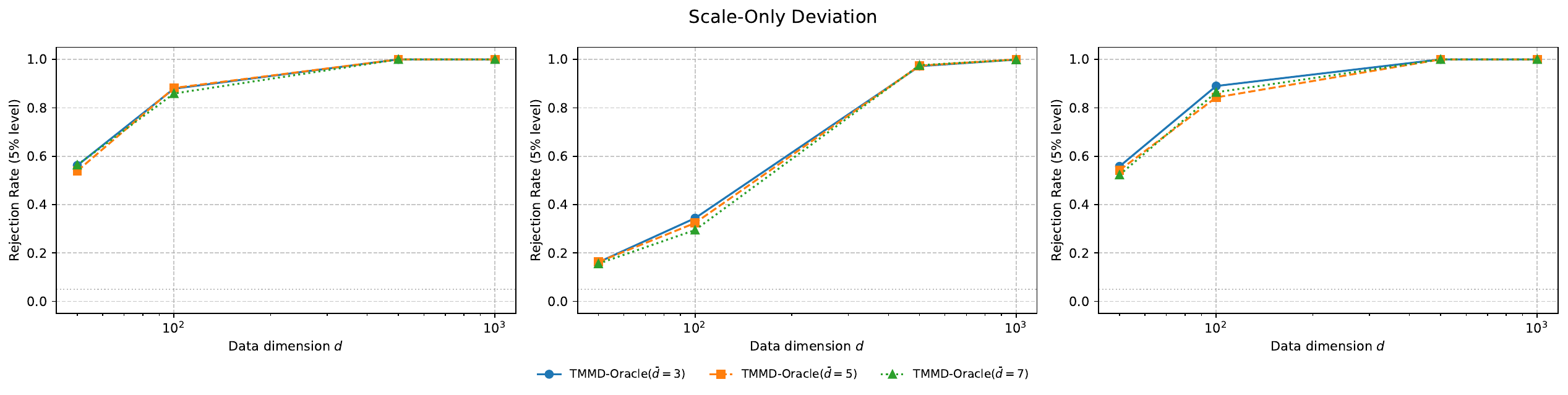}
        \caption{$\mathrm{DGP}_4$}
    \end{subfigure}
    \vspace{0.5cm}
    \begin{subfigure}{0.8\textwidth}
        \centering
        \includegraphics[width=1.0\textwidth]{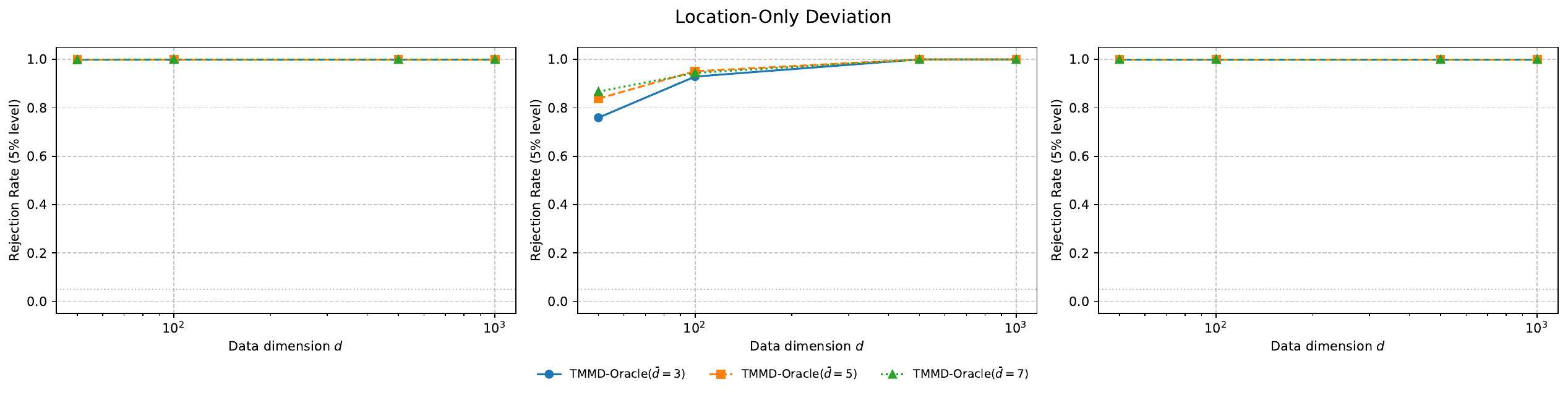}
        \caption{$\mathrm{DGP}_5$}
    \end{subfigure}
    \caption{Empirical power under alternative scenarios with balanced sample size ($n =  m =200$).}
    \label{fig:d_bar_sensitivity_power}
\end{figure}

A critical practical consideration for deploying the data-driven \texttt{TMMD-Oracle} is its sensitivity to the maximum candidate dimension, denoted as the ceiling parameter $\bar{d}$. To evaluate this, we compare the finite-sample performance of \texttt{TMMD-Oracle} across three different search space ceilings: $\bar{d} \in \{3, 5, 7\}$. The results, reported in Figures~\ref{fig:d_bar_sensitivity_size} and ~\ref{fig:d_bar_sensitivity_power}, demonstrate that the test is remarkably robust to the choice of $\bar{d}$.

Under the null distribution (Set 0), the choice of $\bar{d}$ has virtually no impact on size control. Across all four null DGPs and all ambient dimensions, the rejection rates remain tightly clustered around the nominal $0.05$ level regardless of whether the search space is restricted to $\bar{d}=3$ or expanded to $\bar{d}=7$. 

For Sets 1--5, the power of \texttt{TMMD-Oracle} exhibits stability as $\bar{d}$ increases. For instance, under the difficult scale-only deviation ($\sigma^2=0.8$) at $d=1000$, the power is $0.999$, $1.000$, and $0.999$ for $\bar{d} \in \{3, 5, 7\}$, respectively. Similarly, under the challenging $t$-distribution with $df=10$ at $d=1000$, the power ranges narrowly from $0.962$ down to $0.946$ across the three ceiling settings. 

Overall, these results provide strong practical reassurance: practitioners do not need to meticulously tune the ceiling parameter $\bar{d}$. As long as $\bar{d}$ is chosen to be reasonably large enough to encompass the true underlying signal subspace, the \texttt{TMMD-Oracle} seamlessly adapts to the correct truncation dimension.

\subsection{Discussion of Location-Only Deviations}
\label{app:location_only_discussion}
Location-only deviations (Set 5) pose a unique challenge for the \texttt{TMMD(d=1)} test statistic. To understand this behavior, we examine the spectral properties of $ T_{\mathcal{H}} $ under this specific alternative. In particular, with a balanced sample size of $n = m = 100$ and an ambient dimension of $d = 100$, we define $\boldsymbol{P} = \mathcal{N}(0, I_d)$ and $\boldsymbol{Q} = \mathcal{N}(0.6 \mathbf{1}, I_d)$. We generate $1000$ independent samples from each distribution, compute the Gram matrix using the Gaussian kernel, and estimate the eigenvalues. 

\begin{figure}
    \centering
    \includegraphics[width=0.9\textwidth]{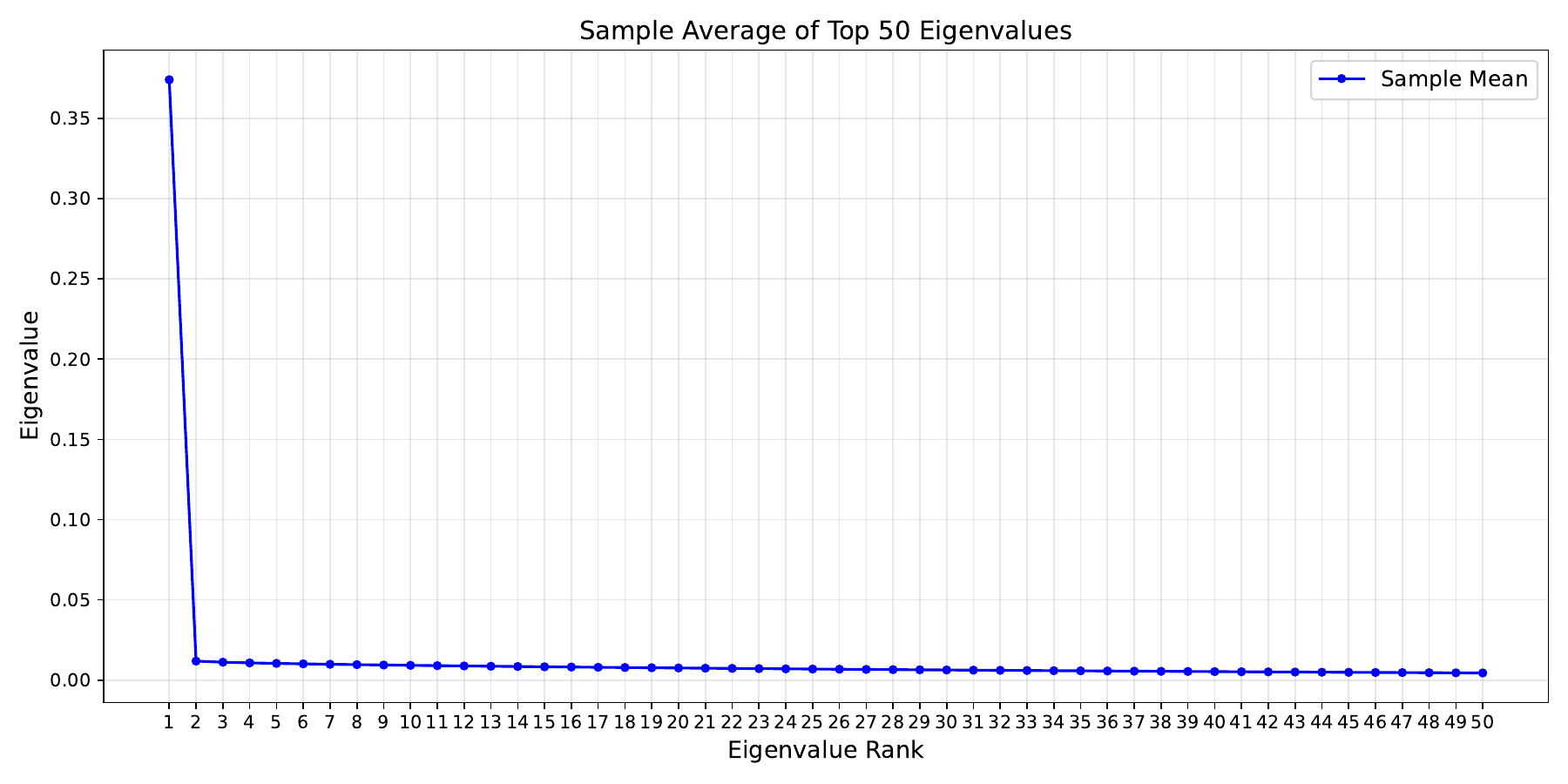}
    \caption{Estimated top 50 eigenvalues of the Gram matrix under the location-only deviation scenario with balanced sample size ($n = m = 100$).}
    \label{fig:location_only_spectrum}
\end{figure}

Figure \ref{fig:location_only_spectrum} plots the averaged top 50 eigenvalues of the Gram matrix $K$. As illustrated, the leading eigenvalue exceeds $0.35$, while the second drops sharply to approximately $0.01$. Notably, the subsequent eigenvalues (ranks 3 to 50) remain at this same level, forming a flat plateau. Because the \texttt{TMMD(d=1)} test statistic projects solely onto the leading eigenfunction, it entirely misses the location signal residing in the second eigenspace, resulting in severe power loss. In contrast, the data-driven \texttt{TMMD-Oracle} successfully identifies the second eigenfunction as the optimal projection direction, thereby fully recovering the location signal and achieving near-perfect power.

The apparent indifference among the 2nd to the 50th eigenvalues indicates that the second largest distinct eigenvalue possesses a large multiplicity. This spectral characteristic explains why test statistics based on full-rank MMD perform well under this scenario: projecting onto the many directions associated with this repeated eigenvalue shares the same spectral gap, meaning no penalty is imposed for including these redundant dimensions. However, the fact that our data-driven algorithm only selects $d = 2$ implies that the procedure is specifically designed for optimal testing, capturing the location signal with a single, most informative projection, rather than for identifying the full multiplicity of the eigenspace.

\section{Multiple Kernel Statistic}

\subsection{Null Distribution of the Aggregated Test Statistic}

A natural extension of our framework is to consider multiple kernels simultaneously. In theory, this allows us to capture a richer class of deviational signals by aggregating information across different directions in different RKHSs. The following theorem characterizes the null distribution of the aggregated test statistic formed by summing the unbiased directional test statistics corresponding to each kernel in a finite collection.

\begin{theorem}
\label{th:multi_kernel_null}
Let $\mathcal{C}$ be a finite collection of kernels satisfying Assumptions 1--5. For each $c \in \mathcal{C}$, let $d_c$ be the truncation dimension, and let $\tilde{D}_{N,c}^{d_c}$ be the corresponding unbiased directional test statistic. Define the aggregated test statistic as
\begin{equation*}
G_N = \sum_{c \in \mathcal{C}} \tilde{D}_{N,c}^{d_c}.
\end{equation*}
Under the null hypothesis $H_0$, we have
\begin{equation*}
G_N \xrightarrow{d} \sum_{k=1}^{D_{\mathcal{C}}} \frac{\omega_k}{p(1-p)} (\chi_{1,k}^2 - 1),
\end{equation*}
where $D_{\mathcal{C}} = \sum_{c \in \mathcal{C}} d_c$, $\{\chi_{1,k}^2\}_{k=1}^{D_{\mathcal{C}}}$ are independent chi-squared random variables with one degree of freedom, and $\omega_1 \ge \cdots \ge \omega_{D_{\mathcal{C}}}$ are the eigenvalues of the joint restricted cross-covariance matrix $\Sigma_{\mathcal{C}} \in \mathbb{R}^{D_{\mathcal{C}} \times D_{\mathcal{C}}}$.

The matrix $\Sigma_{\mathcal{C}}$ is structured in blocks $\Sigma_{\mathcal{C}}^{(c,c')} \in \mathbb{R}^{d_c \times d_{c'}}$ for $c, c' \in \mathcal{C}$. The $(i,j)$-th entry of the block $\Sigma_{\mathcal{C}}^{(c,c')}$ is given by the population cross-covariance of the eigenfunctions:
\begin{equation*}
\left( \Sigma_{\mathcal{C}}^{(c,c')} \right)_{ij} = \operatorname{Cov}_\rho\left(e_{i,c}(Z), e_{j,c'}(Z)\right) = \mathbb{E}_\rho\left[ e_{i,c}(Z) e_{j,c'}(Z) \right] - \mathbb{E}_\rho\left[ e_{i,c}(Z) \right] \mathbb{E}_\rho\left[ e_{j,c'}(Z) \right],
\end{equation*}
where $e_{i,c}$ denotes the $i$-th population eigenfunction of the integral operator associated with kernel $c$.
\end{theorem}

The proof of this result is provided in Appendix~\ref{app:other_proofs}.

\begin{remark}[Structure of the Product RKHS $\mathcal{H}_{\mathcal{C}}$]
\label{rmk:product_rkhs}
Let $\mathcal{C}$ be a finite collection of kernels, and let $\{\mathcal{H}_c\}_{c \in \mathcal{C}}$ be the associated Reproducing Kernel Hilbert Spaces. The direct sum space $\mathcal{H}_{\mathcal{C}} = \bigoplus_{c \in \mathcal{C}} \mathcal{H}_c$ is defined as the set of all tuples of functions $f = (f_c)_{c \in \mathcal{C}}$, where $f_c \in \mathcal{H}_c$ for each $c \in \mathcal{C}$.

The inner product on $\mathcal{H}_{\mathcal{C}}$ is naturally defined as the sum of the individual inner products. For any two elements $f = (f_c)_{c \in \mathcal{C}}$ and $g = (g_c)_{c \in \mathcal{C}}$ in $\mathcal{H}_{\mathcal{C}}$,
\begin{equation*}
\langle f, g \rangle_{\mathcal{H}_{\mathcal{C}}} = \sum_{c \in \mathcal{C}} \langle f_c, g_c \rangle_{\mathcal{H}_c}.
\end{equation*}
Consequently, the induced norm is given by:
\begin{equation*}
\|f\|_{\mathcal{H}_{\mathcal{C}}}^2 = \sum_{c \in \mathcal{C}} \|f_c\|_{\mathcal{H}_c}^2.
\end{equation*}

\textbf{Reproducing Kernel:} The joint feature map $\Phi_{\mathcal{C}}(z) = (k_{c_1}(z, \cdot), k_{c_2}(z, \cdot), \dots)^\top$ maps an observation $z \in \mathcal{Z}$ into $\mathcal{H}_{\mathcal{C}}$. The reproducing property dictates that the kernel of the product space, $k_{\mathcal{C}}(z, z')$, is the inner product of the joint feature maps:
\begin{equation*}
k_{\mathcal{C}}(z, z') = \langle \Phi_{\mathcal{C}}(z), \Phi_{\mathcal{C}}(z') \rangle_{\mathcal{H}_{\mathcal{C}}} = \sum_{c \in \mathcal{C}} \langle k_c(z, \cdot), k_c(z', \cdot) \rangle_{\mathcal{H}_c} = \sum_{c \in \mathcal{C}} k_c(z, z').
\end{equation*}
Thus, the product RKHS $\mathcal{H}_{\mathcal{C}}$ corresponds to the sum kernel $k_{\mathcal{C}} = \sum_{c \in \mathcal{C}} k_c$.

\textbf{Covariance Structure:} The cross-covariance operator in $\mathcal{H}_{\mathcal{C}}$, defined as $\Sigma_{\mathcal{C}}^{\text{op}} = \mathbb{E}_\rho\left[ (\Phi_{\mathcal{C}}(Z) - \mu_{\mathcal{C}}) \otimes (\Phi_{\mathcal{C}}(Z) - \mu_{\mathcal{C}}) \right]$, operates on tuples $f \in \mathcal{H}_{\mathcal{C}}$. Here, the joint feature map is given by $\Phi_{\mathcal{C}}(z) = (k_{c_1}(z, \cdot), k_{c_2}(z, \cdot), \dots)^\top$, and the joint mean embedding is $\mu_{\mathcal{C}} = \mathbb{E}_\rho[\Phi_{\mathcal{C}}(Z)]$. Its action on the $c$-th component involves summing over all components $c'$:
\begin{equation*}
\left( \Sigma_{\mathcal{C}}^{\text{op}} f \right)_c = \sum_{c' \in \mathcal{C}} \Sigma_{c,c'} f_{c'},
\end{equation*}
where $\Sigma_{c,c'} = \mathbb{E}_\rho\left[ (k_c(Z,\cdot) - \mu_c) \otimes (k_{c'}(Z,\cdot) - \mu_{c'}) \right]$ is the cross-covariance operator mapping $\mathcal{H}_{c'}$ to $\mathcal{H}_c$. 
\end{remark}

\begin{remark}[Matrix Representation of the Joint Covariance Operator]
\label{rmk:matrix_representation}

The matrix $\Sigma_{\mathcal{C}}$ defined in Theorem \ref{th:multi_kernel_null} is the finite-dimensional matrix representation of $\Sigma_{\mathcal{C}}^{\text{op}}$ restricted to the truncated eigen-subspaces, with its block entries given by $\left( \Sigma_{\mathcal{C}}^{(c,c')} \right)_{ij} = \langle \Sigma_{c,c'} e_{j,c'}, e_{i,c} \rangle_{\mathcal{H}_c}$. To explicitly see it, we examine the action of the operator on the joint orthonormal basis.

For each kernel $c \in \mathcal{C}$, let $P_{d_c, c} = \sum_{i=1}^{d_c} e_{i,c} \otimes_{\mathcal{H}_c} e_{i,c}$ be the orthogonal projection onto the truncated subspace of $\mathcal{H}_c$. The joint projection operator onto the truncated subspace is $P_{\mathcal{C}} = \bigoplus_{c \in \mathcal{C}} P_{d_c, c}$. The restricted covariance operator is therefore $P_{\mathcal{C}} \Sigma_{\mathcal{C}}^{\text{op}} P_{\mathcal{C}}$.

Let $\mathbf{e}_{i,c} \in \mathcal{H}_{\mathcal{C}}$ denote the canonical embedding of the basis element $e_{i,c} \in \mathcal{H}_c$ into the product space (i.e., a tuple that is zero in all components except the $c$-th component, where it is $e_{i,c}$). The collection $\{\mathbf{e}_{i,c}\}_{c \in \mathcal{C}, i=1,\dots,d_c}$ forms an orthonormal basis for the truncated subspace $\operatorname{Im}(P_{\mathcal{C}})$. The matrix entries of $\Sigma_{\mathcal{C}}$ are defined by the inner products of the restricted operator applied to these basis vectors:
\begin{equation*}
\left( \Sigma_{\mathcal{C}} \right)_{(i,c), (j,c')} = \langle \Sigma_{\mathcal{C}}^{\text{op}} \mathbf{e}_{j,c'}, \mathbf{e}_{i,c} \rangle_{\mathcal{H}_{\mathcal{C}}}.
\end{equation*}

Recall that the action of the joint operator $\Sigma_{\mathcal{C}}^{\text{op}}$ on a tuple $f = (f_{c_1}, f_{c_2}, \dots)^\top$ is defined component-wise by summing across the cross-covariance operators:
\begin{equation*}
\left( \Sigma_{\mathcal{C}}^{\text{op}} f \right)_c = \sum_{c'' \in \mathcal{C}} \Sigma_{c,c''} f_{c''}.
\end{equation*}
Applying this to the basis vector $\mathbf{e}_{j,c'}$ (which has $f_{c''} = \delta_{c''c'} e_{j,c'}$), the $c$-th component of the resulting tuple is:
\begin{equation*}
\left( \Sigma_{\mathcal{C}}^{\text{op}} \mathbf{e}_{j,c'} \right)_c = \sum_{c'' \in \mathcal{C}} \Sigma_{c,c''} \delta_{c''c'} e_{j,c'} = \Sigma_{c,c'} e_{j,c'}.
\end{equation*}
Substituting this back into the inner product, and using the fact that the inner product in $\mathcal{H}_{\mathcal{C}}$ is the sum of the inner products in the individual spaces, we obtain:
\begin{align*}
\langle \Sigma_{\mathcal{C}}^{\text{op}} \mathbf{e}_{j,c'}, \mathbf{e}_{i,c} \rangle_{\mathcal{H}_{\mathcal{C}}} &= \sum_{c'' \in \mathcal{C}} \langle \left( \Sigma_{\mathcal{C}}^{\text{op}} \mathbf{e}_{j,c'} \right)_{c''}, (\mathbf{e}_{i,c})_{c''} \rangle_{\mathcal{H}_{c''}} \\
&= \sum_{c'' \in \mathcal{C}} \langle \Sigma_{c'',c'} e_{j,c'}, \delta_{c''c} e_{i,c} \rangle_{\mathcal{H}_{c''}} \\
&= \langle \Sigma_{c,c'} e_{j,c'}, e_{i,c} \rangle_{\mathcal{H}_c}.
\end{align*}
Finally, by the reproducing property of the cross-covariance operator, this inner product equals the population cross-covariance of the eigenfunctions evaluated under the mixture distribution $\rho$:
\begin{equation*}
\langle \Sigma_{c,c'} e_{j,c'}, e_{i,c} \rangle_{\mathcal{H}_c} = \mathbb{E}_\rho\left[ e_{i,c}(Z) e_{j,c'}(Z) \right] - \mathbb{E}_\rho\left[ e_{i,c}(Z) \right] \mathbb{E}_\rho\left[ e_{j,c'}(Z) \right] = \operatorname{Cov}_\rho\left(e_{i,c}(Z), e_{j,c'}(Z)\right).
\end{equation*}
This establishes the exact equivalence between the abstract operator restriction $P_{\mathcal{C}} \Sigma_{\mathcal{C}}^{\text{op}} P_{\mathcal{C}}$ and the block-structured matrix $\Sigma_{\mathcal{C}}$ defined in Theorem \ref{th:multi_kernel_null}. The diagonal blocks ($c=c'$) capture the variances of the directional components within each kernel, while the off-diagonal blocks ($c \neq c'$) capture the covariances induced by evaluating different kernels on the same data points.
\end{remark}

\begin{remark}[Estimation of the Joint Cross-Covariance Matrix]
\label{rmk:joint_covariance_estimation}
To implement the parametric bootstrap for the aggregated test statistic, one requires a consistent estimator of $\Sigma_{\mathcal{C}}$. For each kernel $c \in \mathcal{C}$, let $\hat{e}_{i,c}$ denote the empirical eigenfunctions obtained via the Nyström extension. For each observation $z_\gamma$ in the poole7d sample ($\gamma = 1, \dots, N$), construct the joint evaluation vector $\mathbf{V}_\gamma \in \mathbb{R}^{D_{\mathcal{C}}}$ by concatenating the leading $d_c$ eigenfunction evaluations across all kernels:
\begin{equation*}
\mathbf{V}_\gamma = \left( \hat{e}_{1,c_1}(z_\gamma), \dots, \hat{e}_{d_{c_1},c_1}(z_\gamma), \hat{e}_{1,c_2}(z_\gamma), \dots, \hat{e}_{d_{c_2},c_2}(z_\gamma), \dots \right)^\top.
\end{equation*}
Let $\bar{\mathbf{V}} = \frac{1}{N} \sum_{\gamma=1}^N \mathbf{V}_\gamma$ be the pooled sample mean of these vectors. The empirical joint cross-covariance matrix is then computed as the standard sample covariance matrix of the centered evaluation vectors:
\begin{equation*}
\hat{\Sigma}_{\mathcal{C}} = \frac{1}{N-1} \sum_{\gamma=1}^N (\mathbf{V}_\gamma - \bar{\mathbf{V}})(\mathbf{V}_\gamma - \bar{\mathbf{V}})^\top \in \mathbb{R}^{D_{\mathcal{C}} \times D_{\mathcal{C}}}.
\end{equation*}
The parametric bootstrap proceeds by generating independent samples $\mathbf{W}^{(b)} \sim \mathcal{N}\left(\mathbf{0}, \frac{1}{\hat{p}(1-\hat{p})}\hat{\Sigma}_{\mathcal{C}}\right)$ for $b = 1, \dots, B$. The bootstrap statistics are calculated as $G_N^{(b)} = \|\mathbf{W}^{(b)}\|_2^2 - \frac{1}{\hat{p}(1-\hat{p})}\operatorname{Tr}(\hat{\Sigma}_{\mathcal{C}})$, and the empirical distribution of $\{G_N^{(b)}\}_{b=1}^B$ serves as the approximated null distribution for $G_N$. 
\end{remark}

\subsection{Additional Simulation Results for Multiple Kernel Statistic}

\begin{figure}
    \centering
    \includegraphics[width=0.9\textwidth]{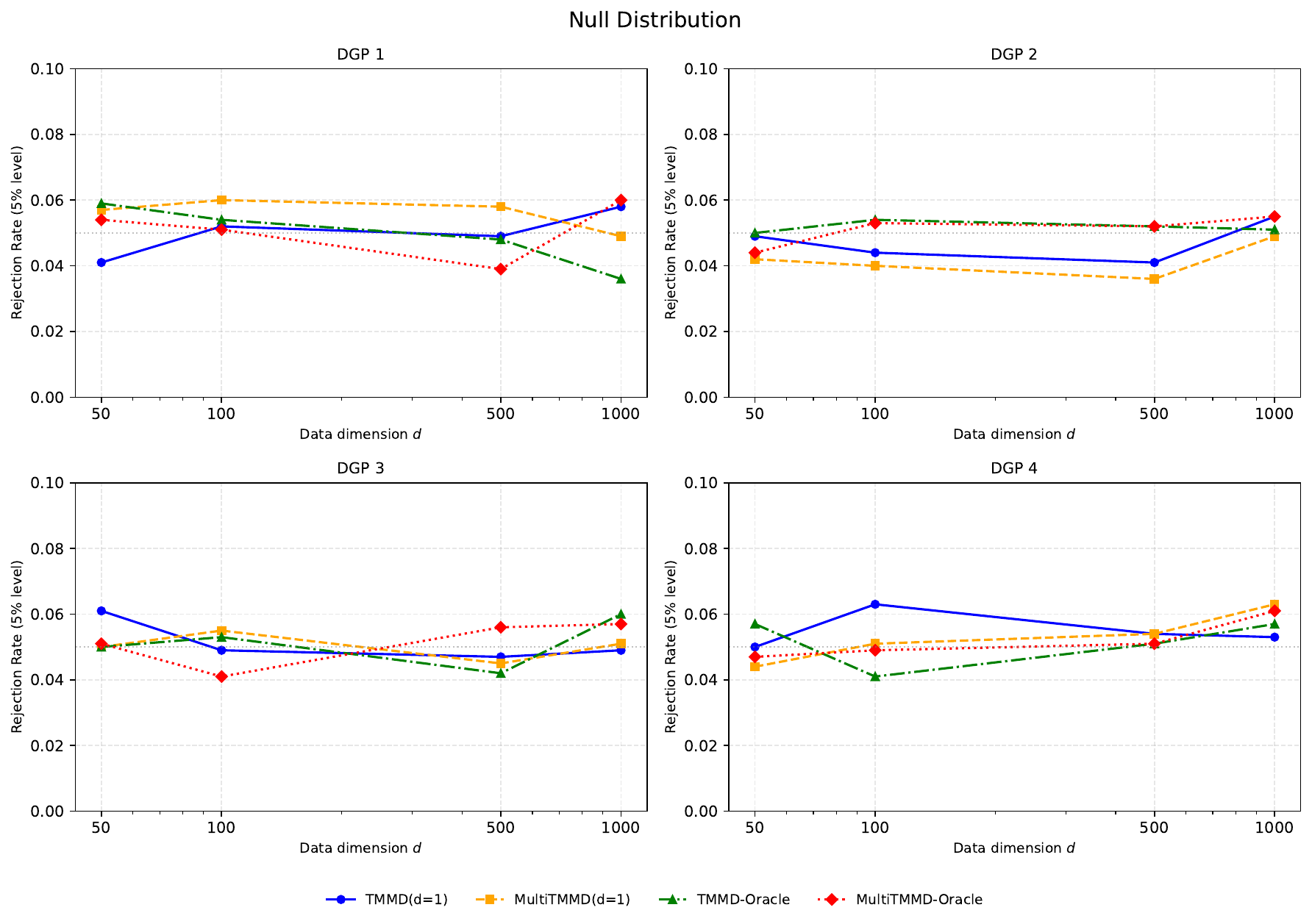}
    \caption{Empirical size control under the null scenarios with balanced sample size.}
    \label{fig:multi_kernel_size}
\end{figure}

\begin{figure}
    \centering
    \begin{subfigure}{0.8\textwidth}
        \centering
        \includegraphics[width=1.0\textwidth]{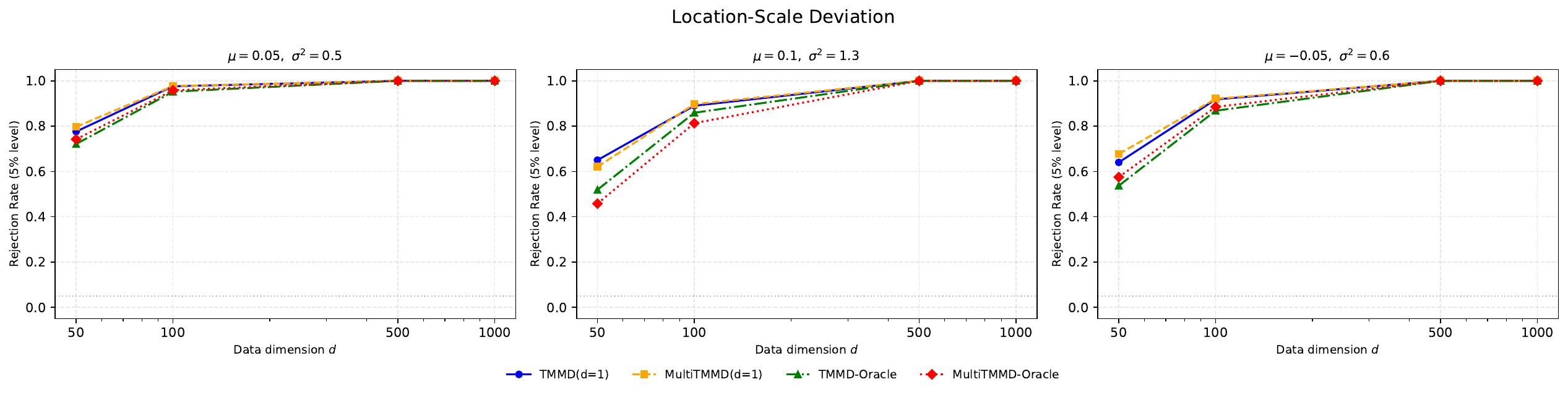}
        \caption{$\mathrm{DGP}_1$}
    \end{subfigure}
    \vspace{0.5cm}
    \begin{subfigure}{0.8\textwidth}
        \centering
        \includegraphics[width=1.0\textwidth]{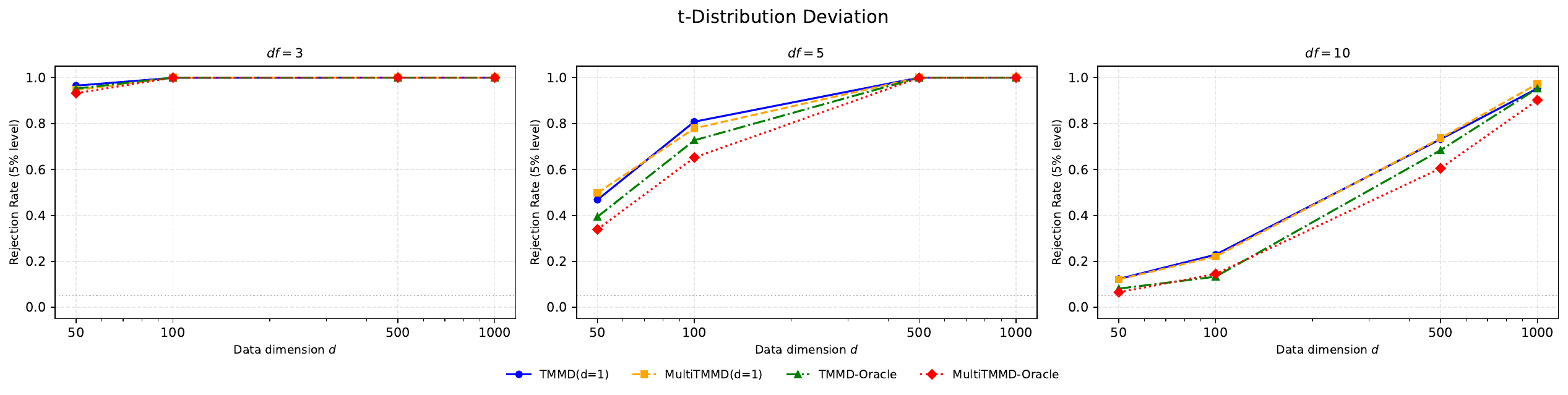}
        \caption{$\mathrm{DGP}_2$}
    \end{subfigure}
    \vspace{0.5cm}
    \begin{subfigure}{0.8\textwidth}
        \centering
        \includegraphics[width=1.0\textwidth]{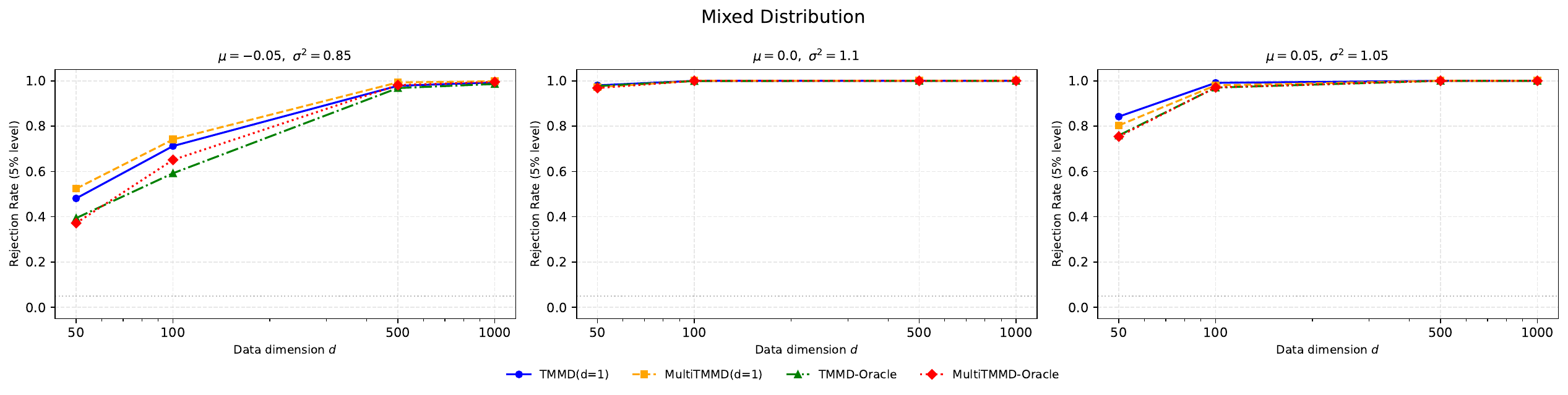}
        \caption{$\mathrm{DGP}_3$}
    \end{subfigure}
    \vspace{0.5cm}
    \begin{subfigure}{0.8\textwidth}
        \centering
        \includegraphics[width=1.0\textwidth]{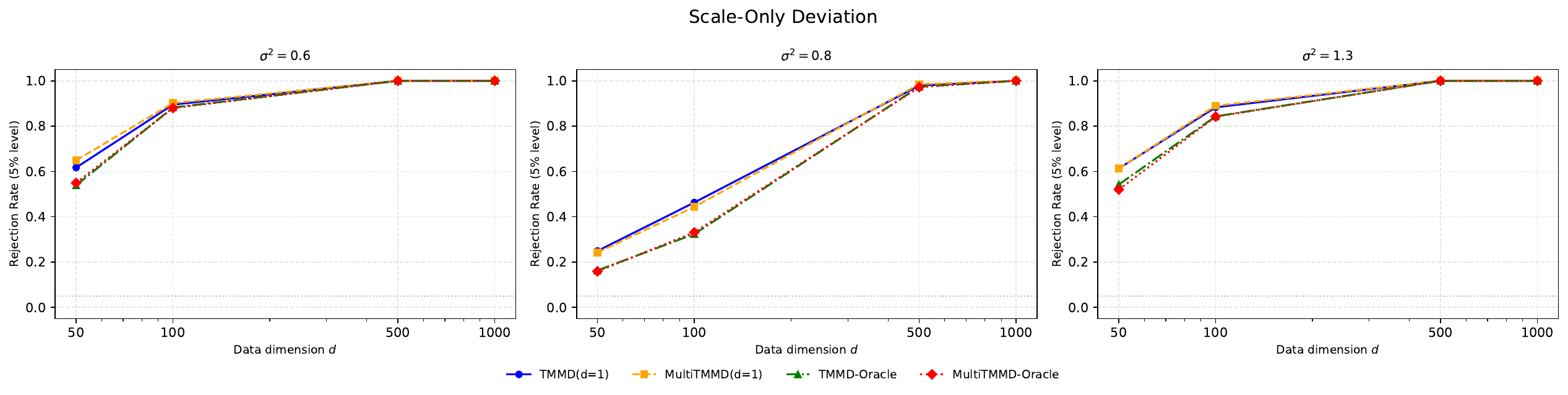}
        \caption{$\mathrm{DGP}_4$}
    \end{subfigure}
    \vspace{0.5cm}
    \begin{subfigure}{0.8\textwidth}
        \centering
        \includegraphics[width=1.0\textwidth]{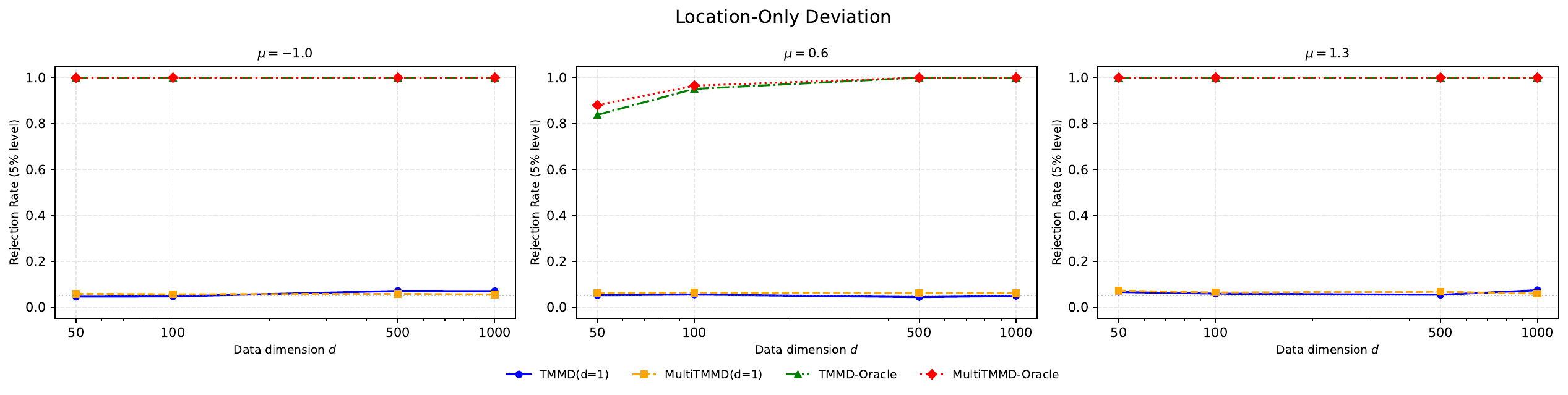}
        \caption{$\mathrm{DGP}_5$}
    \end{subfigure}
    \caption{Empirical power under alternative scenarios with balanced sample size.}
    \label{fig:multi_kernel_power}
\end{figure}

We perform a simulation study to evaluate the finite-sample performance of the aggregated test statistic $G_N$ defined in Theorem \ref{th:multi_kernel_null}. We consider a collection of three kernels: the Gaussian kernel $k_{\mathrm{G}}(x,y)=\exp\!\left(-\|x-y\|_2^2/\sigma\right)$, the Laplacian kernel $k_{\mathrm{L}}(x,y)=\exp\!\left(-\|x-y\|_1/\sigma\right)$, and the inverse multiquadric (IMQ) kernel $k_{\mathrm{IMQ}}(x,y)=\left(\|x-y\|_2^2+\sigma^2\right)^{-1/2}$. For all kernels, we set the bandwidth parameter $\sigma$ to the median pairwise distance among the pooled samples. We compare the performance of the aggregated test statistics, denoted as \texttt{MultiTMMD(d=1)} and \texttt{MultiTMMD-Oracle} against their single kernel counterparts \texttt{TMMD(d=1)} and \texttt{TMMD-Oracle} based solely on the Gaussian kernel. The data-driven selection procedure for the truncation dimension in \texttt{MultiTMMD-Oracle} is based on a grid of $ \bar{d} \times \bar{d} \times \bar{d} $ for the three kernels, with $\bar{d}=5$. Similar to the simulation design in Section \ref{sec:simulation}, we consider the same six sets of DGPs (Set 0--5) and evaluate the empirical size and power under balanced sample sizes ($n = m = 100$) when truncating at $d=1$, and double the sample size ($n = m = 200$) for the data-driven oracle procedure. The results are reported in Figures~\ref{fig:multi_kernel_size} and ~\ref{fig:multi_kernel_power}. The surprising finding is that the aggregated test statistic does not yield any noticeable improvement in power compared to the single-kernel version. In fact, the performance of \texttt{MultiTMMD(d=1)} and \texttt{MultiTMMD-Oracle} is almost identical to that of \texttt{TMMD(d=1)} and \texttt{TMMD-Oracle}, respectively, across all DGPs and ambient dimensions. This suggests that the Gaussian kernel alone is sufficiently powerful to capture the relevant signals in these scenarios, and that adding more kernels does not provide additional discriminatory information. 

\section{Additional Technical Results}
\label{app:additional_tech_results}
\subsection{Useful Bound Results}
The following lemmas come from \cite{rosasco2010learning}. 
\begin{lemma}
    \label{bound_eigenvalues}
    Let $\{\lambda_i\}_{i \in I}$ and $\{\hat{\lambda}_i\}_{i \in I}$ be the at most countable families of eigenvalues of $L_k$ and $K$, respectively, where $\hat{\lambda}_i = 0$ for $i > r$. Then, for any $\tau > 0$, with probability at least $1 - 2e^{-\tau}$, we have
    \begin{align*}
        \sum_{i \in I} \left(\lambda_i - \hat{\lambda}_i\right)^2 &\leq \frac{8 \bar{k}^2 \tau}{N}, \\
        \left|\sum_{i \in I}\left(\lambda_i - \hat{\lambda}_i\right)\right| = \left|\operatorname{Tr}(T_{\mathcal{H}}) - \operatorname{Tr}(T_N)\right| &\leq \frac{2\sqrt{2}\bar{k}\sqrt{\tau}}{\sqrt{N}}, \\
        \sup_{i \in I} \left|\lambda_i - \hat{\lambda}_i \right| &\leq \frac{2\sqrt{2}\bar{k}\sqrt{\tau}}{\sqrt{N}}.
    \end{align*}
\end{lemma}

\begin{lemma}
    \label{lm:TH_and_TN}
    The operators $ T_{\mathcal{H}}$ and $ T_N $ defined in Section~\ref{sec:prelim} are Hilbert-Schmidt. In addition, with confidence $ 1-2\exp(-\tau) $, we have 
    \[ \|T_{\mathcal{H}} - T_N\|_{HS} \le \frac{2\sqrt{2}\bar{k}\sqrt{\tau}}{\sqrt{N}}. \] 
\end{lemma}

\subsection{Useful Perturbation Results}
\label{app:perturbation_results}
We conclude with standard perturbation bounds for compact, self-adjoint operators on a Hilbert space, which form the basis for our kernel sensitivity analysis in Section \ref{sec:kernel_choice} (see \cite{hsing2015theoretical}).

Consider two compact, self-adjoint operators on a Hilbert space $\mathcal{A}$, denoted $A$ and $\tilde{A}$, with the perturbation operator $\delta A = \tilde{A} - A$. The resolvent of $A$ at $z \in \mathbb{C}$ is defined as 
\[ \mathcal{R}(z) = (A - zI)^{-1} = \sum_{i=1}^\infty \frac{1}{\lambda_i - z} \langle \cdot, v_i \rangle v_i, \] 
where $\{\lambda_i, v_i\}_{i=1}^\infty$ are the eigenvalue-eigenfunction pairs of $A$.

Let $\Gamma$ represent the boundary of a closed disk $D$ in the complex plane that contains a subset of eigenvalues $\{\lambda_i: p \le i \le q\}$ strictly in its interior, and no other eigenvalues of $A$. The orthogonal projection operator $P = \sum_{i=p}^q v_i \otimes v_i$ onto the corresponding eigenspace can be recovered via a contour integral of the resolvent:
\[ P = -\frac{1}{2\pi i} \oint_{\Gamma} \mathcal{R}(z) \, dz, \] 
where $ \oint_{\Gamma} $ denotes the contour integral along $\Gamma$ in the counter-clockwise direction. 

For a non-zero eigenvalue $\lambda_i$ with multiplicity $m_i$, the projection onto its eigenspace is $P_i = \sum_{j=1}^{m_i} v_{i,j} \otimes v_{i,j}$. To systematically pair eigenvalues between $A$ and $\tilde{A}$, let $\lambda_1 > \lambda_2 > \cdots$ denote the distinct non-zero eigenvalues of $A$ with multiplicities $m_1, m_2, \ldots$. Let $\tilde{\lambda}_1 \ge \tilde{\lambda}_2 \ge \cdots$ be the repeated eigenvalues of $\tilde{A}$. Defining the cumulative sums $M_i = \sum_{j=1}^i m_j$, we associate the projection $\tilde{P}_i = \sum_{j = M_{i-1}+1}^{M_i} \tilde{v}_j \otimes \tilde{v}_j$ with the perturbed counterparts of $\lambda_i$.     

\begin{theorem}
    \label{th:perturbation_space}
    Let $\lambda_1 > \lambda_2 > \cdots$ be the distinct eigenvalues of a self-adjoint and compact operator $A$. Take $\Gamma$ as the circle centered at $\lambda_i$ with radius $\eta_i = \frac{1}{2} \min_{l \neq i}|\lambda_l - \lambda_i|$. Assume the perturbation satisfies $\| \delta A \|_{\mathrm{op}} < \eta_i$. Then for the projection operator $\tilde{P}_i$ defined above, we have 
    \[ \tilde{P}_i - P_i = \mathcal{S}_i (\delta A) P_i + P_i (\delta A) \mathcal{S}_i + \frac{1}{2\pi i} \oint_{\Gamma} \mathcal{M}(z) \, dz, \]
    where 
    \[ \mathcal{S}_i = \sum_{j \neq i} \frac{1}{\lambda_j - \lambda_i} P_j, \quad \text{and} \quad \mathcal{M}(z) = \mathcal{R}(z) \sum_{j=2}^\infty \left(-\delta A \, \mathcal{R}(z)\right)^j. \]    
    Furthermore, the following norm bounds hold:
    \[ \left\| \tilde{P}_i - P_i - (\mathcal{S}_i (\delta A) P_i + P_i (\delta A) \mathcal{S}_i) \right\|_{\mathrm{op}} < \frac{\bigl(\|\delta A \|_{\mathrm{op}}/\eta_i\bigr)^2}{1-\bigl(\|\delta A \|_{\mathrm{op}}/\eta_i\bigr)^2}, \]
    \[ \left\|\tilde{P}_i - P_i\right\|_{\mathrm{op}} < \frac{\|\delta A \|_{\mathrm{op}}/\eta_i}{1-\|\delta A \|_{\mathrm{op}}/\eta_i}. \]
\end{theorem}

\begin{theorem}
    \label{th:perturbation_eigenvalue}
    Let $ A $, $\{\lambda_i\}$, $P_i$, and $\tilde{P}_i$ be defined as in Theorem~\ref{th:perturbation_space}. Then for each $j = M_{i-1}+1, \ldots, M_i$, the perturbed eigenvalue $\tilde{\lambda}_j$ satisfies 
    \[ \tilde{\lambda}_j - \lambda_i = \langle (\delta A) \tilde{v}_j, \tilde{v}_j \rangle + \langle (\tilde{P}_i - P_i) (A - \lambda_i I)(\tilde{P}_i - P_i) \tilde{v}_j, \tilde{v}_j \rangle, \]
    where $I$ is the identity operator.
\end{theorem}

\subsection{Orthogonal Transformations}
\label{orthogonal transformation}
The alignment between the empirical and population subspaces ($ \hat{E} $ and $ E $, respectively) can be made explicit via an orthogonal transformation. Define the $d \times d$ matrix $M$ by $M_{ij} = \langle \hat{e}_i, e_j \rangle_{\mathcal{H}}$, and consider two functions $ f \in \hat{E} \subset \mathcal{H} $ and $ g \in E \subset \mathcal{H}$ expressed in the respective bases:
\[ f = \sum_{i=1}^d v_i \hat{e}_i, \quad g = \sum_{j=1}^d u_j e_j. \]
We normalize these functions to have unit norm, which implies $\|v\|_2 = \|u\|_2 = 1$. The cosine of the angle between these two functions is given by their inner product: $ \langle f, g \rangle_{\mathcal{H}} = \sum_{i,j=1}^d v_i u_j \langle \hat{e}_i, e_j \rangle_{\mathcal{H}} = u^\top M v $. From this perspective, the singular values $ \{\sigma_i\}_{i=1}^d$ of $M$ represent the cosines of the principal angles between the subspaces $\hat{E}$ and $E$, and consequently, we have $ \sigma_i^2 \in [0,1] $ for all $i=1,\ldots,d$. 

Furthermore, let $P_d$ and $\hat{P}_d$ denote the orthogonal projection operators onto the population and empirical subspaces, respectively. The matrix representation of the composition $P_d \hat{P}_d$ with respect to the base $ \{\hat{e}_i\}_{i=1}^d $ can be found by applying the projection operators to the basis functions:
\begin{align*}
    \langle \hat{e}_i, P_d \hat{P}_d \hat{e}_j \rangle_{\mathcal{H}} &= \langle \hat{e}_i, P_d \hat{e}_j \rangle_{\mathcal{H}} = \sum_{k=1}^d \langle \hat{e}_i, e_k \rangle_{\mathcal{H}} \langle e_k, \hat{e}_j \rangle_{\mathcal{H}} = (M^\top M)_{ij}.
\end{align*} 
Thus, $ M^\top M $ is the matrix representation of $P_d \hat{P}_d$ in the empirical basis, and its eigenvalues are precisely $\{\sigma_i^2\}_{i=1}^d$ by Theorem~\ref{th:spectral_equivalence}. 

The squared Hilbert--Schmidt norm of the difference between the projection operators expands as
\begin{align*}
    \left\|P_d - \hat{P}_d\right\|_{\mathrm{HS}}^2 &= \operatorname{Tr}\left(\left(P_d - \hat{P}_d\right)^2\right) \\
    &= \operatorname{Tr}(P_d) + \operatorname{Tr}(\hat{P}_d) - 2 \operatorname{Tr}(P_d\hat{P}_d) \\
    &= 2d - 2 \operatorname{Tr}(M^\top M) = 2\sum_{i=1}^d (1 - \sigma_i^2).
\end{align*}
By Lemma \ref{bound_projections}, $\left\|P_d - \hat{P}_d\right\|_{\mathrm{HS}}^2 = O_p(N^{-1})$, which implies $\sum_{i=1}^d \sigma_i^2 = d - O_p(N^{-1})$.

Let the singular value decomposition of $M^\top$ be $U \Sigma V^\top$, where $U, V \in \mathbb{R}^{d \times d}$ are orthogonal and $\Sigma = \operatorname{diag}(\sigma_1, \ldots, \sigma_d)$. We construct a sequence of random orthogonal matrices $O_N = UV^\top$. The rotated empirical eigenfunctions $\tilde{e}_i = \sum_{j=1}^d (O_N)_{ij} \hat{e}_j$ then satisfy
\begin{align*}
    \langle e_i, \tilde{e}_j \rangle_{\mathcal{H}} &= \sum_{k=1}^d (O_N)_{kj} \langle e_i, \hat{e}_k \rangle_{\mathcal{H}} = \sum_{k=1}^d (O_N)_{kj} M_{ik} = \left(M O_N\right)_{ij}.
\end{align*}
Substituting $O_N = UV^\top$ and $M^\top = U\Sigma V^\top$ yields $M O_N = V\Sigma V^\top$, which is symmetric and positive semi-definite. Consequently,
\begin{align*}
    \sum_{i=1}^d \| e_i - \tilde{e}_i \|_{\mathcal{H}}^2 &= 2\left(d - \operatorname{Tr}(V \Sigma V^\top)\right) = 2\left(d - \sum_{i=1}^d \sigma_i\right) \\
    &\le 2\left(d - \sum_{i=1}^d \sigma_i^2\right) = O_p(N^{-1}).
\end{align*}    
Since $ d $ is finite, this implies $\| e_i - \tilde{e}_i \|_{\mathcal{H}}^2 = O_p(N^{-1})$ for each $i=1,\ldots,d$. Thus, the rotated empirical eigenfunctions $\{\tilde{e}_i\}_{i=1}^d$ are consistent estimators of the population eigenfunctions $\{e_i\}_{i=1}^d$ in the $\mathcal{H}$-norm.
 
Finally, because orthogonal transformations preserve the Euclidean norm, the squared norm of the projection of any $f \in \mathcal{H}$ onto the empirical subspace is invariant to this rotation:
    \[ \sum_{i=1}^d \langle f, \hat{e}_i \rangle_{\mathcal{H}}^2 = \sum_{i=1}^d \langle f, \tilde{e}_i \rangle_{\mathcal{H}}^2. \] 

\subsection{Trace Convergence}
The following lemma is used in the proof of Theorem \ref{null_distribution_test_statistic} to establish the convergence of the trace correction terms.
\begin{lemma}
\label{lm:trace_convergence}
    Let $d$ be the sum of the multiplicities of the first $v$ distinct eigenvalues of the integral operator $L_k$. Under the null hypothesis and the conditions of Lemma \ref{bound_projections}, we have:
        \[ \sum_{i=1}^{d} \frac{1}{n} \left( (H_N \boldsymbol{\hat{e}}_i)_X \right)^\top \left( (H_N \boldsymbol{\hat{e}}_i)_X \right) \xrightarrow{p} \sum_{i=1}^{d} \operatorname{Var}(e_i(Z)), \]
    and
        \[ \sum_{i=1}^{d} \frac{1}{m} \left( (H_N \boldsymbol{\hat{e}}_i)_Y \right)^\top \left( (H_N \boldsymbol{\hat{e}}_i)_Y \right) \xrightarrow{p} \sum_{i=1}^{d} \operatorname{Var}(e_i(Z)). \]
\end{lemma}

The proof is provided in Appendix~\ref{app:other_proofs}.

\subsection{Spectral Gap and Truncation Dimension}
\label{app:spectral_gap}
Lemma~\ref{bound_projections} provides a general relation between sample size and the spectral gap (and hence the truncation dimension). In this section, we restrict our attention to specific kernels to derive a more precise relationship between the sample size and the truncation dimension.

Let $k(s,t)$ be a smooth radial kernel defined on $\mathbb{R}^q$ with $q$ fixed. Specifically, let $k(s,t) = h(\|s-t\|)$ and define $g(\cdot) = h(\sqrt{\cdot})$. Assume that $|g^{(l)}(r)| \leq l! M^l$ for all $l \geq 0$ and $r>0$, where $g^{(l)}$ denotes the $l$-th order derivative of $g$, and $M$ is a positive constant. Two prominent kernels satisfying this condition are the Gaussian kernel and the inverse multiquadric kernel. The following lemma characterizes the eigenvalue decay rate of the integral operator $L_k$ (and $\Sigma$) associated with such kernels.

\begin{lemma}[Theorem 5 in \cite{belkin2018approximation}]
\label{lemma:eigen_decay}
Let $\bar{k} = \sup_{z \in \mathcal{Z} \subset \mathbb{R}^q}k(z,z)$, and let $k$ be a kernel satisfying the conditions mentioned above. Then for some constants $C, C'>0$, we have 
 $$ 
    \lambda_d \leq \sqrt{\bar{k}} C \exp(-C' d^{1/q}), \quad \text{for all } d \geq 1.
 $$ 
\end{lemma}

\begin{remark}
    Note that all quantities in Lemma~\ref{lemma:eigen_decay} are independent of the probability measure $\mathbb{P}$. In particular, when $\mathbb{P}$ is a discrete measure, the integral operator $L_k$ can be represented as a finite-dimensional matrix. Thus, this result provides a uniform bound on the eigenvalue decay rate that is independent of the matrix size (i.e., the sample size). 
\end{remark}

Assuming the eigenvalues are simple, the spectral gap is given by $\eta_d = \frac{1}{2} (\lambda_d - \lambda_{d+1})$. Because $\exp(-C' d^{1/q})$ is convex and decreasing in $d$, the spectral condition in Theorem~\ref{th:perturbation_space} ($\|\delta A\|_{\text{op}} < \eta_d$) is satisfied provided\footnote{This follows from approximating the spectral gap via the derivative of the eigenvalue bound: $\eta_d \approx \frac{1}{2} |\frac{d}{dd} \lambda_d| > O_p(N^{-1/2})$.}  
\[ C' d^{1/q} + \frac{q-1}{q} \log d < \frac{1}{2} \log N - \log q + M_0, \]
where $M_0$ is a constant depending on $C, C'$, and $\bar{k}$. Solving explicitly for $d$ in terms of $N$ involves a transcendental equation. However, intuition can be gained by considering the univariate case $q=1$, where the condition simplifies to
 $$ 
    d < \frac{1}{2C'} \log N + M_0.
 $$ 
This implies that even when $q=1$, the required sample size $N$ must grow exponentially with the truncation dimension $d$, which is practically demanding. In our simulation studies, the total sample size is $N=200$; thus, setting the truncation dimension to $d=1$ for the \texttt{TMMD(d=1)} test statistic and $\bar{d}=5$ for the \texttt{TMMD-Oracle} test statistic are reasonable choices that satisfy the spectral condition with high probability.

\begin{remark}
    One can also observe the ``curse of dimensionality'' through the lens of eigenvalue decay. As the ambient dimension $q$ increases, the eigenvalues $\lambda_d$ tend to decay more slowly, resulting in a smaller spectral gap $\eta_d$. Consequently, it becomes increasingly challenging to satisfy the spectral condition, even for small truncation dimensions $d$.
\end{remark}

\subsection{Matrix Representation of Operators}
\label{app:matrix_representation}
To bridge the gap between abstract operators on Hilbert spaces and finite-dimensional matrices, we rely on the concept of a matrix representation.

\begin{definition}[Matrix Representation]
    Let $\mathcal{V}$ be a $d$-dimensional subspace of a Hilbert space $\mathcal{H}$, equipped with an orthonormal basis $\mathcal{B} = \{v_1, \dots, v_d\}$. Let $A : \mathcal{V} \to \mathcal{V}$ be a linear operator defined on this subspace. The \textit{matrix representation} of $A$ with respect to the basis $\mathcal{B}$, denoted $[A]_{\mathcal{B}} \in \mathbb{R}^{d \times d}$, is defined as the unique matrix whose $(i,j)$-th entry is given by:
    \[ \left( [A]_{\mathcal{B}} \right)_{ij} = \langle v_i, A v_j \rangle_{\mathcal{H}}, \quad i, j = 1, \dots, d. \]
\end{definition}

The mapping $A \mapsto [A]_{\mathcal{B}}$ is an algebra isomorphism. Specifically, it preserves linear combinations and operator compositions: for any operators $A_1, A_2$ on $\mathcal{V}$ and scalars $c_1, c_2$, we have $[c_1 A_1 + c_2 A_2]_{\mathcal{B}} = c_1 [A_1]_{\mathcal{B}} + c_2 [A_2]_{\mathcal{B}}$, and $[A_1 A_2]_{\mathcal{B}} = [A_1]_{\mathcal{B}} [A_2]_{\mathcal{B}}$.

The critical utility of the matrix representation is that it preserves the spectral structure of the operator.

\begin{theorem}
    \label{th:spectral_equivalence}
    Let $A : \mathcal{V} \to \mathcal{V}$ be a linear operator and $[A]_{\mathcal{B}}$ be its matrix representation in some orthonormal basis $\mathcal{B}$. The eigenvalues of the operator $A$ are exactly the eigenvalues of the matrix $[A]_{\mathcal{B}}$.
\end{theorem}

The proof is provided in Appendix~\ref{app:other_proofs}.

In the context of our perturbation analysis, we apply this framework to the subspace $\mathcal{V} = \text{span}\{\phi_1(\theta), \dots, \phi_d(\theta)\}$ equipped with the orthonormal basis $\mathcal{B} = \{\phi_i(\theta)\}_{i=1}^d$. 

When we define the matrix $M(\theta)$ as the matrix representation of the restricted operator $P_d(\theta) \boldsymbol{C}_{\theta} P_d(\theta)$, we are explicitly setting:
\[ M_{ij}(\theta) = \left( [P_d(\theta) \boldsymbol{C}_{\theta} P_d(\theta)]_{\mathcal{B}} \right)_{ij} = \langle \phi_i(\theta), P_d(\theta) \boldsymbol{C}_{\theta} P_d(\theta) \phi_j(\theta) \rangle_{\rho}. \]
Similarly, because the basis vectors $\phi_i(\theta)$ are eigenvectors of the operator $\Lambda_d^{-1/2}(\theta)$, its matrix representation in $\mathcal{B}$ is simply the diagonal matrix $[\Lambda_d^{-1/2}(\theta)]_{\mathcal{B}}=\text{diag}(\lambda_1^{-1/2}, \dots, \lambda_d^{-1/2})$. 

By the algebra isomorphism property of matrix representations, the matrix representation of the composite operator $T_\theta = \Lambda_d^{-1/2}(\theta) P_d(\theta) \boldsymbol{C}_{\theta} P_d(\theta) \Lambda_d^{-1/2}(\theta)$ is precisely the product of the individual matrix representations:
\[ [T_\theta]_{\mathcal{B}} = [\Lambda_d^{-1/2}(\theta)]_{\mathcal{B}} [P_d(\theta) \boldsymbol{C}_{\theta} P_d(\theta)]_{\mathcal{B}} [\Lambda_d^{-1/2}(\theta)]_{\mathcal{B}} = [\Lambda_d^{-1/2}(\theta)]_{\mathcal{B}} M(\theta) [\Lambda_d^{-1/2}(\theta)]_{\mathcal{B}}. \]
From Proposition~\ref{prop: l_2_covariance}, we algebraically derived that $[\Lambda_d^{-1/2}(\theta)]_{\mathcal{B}} M(\theta) [\Lambda_d^{-1/2}(\theta)]_{\mathcal{B}} = \Sigma_d(\theta)$. Therefore, $\Sigma_d(\theta)$ is exactly the matrix representation of $T_\theta$. By the Spectral Equivalence theorem, the eigenvalues of the operator $T_\theta$ are precisely the eigenvalues of $\Sigma_d(\theta)$, which are the asymptotic variance weights $\sigma_i(\theta)$.

\section{Other Proofs}
\label{app:other_proofs}

\subsection{Proof of Lemma \ref{lm:datadriven}}

Let $\mathcal{D} = \{1, 2, \dots, \bar{d}\}$ denote the finite set of candidate truncation levels. For any fixed $d \in \mathcal{D}$, define the unconditional rejection probability using dimension $d$ on the test set as 
\begin{equation*}
    p_N(d) := \mathbb{P}_{H_0}\bigl( \tilde{D}_N^d > \hat{c}_d^{(\text{test})}(1-\alpha) \bigr),
\end{equation*}
where $\hat{c}_d^{(\text{test})}(1-\alpha)$ is the $1-\alpha$ quantile of the parametric bootstrap distribution computed from $\mathcal{D}_{\text{test}}$. Note that the event $\{P_{\text{final}} \le \alpha\}$ is equivalent to the event $\{\tilde{D}_N^d \ge \hat{c}_d^{(\text{test})}(1-\alpha)\}$. Because the parametric bootstrap distribution is a continuous weighted sum of $\chi^2_1$ random variables (as the estimated covariance matrix $\hat{\Sigma}_d$ is almost surely positive definite), the probability of exact equality is asymptotically negligible, and thus we may work with the strict inequality without loss of generality.

Under the null hypothesis, Theorem~\ref{null_distribution_test_statistic} establishes that $\tilde{D}_N^d$ converges in distribution to a continuous weighted sum of centered $\chi^2_1$ random variables. Simultaneously, the consistency of the empirical covariance matrix $\hat{\Sigma}_d$ restricted to the first $d$ components ensures that the bootstrap critical value $\hat{c}_d^{(\text{test})}(1-\alpha)$ converges in probability to the true asymptotic $(1-\alpha)$-quantile. By Slutsky's lemma, it follows that 
\begin{equation*}
    \lim_{N \to \infty} p_N(d) = \alpha
\end{equation*}
for each fixed $d \in \mathcal{D}$. Because the candidate set $\mathcal{D}$ is finite and does not grow with the sample size, this pointwise convergence implies uniform convergence over the set of candidate dimensions:
\begin{equation*}
    \max_{d \in \mathcal{D}} \bigl| p_N(d) - \alpha \bigr| \xrightarrow{N \to \infty} 0.
\end{equation*}

Now, consider the data-driven selection $\hat{d}$. Let $p_{N'}(d) := \mathbb{P}(\hat{d} = d)$ denote the marginal probability that the selection rule chooses dimension $d$, which may depend on the training sample size $N'$. By the law of total probability, the unconditional rejection probability of the final test is
\begin{equation*}
    \mathbb{P}_{H_0}\bigl( P_{\text{final}} \le \alpha \bigr) 
    = \sum_{d \in \mathcal{D}} \mathbb{P}_{H_0}\bigl( \tilde{D}_N^d > \hat{c}_d^{(\text{test})}(1-\alpha) \;\big|\; \hat{d} = d \bigr) \, p_{N'}(d).
\end{equation*}

Crucially, by assumption, $\hat{d}$ is a measurable function solely of the training data $\mathcal{D}_{\text{train}}$, whereas the test statistic $\tilde{D}_N^d$ and its bootstrap critical value $\hat{c}_d^{(\text{test})}(1-\alpha)$ are computed entirely from the independent test data $\mathcal{D}_{\text{test}}$. Therefore, the event $\{\hat{d} = d\}$ is statistically independent of the event $\{\tilde{D}_N^d > \hat{c}_d^{(\text{test})}(1-\alpha)\}$. It follows that the conditional probability reduces to the unconditional one:
\begin{equation*}
    \mathbb{P}_{H_0}\bigl( \tilde{D}_N^d > \hat{c}_d^{(\text{test})}(1-\alpha) \;\big|\; \hat{d} = d \bigr) = \mathbb{P}_{H_0}\bigl( \tilde{D}_N^d > \hat{c}_d^{(\text{test})}(1-\alpha) \bigr) = p_N(d).
\end{equation*}

Substituting this independence result back into the summation yields
\begin{equation*}
    \mathbb{P}_{H_0}\bigl( P_{\text{final}} \le \alpha \bigr) = \sum_{d \in \mathcal{D}} p_N(d) \, p_{N'}(d).
\end{equation*}

To establish the limit, fix an arbitrary $\epsilon > 0$. By the uniform convergence of $p_N(d)$, there exists $N_0 \in \mathbb{N}$ such that for all $N \ge N_0$, we have the uniform bound $\alpha - \epsilon \le p_N(d) \le \alpha + \epsilon$ for all $d \in \mathcal{D}$. Applying these bounds to the summation, and recognizing that $\sum_{d \in \mathcal{D}} p_{N'}(d) = 1$, we obtain
\begin{equation*}
    (\alpha - \epsilon) \sum_{d \in \mathcal{D}} p_{N'}(d) \le \sum_{d \in \mathcal{D}} p_N(d) \, p_{N'}(d) \le (\alpha + \epsilon) \sum_{d \in \mathcal{D}} p_{N'}(d),
\end{equation*}
which simplifies to $\alpha - \epsilon \le \mathbb{P}_{H_0}(P_{\text{final}} \le \alpha) \le \alpha + \epsilon$. Because $\epsilon > 0$ was arbitrary, taking the limit as $N', N \to \infty$ yields
\begin{equation*}
    \lim_{N', N \to \infty} \mathbb{P}_{H_0}\bigl( P_{\text{final}} \le \alpha \bigr) = \alpha.
\end{equation*}
This completes the proof.

\subsection{Proof of Lemma~\ref{lm:d_consistency}}
Under the fixed alternative hypothesis, by Lemma~\ref{residual} and Theorem~\ref{th:fixed-alternative}, we have:
\[ \frac{1}{N'} D_{N'}^d = \sum_{i=1}^d d_i^2 +o_p(1). \]
The unbiased statistic $ \tilde{D}_{N'}^d $ differs from $ D_{N'}^d $ by trace correction terms that are $ O_p(1) $. Hence 
\[ \frac{1}{N'} \tilde{D}_{N'}^d = \sum_{i=1}^d d_i^2 +o_p(1). \] 
Since $ \bar{d} $ is finite, the convergence is uniform over $ \mathcal{D}=\{1,2,\ldots,\bar{d}\} $. Furthermore, it is easy to check that uniformly in $ d \in \mathcal{D} $, 
\[ \operatorname{tr}(\hat{\Gamma}_d) \xrightarrow{p} \operatorname{tr}(\Gamma_d) >0. \] 
Thus, uniformly in $ d \in \mathcal{D} $, we have
\[ Q_{N'}(d) \xrightarrow{p} \frac{\sum_{i=1}^2 d_i^2}{\operatorname{tr}(\Gamma_d)}, \]
which implies that 
\[ \max_{d \in \mathcal{D}} |Q_{N'}(d)-Q(d)| \xrightarrow{p}0. \]
Since $ d^* $ is the unique maximizer of $ Q(d) $, there is a positive gap between the maximum value and the second largest value:
\[ \Delta = \min_{d\in \mathcal{D}, d\neq d^*}(Q(d) - Q(d^*))>0. \] 
For every $ d \neq d^* $, one must have $ Q(d) \le Q(d^*) - \Delta $. 

Next, define an event $ E_{N'}=\{\max_{d\in \mathcal{D}} |Q_{N'}(d)-Q(d)| < \Delta/2\} $. By the uniform convergence, we have $ \mathbb{P}(E_{N'}) \to 1 $. On the event $ E_{N'} $, for every $ d \neq d^* $, we have
\[ |Q_{N'}(d)-Q(d)| < \Delta/2 \implies Q_{N'}(d) < Q(d) + \Delta/2 \le (Q(d^*) - \Delta) + \Delta/2 = Q(d^*) - \Delta/2. \]  
On the other hand, for $ d = d^* $, on the event $ E_{N'} $, we have
\[ |Q_{N'}(d^*)-Q(d^*)| < \Delta/2 \implies Q_{N'}(d^*) > Q(d^*) - \Delta/2. \]
Combining the above inequalities, we conclude that on the event $ E_{N'} $, for every $ d \neq d^* $, we have 
\[ Q_{N'}(d^*) > Q(d^*) - \Delta/2 >  Q_{N'}(d).\]
Thus, $ d^* $ is the unique maximizer of $ Q_{N'}(d) $ on the event $ E_{N'} $. 

Finally, note that $ \mathbb{P}(\hat{d} = d^*) \ge \mathbb{P}(E_{N'}) \to 1 $. Hence, $ \hat{d} $ is a consistent estimator of $ d^* $.

\subsection{Proof of Theorem \ref{th:multi_kernel_null}}
The proof extends the single-kernel methodology by mapping the problem into a product RKHS, which naturally captures the dependencies induced by evaluating multiple kernels on the identical pooled sample.

Define the product RKHS $\mathcal{H}_{\mathcal{C}} = \bigoplus_{c \in \mathcal{C}} \mathcal{H}_c$, equipped with the standard inner product. The joint feature map is given by $\Phi_{\mathcal{C}}(z) = (k_{c_1}(z, \cdot), k_{c_2}(z, \cdot), \dots)^\top$. Under $H_0$, the joint mean embedding is $\mu_{\mathcal{C}} = \mathbb{E}_\rho[\Phi_{\mathcal{C}}(Z)]$, and the empirical joint mean difference for the two samples is $\hat{\mu}_{\mathcal{C}} - \hat{\nu}_{\mathcal{C}}$. By the Functional Central Limit Theorem in the product space, we have
\[ \sqrt{N} \left( (\hat{\mu}_{\mathcal{C}} - \mu_{\mathcal{C}}) - (\hat{\nu}_{\mathcal{C}} - \mu_{\mathcal{C}}) \right) \xrightarrow{d} \xi_{\mathcal{C}}, \]
where $\xi_{\mathcal{C}}$ is a centered Gaussian random element in $\mathcal{H}_{\mathcal{C}}$ whose covariance operator corresponds to the matrix $\Sigma_{\mathcal{C}}$.

To handle the estimation error of the empirical eigenfunctions, we invoke the rotation argument from Theorem \ref{null_distribution_multiple}. For each kernel $c$, there exist random orthogonal matrices $O_{N,c}$ such that the rotated empirical eigenfunctions $\tilde{e}_{i,c} = \sum_{j=1}^{d_c} (O_{N,c})_{ji} \hat{e}_{j,c}$ satisfy $\|\tilde{e}_{i,c} - e_{i,c}\|_{\mathcal{H}_c} = O_p(N^{-1/2})$. Define the vector of joint directional components:
\[ \mathbf{Z}_N = \left( \langle \hat{\mu}_P - \hat{\mu}_Q, \tilde{e}_{1,c_1} \rangle_{\mathcal{H}_{c_1}}, \dots, \langle \hat{\mu}_P - \hat{\mu}_Q, \tilde{e}_{d_{c_1},c_1} \rangle_{\mathcal{H}_{c_1}}, \langle \hat{\mu}_P - \hat{\mu}_Q, \tilde{e}_{1,c_2} \rangle_{\mathcal{H}_{c_2}}, \dots \right)^\top \in \mathbb{R}^{D_{\mathcal{C}}}. \]
Using the same perturbation arguments as in Lemma \ref{residual} and Theorem \ref{null_distribution_multiple}, replacing $\tilde{e}_{i,c}$ with $e_{i,c}$ introduces an $o_p(1)$ error in the scaled inner products. By the continuous mapping theorem applied to the joint CLT, we establish the joint asymptotic normality:
\[ \sqrt{N} \mathbf{Z}_N \xrightarrow{d} \mathcal{N}\left( \mathbf{0}, \frac{1}{p(1-p)} \Sigma_{\mathcal{C}} \right). \]
The off-diagonal blocks of $\Sigma_{\mathcal{C}}$ correctly capture the asymptotic covariance between the directional components of different kernels, which arises from the shared sample points.

By the orthogonal invariance of the norm, the sum of the biased test statistics equals the squared Euclidean norm of the joint vector:
\[ \sum_{c \in \mathcal{C}} D_{N,c}^{d_c} = N \sum_{c \in \mathcal{C}} \sum_{i=1}^{d_c} \langle \hat{\mu}_P - \hat{\mu}_Q, \hat{e}_{i,c} \rangle_{\mathcal{H}_c}^2 = N \|\mathbf{Z}_N\|_2^2. \]
Applying the continuous mapping theorem to the function $x \mapsto x^\top x$, the biased sum converges to:
\[ \sum_{c \in \mathcal{C}} D_{N,c}^{d_c} \xrightarrow{d} \frac{1}{p(1-p)} \mathbf{W}^\top \mathbf{W}, \quad \text{where } \mathbf{W} \sim \mathcal{N}(\mathbf{0}, \Sigma_{\mathcal{C}}). \]
Because $\Sigma_{\mathcal{C}}$ is symmetric and positive semi-definite, it admits the spectral decomposition $\Sigma_{\mathcal{C}} = U \Lambda U^\top$, where $\Lambda = \operatorname{diag}(\omega_1, \dots, \omega_{D_{\mathcal{C}}})$. Thus, $\mathbf{W}^\top \mathbf{W} \overset{d}{=} \sum_{k=1}^{D_{\mathcal{C}}} \omega_k \chi_{1,k}^2$.

For the unbiased statistic $G_N = \sum_{c \in \mathcal{C}} \tilde{D}_{N,c}^{d_c}$, we must subtract the trace correction terms for each kernel. By Lemma \ref{lm:trace_convergence} and Theorem \ref{null_distribution_test_statistic}, the trace term for kernel $c$ converges in probability to $\frac{1}{p(1-p)} \operatorname{Tr}(\Sigma_{\mathcal{C}}^{(c,c)})$. Summing over all kernels yields:
\[ \sum_{c \in \mathcal{C}} \frac{1}{p(1-p)} \operatorname{Tr}(\Sigma_{\mathcal{C}}^{(c,c)}) = \frac{1}{p(1-p)} \operatorname{Tr}(\Sigma_{\mathcal{C}}) = \frac{1}{p(1-p)} \sum_{k=1}^{D_{\mathcal{C}}} \omega_k. \]
Subtracting this asymptotic expectation of the biased statistics via Slutsky's theorem, we obtain the limiting distribution of the aggregated unbiased statistic:
\[ G_N \xrightarrow{d} \frac{1}{p(1-p)} \left( \sum_{k=1}^{D_{\mathcal{C}}} \omega_k \chi_{1,k}^2 - \sum_{k=1}^{D_{\mathcal{C}}} \omega_k \right) = \sum_{k=1}^{D_{\mathcal{C}}} \frac{\omega_k}{p(1-p)} (\chi_{1,k}^2 - 1). \]
This completes the proof.

\subsection{Proof of Lemma \ref{lm:trace_convergence}}
We present the detailed proof for the $X$-sample; the convergence for the $Y$-sample follows by an entirely analogous argument. 

Recall that $H_N = I_N - \frac{1}{N} \mathbf{1}_N \mathbf{1}_N^\top$ denotes the centering matrix. For each $i \in \{1, \dots, d\}$, let $\boldsymbol{\hat{e}}_i = (\hat{e}_i(z_1), \dots, \hat{e}_i(z_N))^\top \in \mathbb{R}^N$ be the vector of empirical eigenfunction evaluations over the pooled sample. The subscript $X$ denotes the operation of extracting the first $n$ entries of a vector, corresponding to the $X$-sample. Thus, the $\alpha$-th component of the centered vector $(H_N \boldsymbol{\hat{e}}_i)_X$ is explicitly given by:
\begin{equation*}
\left[ \boldsymbol{\hat{e}}_i - \frac{1}{N} \mathbf{1}_N \mathbf{1}_N^\top \boldsymbol{\hat{e}}_i \right]_\alpha = \hat{e}_i(x_\alpha) - \bar{e}_i,
\end{equation*}
where $\bar{e}_i = \frac{1}{N} \sum_{\gamma=1}^N \hat{e}_i(z_\gamma)$ is the empirical mean of the $i$-th eigenfunction over the entire pooled sample. Furthermore, define the $X$-sample specific empirical mean as $\hat{\mu}_{i,X} = \frac{1}{n} \sum_{\alpha=1}^n \hat{e}_i(x_\alpha)$.

To establish the convergence of the centered quadratic form, we employ a standard sample variance decomposition. For any fixed $i$, adding and subtracting $\hat{\mu}_{i,X}$ yields:
\begin{align*}
\sum_{\alpha=1}^n (\hat{e}_i(x_\alpha) - \bar{e}_i)^2 &= \sum_{\alpha=1}^n \left( (\hat{e}_i(x_\alpha) - \hat{\mu}_{i,X}) + (\hat{\mu}_{i,X} - \bar{e}_i) \right)^2 \\
&= \sum_{\alpha=1}^n (\hat{e}_i(x_\alpha) - \hat{\mu}_{i,X})^2 + n(\hat{\mu}_{i,X} - \bar{e}_i)^2,
\end{align*}
where the cross-term vanishes due to the fact that $\sum_{\alpha=1}^n (\hat{e}_i(x_\alpha) - \hat{\mu}_{i,X}) = 0$. Dividing by $n$ and summing over $i = 1, \dots, d$, our target expression cleanly decomposes into two tractable terms:
\[ 
\sum_{i=1}^d \frac{1}{n} \left( (H_N \boldsymbol{\hat{e}}_i)_X \right)^\top \left( (H_N \boldsymbol{\hat{e}}_i)_X \right) = \underbrace{\sum_{i=1}^d \frac{1}{n} \sum_{\alpha=1}^n (\hat{e}_i(x_\alpha) - \hat{\mu}_{i,X})^2}_{A_N} + \underbrace{\sum_{i=1}^d (\hat{\mu}_{i,X} - \bar{e}_i)^2}_{B_N}. \]
We now analyze the asymptotic behavior of $A_N$ and $B_N$ separately.

\textbf{Convergence of $A_N$.} 
Expanding the square within $A_N$, we obtain $A_N = \sum_{i=1}^d \frac{1}{n} \sum_{\alpha=1}^n \hat{e}_i(x_\alpha)^2 - \sum_{i=1}^d \hat{\mu}_{i,X}^2$. We evaluate the limits of these two sub-components. 

First, consider $\sum_{i=1}^d \frac{1}{n} \sum_{\alpha=1}^n \hat{e}_i(x_\alpha)^2$. Because orthogonal transformations preserve the Euclidean norm, if we let $\tilde{e}_i = \sum_{j=1}^d (O_N)_{ij} \hat{e}_j$ denote the rotated empirical eigenfunctions (which satisfy $\|\tilde{e}_i - e_i\|_H = O_p(N^{-1/2})$), we have:
\begin{equation*}
\sum_{i=1}^d \hat{e}_i(x_\alpha)^2 = \sum_{i=1}^d \tilde{e}_i(x_\alpha)^2 = \|\hat{P}_d k(x_\alpha, \cdot)\|_H^2,
\end{equation*}
where $\hat{P}_d = \sum_{i=1}^d \hat{e}_i \otimes_H \hat{e}_i$ is the empirical projection operator. Similarly, the population counterpart is $\sum_{i=1}^d e_i(x_\alpha)^2 = \|P_d k(x_\alpha, \cdot)\|_H^2$.

To bound the difference between the empirical and population averages, we utilize the algebraic identity for orthogonal projections: $\|Qf\|^2 - \|Pf\|^2 = \langle (Q-P)f, (Q+P)f \rangle$. Applying the Cauchy-Schwarz inequality, the boundedness of the kernel $\|k(x_\alpha, \cdot)\|_H \le \sqrt{\bar{k}}$, and the fact that $\|\hat{P}_d + P_d\|_{op} \le 2$, we obtain:
\begin{equation*}
\left| \|\hat{P}_d k(x_\alpha, \cdot)\|_H^2 - \|P_d k(x_\alpha, \cdot)\|_H^2 \right| \le 2\bar{k} \|\hat{P}_d - P_d\|_{op}.
\end{equation*}
Averaging this bound over the $X$-sample gives:
\begin{equation*}
\left| \frac{1}{n} \sum_{\alpha=1}^n \sum_{i=1}^d \hat{e}_i(x_\alpha)^2 - \frac{1}{n} \sum_{\alpha=1}^n \sum_{i=1}^d e_i(x_\alpha)^2 \right| \le 2\bar{k} \|\hat{P}_d - P_d\|_{op}.
\end{equation*}
By Lemma \ref{bound_projections}, the Hilbert-Schmidt norm converges as $\|\hat{P}_d - P_d\|_{HS} = O_p(N^{-1/2})$. Because the operator norm is upper-bounded by the Hilbert-Schmidt norm, it follows that $\|\hat{P}_d - P_d\|_{op} \xrightarrow{p} 0$. By the standard Law of Large Numbers (where the moment $\mathbb{E}[e_i(X)^2]$ is finite due to the kernel boundedness), the population average converges to its expectation:
\begin{equation*}
\frac{1}{n} \sum_{\alpha=1}^n \|P_d k(x_\alpha, \cdot)\|_H^2 = \frac{1}{n} \sum_{\alpha=1}^n \sum_{i=1}^d e_i(x_\alpha)^2 \xrightarrow{p} \sum_{i=1}^d \mathbb{E}[e_i(X)^2].
\end{equation*}
An application of Slutsky's theorem thus yields:
\begin{equation}
\label{eq:step1_raw_moment}
\sum_{i=1}^d \frac{1}{n} \sum_{\alpha=1}^n \hat{e}_i(x_\alpha)^2 \xrightarrow{p} \sum_{i=1}^d \mathbb{E}[e_i(X)^2].
\end{equation}

Next, we analyze $\sum_{i=1}^d \hat{\mu}_{i,X}^2$. By orthogonal invariance, this equals $\sum_{i=1}^d \tilde{\mu}_{i,X}^2$, where $\tilde{\mu}_{i,X} = \frac{1}{n} \sum_{\alpha=1}^n \tilde{e}_i(x_\alpha)$. Define the $X$-sample mean of the population eigenfunctions as $\bar{e}_{i,X} = \frac{1}{n} \sum_{\alpha=1}^n e_i(x_\alpha)$. Using the reproducing property of the RKHS, the point-wise difference is bounded by:
\begin{equation*}
|\tilde{e}_i(x_\alpha) - e_i(x_\alpha)| = |\langle k(x_\alpha, \cdot), \tilde{e}_i - e_i \rangle_H| \le \sqrt{\bar{k}} \|\tilde{e}_i - e_i\|_H = O_p(N^{-1/2}).
\end{equation*}
Averaging over the $n$ samples implies $\tilde{\mu}_{i,X} - \bar{e}_{i,X} \xrightarrow{p} 0$. By the LLN, $\bar{e}_{i,X} \xrightarrow{p} \mathbb{E}[e_i(X)]$, and hence $\tilde{\mu}_{i,X} \xrightarrow{p} \mathbb{E}[e_i(X)]$. Because $d$ is fixed, the Continuous Mapping Theorem ensures:
\begin{equation}
\label{eq:step1_mean_squared}
\sum_{i=1}^d \hat{\mu}_{i,X}^2 = \sum_{i=1}^d \tilde{\mu}_{i,X}^2 \xrightarrow{p} \sum_{i=1}^d \mathbb{E}[e_i(X)]^2.
\end{equation}
Combining Equations \eqref{eq:step1_raw_moment} and \eqref{eq:step1_mean_squared} via Slutsky's theorem, we conclude that:
\[ A_N \xrightarrow{p} \sum_{i=1}^d \mathbb{E}[e_i(X)^2] - \sum_{i=1}^d \mathbb{E}[e_i(X)]^2 = \sum_{i=1}^d \operatorname{Var}(e_i(X)). \]

\textbf{Convergence of $B_N$.} 
Recall that $B_N = \sum_{i=1}^d (\hat{\mu}_{i,X} - \bar{e}_i)^2$. Exploiting orthogonal invariance once more, we equivalently write this as $\sum_{i=1}^d (\tilde{\mu}_{i,X} - \tilde{\bar{e}}_i)^2$, where $\tilde{\bar{e}}_i = \frac{1}{N} \sum_{\gamma=1}^N \tilde{e}_i(z_\gamma)$ is the pooled sample mean of the rotated eigenfunctions. 
Decomposing $\tilde{\bar{e}}_i$ into its constituent $X$ and $Y$ components yields:
\begin{equation*}
\tilde{\bar{e}}_i = \frac{n}{N} \tilde{\mu}_{i,X} + \frac{m}{N} \tilde{\mu}_{i,Y},
\end{equation*}
where $\tilde{\mu}_{i,Y} = \frac{1}{m} \sum_{\beta=1}^m \tilde{e}_i(y_\beta)$. The difference between the $X$-specific mean and the pooled mean is therefore:
\begin{equation*}
\tilde{\mu}_{i,X} - \tilde{\bar{e}}_i = \tilde{\mu}_{i,X} - \left( \frac{n}{N} \tilde{\mu}_{i,X} + \frac{m}{N} \tilde{\mu}_{i,Y} \right) = \frac{m}{N} (\tilde{\mu}_{i,X} - \tilde{\mu}_{i,Y}).
\end{equation*}
Under the null hypothesis $H_0 : P = Q = \rho$, the random variables $X$ and $Y$ are identically distributed. Applying the identical point-wise bounding argument used before to the $Y$-sample establishes that $\tilde{\mu}_{i,Y} \xrightarrow{p} \mathbb{E}[e_i(Y)] = \mathbb{E}[e_i(X)]$. Consequently, the difference of the sample means converges to zero in probability:
\begin{equation*}
\tilde{\mu}_{i,X} - \tilde{\mu}_{i,Y} \xrightarrow{p} \mathbb{E}[e_i(X)] - \mathbb{E}[e_i(Y)] = 0.
\end{equation*}
Since the deterministic scaling factor $m/N$ is bounded by $1$, it follows that $\tilde{\mu}_{i,X} - \tilde{\bar{e}}_i \xrightarrow{p} 0$. A final application of the Continuous Mapping Theorem to the finite sum provides:
\[ B_N = \sum_{i=1}^d (\tilde{\mu}_{i,X} - \tilde{\bar{e}}_i)^2 \xrightarrow{p} \sum_{i=1}^d 0^2 = 0. \]

\subsection{Proof of Theorem~\ref{th:spectral_equivalence}}
Suppose $\lambda$ is an eigenvalue of $A$ with corresponding non-zero eigenvector $f \in \mathcal{V}$. Since $\mathcal{B}$ spans $\mathcal{V}$, we can write $f = \sum_{k=1}^d x_k v_k$ for some coefficient vector $\mathbf{x} = (x_1, \dots, x_d)^\top \neq \mathbf{0}$. The eigenvalue equation $A f = \lambda f$ becomes:
    \[ A \left( \sum_{j=1}^d x_j v_j \right) = \lambda \sum_{i=1}^d x_i v_i \implies \sum_{j=1}^d x_j (A v_j) = \sum_{j=1}^d \lambda x_j v_j. \]
    Taking the inner product of both sides with $v_i$ and using the orthonormality of $\mathcal{B}$ ($\langle v_i, v_j \rangle = \delta_{ij}$), we isolate the $i$-th component:
    \[ \sum_{j=1}^d x_j \langle v_i, A v_j \rangle = \lambda x_i. \]
    By definition, $\langle v_i, A v_j \rangle = ([A]_{\mathcal{B}})_{ij}$. Therefore, the equation is equivalent to the matrix-vector equation:
    \[ [A]_{\mathcal{B}} \mathbf{x} = \lambda \mathbf{x}. \]
    Since $\mathbf{x} \neq \mathbf{0}$, this demonstrates that $\lambda$ is an eigenvalue of the matrix $[A]_{\mathcal{B}}$ with eigenvector $\mathbf{x}$. The converse follows by reversing the steps.

\section{Comparison with Spectral Regularization Approaches}
\label{app:comparison_hagrass}

A shared motivation in recent kernel two-sample testing literature is the sub-optimality of the standard Maximum Mean Discrepancy (MMD) test. Because the MMD aggregates squared norms across all RKHS directions, it inevitably includes trailing directional components that are poorly estimated in finite samples, thereby degrading test power. Both our proposed method and the spectral regularization framework of \cite{hagrass2024spectral} address this limitation by leveraging the spectral decomposition of the kernel integral operator to emphasize well-estimated leading eigen-directions. However, the two methodologies diverge fundamentally in their modification strategies, theoretical objectives, and computational implementations.

The primary methodological distinction lies in the mechanism of spectral modification. \cite{hagrass2024spectral} propose a \emph{spectral regularization} framework that modifies the MMD via a regularizer $g_\lambda(x)$ (e.g., Tikhonov or Showalter regularization). This approach smoothly down-weights the contribution of smaller eigenvalues and explicitly incorporates the pooled covariance operator, yielding a statistic analogous to a kernelized Hotelling's $T^2$. In contrast, our method employs \emph{spectral truncation}, strictly retaining only the leading $d$ directional components and discarding the remainder. While regularization preserves contributions from all components with smooth decay, our truncation approach enforces a hard bias-variance trade-off: it entirely eliminates the noise floor associated with the spectral tail, directly optimizing the finite-sample signal-to-noise ratio (SNR) when the distributional difference is concentrated in the leading eigen-directions.

These distinct strategies necessitate different theoretical lenses. The spectral regularization framework focuses on \emph{minimax optimality} and separation boundaries in Hellinger distance. \cite{hagrass2024spectral} rigorously demonstrate that the standard MMD test is not minimax optimal and prove that their regularized test achieves the minimax optimal separation rate over specific smoothness classes defined by the range of the integral operator. Conversely, our theoretical analysis centers on the \emph{finite-sample power} and the precise asymptotic distributions of the truncated directional components. By characterizing the null distribution as a weighted sum of independent chi-squared variables and analyzing power under both fixed and local alternatives, we explicitly quantify the variance reduction achieved by discarding the noisy trailing components. This distinction highlights a trade-off: regularization provides optimal detection rates under smoothness constraints, whereas truncation maximizes finite-sample power by aggressively discarding high-variance dimensions.

Furthermore, the approaches differ significantly in the determination of critical values and computational scaling. Because the asymptotic null distribution of the regularized statistic is complex, \cite{hagrass2024spectral} rely on permutation tests to control the Type-I error. While permutation tests are distribution-free, they incur a substantial computational burden. Our truncation approach circumvents this by deriving the explicit asymptotic null distribution, which facilitates a computationally efficient \emph{parametric bootstrap} procedure. Additionally, because our statistic requires only the top $d$ eigen-pairs of the Gram matrix, the computational complexity reduces from $O(N^3)$ to $O(N^2 d)$.

Finally, while spectral regularization employs aggregation over regularization parameters and kernels to achieve adaptivity, our truncation framework offers inherent robustness to the choice of kernel hyperparameters. Through perturbation analysis, we establish that the population-level directional contributions and the finite-sample test statistic are stable under small perturbations of the kernel parameter, provided a sufficient spectral gap exists. This stability implies that by restricting the test to the leading spectral components, our method naturally mitigates the sensitivity to kernel bandwidth that plagues full-rank MMD tests, offering a distinct mechanism for robustness compared to kernel aggregation.

\end{document}